\documentclass[a4paper,fleqn,usenatbib]{mnras}

\usepackage[T1]{fontenc}
\usepackage{ae,aecompl}
\usepackage{graphicx}	      
\usepackage{amsmath}	      
\usepackage{amssymb}	      
\usepackage{array}

\newcommand {\pc}             {\,\rm pc}
\newcommand {\kpc}            {\,\rm kpc}
\newcommand {\Mpc}            {\,\rm Mpc}
\newcommand {\Msun}           {\,\rm{M}_{\sun}}
\newcommand {\Myr}            {\,\rm Myr}
\newcommand {\Gyr}            {\,\rm Gyr}
\newcommand {\kelvin}         {\,\rm K}
\newcommand {\kms}            {\,\rm km/s}
\newcommand {\CellsPerSec}    {\,\rm{cells/s}}
\newcommand {\yt}             {\texttt{yt}}
\newcommand {\libyt}          {\texttt{libyt}}
\newcommand {\gamerone}       {\textsc{gamer-1}}
\newcommand {\gamer}          {\textsc{gamer-2}}
\newcommand {\flash}          {\textsc{flash}}
\newcommand {\enzo}           {\textsc{e}nzo}
\newcommand {\athena}         {\textsc{athena}}
\newcommand {\arepo}          {\textsc{arepo}}
\newcommand {\grackle}        {\textsc{grackle}}
\newcommand {\about}          {{\sim}}
\newcommand {\Nnode}          {N_{\rm node}}
\newcommand {\Nnoderef}       {N_{\rm node,ref}}
\newcommand {\Nbuf}           {N_{\rm buf}}
\newcommand {\Niter}          {N_{\rm iter}}

\newcommand {\sref}[1]        {Section~\ref{#1}}
\newcommand {\fref}[1]        {Fig.~\ref{#1}}
\newcommand {\tref}[1]        {Table~\ref{#1}}
\newcommand {\eref}[1]        {Equation~(\ref{#1})}

\newcommand {\be}             {\begin{equation}}
\newcommand {\ee}             {\end{equation}}

\title[GAMER-2]{GAMER-2: a GPU-accelerated adaptive mesh refinement code --
accuracy, performance, and scalability}

\author[H.-Y. Schive et al.]
  {Hsi-Yu Schive$^{1,7}$\thanks{E-mail: hyschive@gmail.com},
   John A. ZuHone$^{2}$,
   Nathan J. Goldbaum$^{1}$,
   Matthew J. Turk$^{3,4}$, \newauthor
   Massimo Gaspari$^{5}$\thanks{{\it Einstein} and {\it Spitzer} Fellow}
   and Chin-Yu Cheng$^{6}$
\\
$^{1}$National Center for Supercomputing Applications, University of
Illinois, Urbana-Champaign, IL, 61820, USA\\
$^{2}$Harvard-Smithsonian Center for Astrophysics, 60 Garden St.,
Cambridge, MA 02138, USA\\
$^{3}$School of Information Sciences, University of Illinois,
Urbana-Champaign, IL, 61820, USA\\
$^{4}$Department of Astronomy, University of Illinois,
Urbana-Champaign, IL, 61820, USA\\
$^{5}$Department of Astrophysical Sciences, Princeton University,
4 Ivy Lane, Princeton, NJ 08544-1001, USA\\
$^{6}$Department of Electrical and Computer Engineering, University of
Illinois, Urbana-Champaign, IL, 61820, USA\\
$^{7}$Institute of Astrophysics, National Taiwan University, 10617 Taipei, Taiwan
}

\date{Accepted XXX. Received YYY; in original form ZZZ}

\pubyear{2017}
\begin{document}
\label{firstpage}
\pagerange{\pageref{firstpage}--\pageref{lastpage}}
\maketitle

\begin{abstract}
We present \textsc{gamer-2}, a GPU-accelerated adaptive mesh
refinement (AMR) code for astrophysics. It provides a rich set of
features, including adaptive time-stepping, several hydrodynamic schemes,
magnetohydrodynamics, self-gravity, particles,
star formation, chemistry and radiative processes with
\textsc{grackle}, data analysis with \texttt{yt}, and memory pool for efficient object allocation.
\textsc{gamer-2} is fully bitwise reproducible. For the performance optimization, it adopts
hybrid OpenMP/MPI/GPU parallelization and utilizes overlapping
CPU computation, GPU computation, and CPU-GPU communication.
Load balancing is achieved using a Hilbert space-filling curve on a
level-by-level basis without the need to duplicate the entire AMR
hierarchy on each MPI process.
To provide convincing demonstrations of the accuracy and performance
of \textsc{gamer-2}, we directly compare with \textsc{e}nzo on
isolated disk galaxy simulations and with \textsc{flash} on galaxy
cluster merger simulations. We show that the physical results
obtained by different codes are in very good agreement, and
\textsc{gamer-2} outperforms \textsc{e}nzo and \textsc{flash} by
nearly one and two orders of magnitude, respectively, on the
Blue Waters supercomputers using $1-256$ nodes. More importantly,
\textsc{gamer-2} exhibits similar or even better parallel scalability
compared to the other two codes. We also demonstrate good weak and
strong scaling using up to 4096 GPUs and 65,536 CPU cores,
and achieve a uniform resolution as high as $10{,}240^3$ cells.
Furthermore, \textsc{gamer-2} can be adopted as an AMR+GPUs framework
and has been extensively used for the wave dark matter ($\psi$DM)
simulations.
\textsc{gamer-2} is open source
(available at \url{https://github.com/gamer-project/gamer})
and new contributions are welcome.
\end{abstract}

\begin{keywords}
methods: numerical -- hydrodynamics
\end{keywords}

\section{Introduction}
\label{sec:intro}

Many problems in computational astrophysics require resolving
structures at a wide range of spatial scales. For this reason, the
adaptive mesh refinement (AMR) method \citep{Berger1984,Berger1989}
has played an indispensable role by enabling high dynamic range
simulations of astrophysical phenomena. The fundamental principle in
AMR is to allow the simulation resolution, in both space and time, to
adaptively and locally adjust so as to concentrate computational
resources on regions requiring higher resolution. It is achieved by
first covering the entire computational domain with a
uniform-resolution grid, and then adding hierarchies of nested refined
grid patches (also referred to as `patches', `grids', or `blocks') with
decreasing cell spacing over subvolumes of interest.

There have been many hydrodynamic AMR codes for astrophysics
\citep[e.g.][]{Kravtsov1997,FLASH,Teyssier2002,Cunningham2009,
Schive2010,Almgren2010,Mignone2012,Almgren2013,Enzo,White2016}.
Among these, the AMR implementations can be broadly classified into
three categories based on the basic units adopted for grid refinement.
The first category uses rectangular patches of arbitrary aspect ratios
as the refinement units, where different patches can have different
sizes and shapes; for example, \enzo\ \citep{Enzo} utilizes this scheme. The second category performs
refinement on a cell-by-cell basis,
for example, \textsc{art} \citep{Kravtsov1997} and \textsc{ramses}
\citep{Teyssier2002}. The third category
performs refinement on patches of fixed size (e.g.\ $8^3$ cells); in
other words, all patches are restricted to be geometrically similar
to each other. Examples include \flash\ \citep{FLASH},
\textsc{gamer} \citep{Schive2010}, and \textsc{athena++}
\citep{Athena,White2016}.
In addition to AMR, it is also possible to achieve high resolution with
Lagrangian particles \citep[e.g. \textsc{gadget-2},][]{GADGET},
a moving unstructured mesh \citep[e.g. \arepo,][]{AREPO},
or a meshless method \citep[e.g. \textsc{gizmo},][]{GIZMO}.

Use of graphic processing units (GPUs) has recently become a promising
technique to achieve substantial speedups in simulation
performance. However, compared to the uniform-resolution approaches,
it remains extremely challenging for AMR codes to efficiently exploit
the petascale computing power in heterogeneous CPU/GPU supercomputers,
mainly due to the complicated AMR data structure, load imbalance,
and the great amount of work required to isolate and convert existing
physics modules to run with high-efficiency on GPUs. So far, only a
few astrophysical AMR codes have taken advantage of GPU
acceleration. For example, both \enzo\ \citep{Enzo} and
\textsc{ramses} \citep{Teyssier2002} have ported the hydrodynamic and
magnetohydrodynamic solvers to GPUs \citep{Wang2010,Kestener2010}, and
\flash\ has ported an octree Poisson solver to GPUs \citep{Lukat2016}.
\textsc{gamer} \citep[{\bf G}PU-accelerated {\bf A}daptive
 {\bf ME}sh {\bf R}efinement,][hereafter referred to as
 \gamerone]{Schive2010}, is the first astrophysical AMR code designed from scratch to
exploit GPU acceleration. It supports both GPU hydrodynamic and
Poisson solvers. However, the physical modules supported in
\gamerone\ are much more limited compared to other widely-adopted AMR
codes, which restricts possible applications of the code.

Here we present \gamer, a significant revision of \gamerone\
that includes much richer functionality, including adaptive time-stepping,
several hydrodynamic schemes, magnetohydrodynamics, dual energy
formalism, self-gravity, particles, star formation, chemistry and
radiative processes with the \textsc{grackle} library, data analysis
with the \texttt{yt} package, memory pool, bitwise reproducibility,
test problem infrastructure, and the capability of being used as an
AMR+GPUs framework.
It also incorporates significant improvements in accuracy, stability,
performance, and scalability.

For a GPU-accelerated
astrophysical AMR code there are at least three questions to be
addressed in order to make a fair performance comparison with a
CPU-only AMR code. First, does it sacrifice accuracy for performance?
Second, does it still outperform CPUs by a large margin when enabling
a rich set of physical modules? Third, how does the performance scale
with the number of GPUs especially when
running extremely large parallel simulations?
To provide clear and convincing answers to these questions, we directly
compare \gamer\ with two widely adopted AMR codes, \flash\ and
\enzo, based on realistic astrophysical applications, namely,
binary cluster merger simulations and isolated disk galaxy simulations,
where we enable GPU acceleration for \enzo\ as well.
These comparison simulations show that the physical
results obtained by different codes are in very good agreement,
and \gamer\ outperforms \enzo\ and \flash\ by nearly one and two
orders of magnitude, respectively, on the Blue Waters
supercomputer\footnote{https://bluewaters.ncsa.illinois.edu.}
using $1-256$ nodes. We also demonstrate good weak and strong
scaling in \gamer\ using up to 4096 GPUs and 65,536 CPU cores.

This paper is structured as follows. We describe the numerical
algorithms in \sref{sec:algorithm} and the performance optimizations
in \sref{sec:optimization}. \sref{sec:test} shows the code tests,
especially focusing on the comparison simulations with
\gamer, \enzo, and \flash. Finally, we summarize our results
and discuss future work in \sref{sec:summary}.

\section{Numerical Algorithms}
\label{sec:algorithm}

We provide in this section an overview of the numerical algorithms
implemented in \gamer, including the AMR structure, hydrodynamic
and gravity solvers, particle integration, and other miscellaneous
features. Detailed descriptions of the performance optimizations and
parallelization are given in \sref{sec:optimization}.

\subsection{Adaptive mesh refinement}
\label{subsec:AMR}

The AMR structure in \gamer\ is very similar to that in the original
\gamerone\ code, and therefore we only provide a
short summary here.
Note that in this paper we use the terms `patch', `grid', and
`block' interchangeably.
\gamer\ adopts a block-structured AMR where the
simulation domain is covered by a hierarchy of patches
with various resolutions. Patches with the
same resolution are referred to as being on the same AMR level, where
the root level, $l=0$, has the coarsest resolution. The resolution
ratio between two adjacent levels is currently fixed to 2. The levels
of any two nearby patches can differ by at most 1, so the \gamer\ AMR
hierarchy is always properly nested.

All patches are restricted to be geometrically similar to each other,
which is similar to the AMR implementation in \flash.
We assume a fixed patch size of $8^3$ cells throughout this paper
unless otherwise specified. The AMR hierarchy is manipulated by an
octree data structure. A patch on level $l$ can have zero or 8
child patches on level $l+1$ that cover the same physical
domain as their parent patch, and up to 26 sibling
patches (including those along the diagonal directions) on the same
level. For convenience, all patches store the patch identification numbers of their
parent, child, and sibling patches (if any exist).

\begin{figure*}
\centering
\includegraphics[width=13cm]{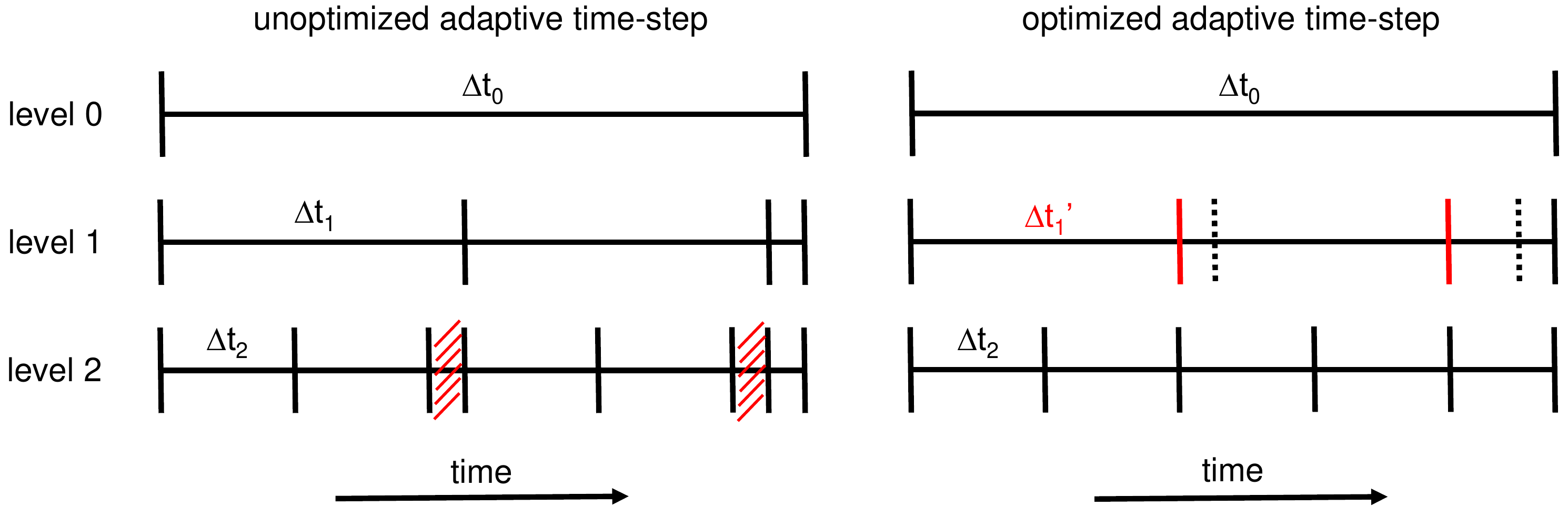}
\caption{
Illustration of the adaptive time-step integration. The left panel
shows the unoptimized scheme where the diagonal hatching highlights
the extra small time-steps on level 2 required to synchronize with
level 1. The right panel shows the optimized scheme where the
time-steps on level 1, $\Delta t_1'$, are reduced by a small fraction
in order to eliminate the extra synchronization time-steps on level 2.
The vertical dotted lines represent the original time-steps on
level 1. Note that the optimized scheme reduces the total number of
updates on level 2 from 7 and 5, while that on levels 0 and 1 remain
unchanged.
}
\label{fig:adaptive_dt}
\end{figure*}

For advancing patches on different levels, \gamer\ supports both the
shared time-step and adaptive time-step integrations. For the former,
all patches in the simulations are restricted to have the same
evolution time-step, which is generally more robust but less efficient
as patches on lower levels might have unnecessarily small
time-steps. This fixed time-step scheme is used by the \flash\ code.
For the latter, patches on higher levels can have smaller time-steps.
Furthermore, \gamer\ does not require the time-step ratio between two
adjacent levels to be a constant, which is similar to the implementation
in \enzo. Different levels are advanced in a way similar to the W-cycle
in a multigrid scheme, where the lower levels are advanced first and
then wait until they are synchronized with higher levels. This approach
can improve performance notably, especially when higher refinement
levels only cover small subvolumes of the simulation domain.

For the adaptive time-step integration, in order to synchronize two
adjacent levels, the finer level sometimes requires an extra update
with an extremely small time-step, which can have a non-negligible
impact on the overall performance. To alleviate this issue, we allow
the time-step on level $l$, $\Delta t_l$, to increase by a small
fraction (typically $C_{\rm inc} \about 10\%$) if that helps synchronize with
level ${l-1}$. Moreover, and less intuitively, we also allow $\Delta
t_l$ to \emph{decrease} by a small fraction (typically $C_{\rm dec} \about 10\%$
as well) if it could remove that extra small time-step on level
$l+1$. See \fref{fig:adaptive_dt} for an illustration of the second
optimization.
The procedure for computing the final optimum $\Delta t_l$ can be
described by the following pseudo-code:
\begin{verbatim}
// "lv" is the abbreviation of "level"
dt(lv) = ComputeTimeStep(lv)
dt_inc = t(lv-1) - t(lv)
dt_dec = 2*dt(lv+1)
// synchronize lv and lv-1
if (1+C_inc)*dt(lv)>dt_inc
   dt(lv) = dt_inc
// synchronize lv and lv+1
else if dt(lv)>dt_dec and (1-C_dec)*dt(lv)<dt_dec
   dt(lv) = dt_dec
\end{verbatim}
Note that $\Delta t_{l+1}$ is taken from the previous update under
the assumption that this value does not change too quickly, and the
coefficient 2 for computing $\Delta t_{\rm dec}$ further assumes that
$\Delta t_l$ is linearly proportional to the cell spacing on
level $l$. Also note that this procedure can be directly applied to
simulations with multiple levels.
In principle, allowing $\Delta t$ to vary by a larger fraction
(by increasing $C_{\rm inc}$ or $C_{\rm dec}$)
can
improve performance further, but it could also deteriorate accuracy
and stability.

A patch can be checked for refinement when synchronized with the child
level. \gamer\ supports a variety of refinement criteria, including,
for example, the amplitude, first derivative and second derivative
of simulation variables, vorticity, and Jeans length. The second
derivative criterion follows the error estimator suggested by
\citet{Lohner1987} and implemented in
\flash.
For the simulations with particles, one can also check the number of
particles in a patch, the number of particles in a cell, and the
particle mass in a cell. The refinement thresholds on different
levels can be set independently. A patch is flagged for refinement if
any of its cells satisfy the refinement criteria.
In addition, we add $(1+2N_{\rm buf})^3-1$ flag buffers,
where $N_{\rm buf}=0-8$,
around each flagged cell and refine the corresponding sibling patches
as well if any of these flag buffers extend across the patch border.
See Fig. 2 in \citet{Schive2010} for an illustration.

\gamer\ does not implement explicit grid derefinement criteria.
A patch is removed if its parent patch does not satisfy any
refinement criteria and if it has no child patches. In addition,
a patch is not refined if it would violate the proper-nesting
constraint. This approach makes grid refinement significantly more
efficient, since only one refinement level is altered at a time and
there is no need to reconstruct the entire AMR hierarchy, which
would otherwise be very expensive. However, it could also cause an
issue of insufficient resolution since patches might not be refined
in time due to the proper-nesting check. To solve this issue,
we typically set $N_{\rm buf}=8$ on all levels except for
$l_{\rm max}-1$, where $l_{\rm max}$ is the maximum refinement level,
so that patches on lower levels are always pre-allocated.
Having $N_{\rm buf}=8$ on level $l_{\rm max}-1$ is unnecessary
as no patches will be allocated above level $l_{\rm max}$, and thus
we typically set $N_{\rm buf}=2-4$ on level $l_{\rm max}-1$ just to
prevent the phenomenon of interest from leaving the highest resolution
regions. As will be demonstrated in \sref{sec:test}, this approach,
although not optimal, is found to do a reasonable job.

\subsection{Hydrodynamics}
\label{subsec:hydro}

\gamer\ supports four hydrodynamic schemes: the relaxing total variation
diminishing scheme \citep[RTVD;][]{Jin1995}, the MUSCL-Hancock scheme
\citep[MHM; for an introduction, see][]{Toro2009}, a variant of the
MUSCL-Hancock scheme \citep[VL;][]{Falle1991,vanLeer2006}, and the
corner transport upwind scheme \citep[CTU;][]{Colella1990}.
There are several variants of the CTU scheme in the literature, and
we adopt the one requiring 6 Riemann solvers per cell per time-step,
similar to that implemented in \athena\ \citep{Athena}.
The detailed implementation of these schemes have been described
previously \citep{Trac2003,Athena,Stone2009,Schive2010,Schive2012},
which we do not repeat here.
\gamer\ also supports magnetohydrodynamics\footnote{The MHD extension
has not been made publicly available yet but will be released soon.}
\citep{Zhang2018}
using the CTU scheme and the constrained transport
\citep[CT;][]{Evans1988} technique to ensure the divergence-free
constraint on the magnetic field, for which we closely follow the
implementation in \athena.

The RTVD scheme is dimensionally split
and Riemann-solver-free, and the other three schemes are dimensionally
unsplit and Riemann-solver-based. \gamer\ supports four Riemann solvers
for hydrodynamics, namely, exact solver based on \citet{Toro2009},
Roe's solver \citep{Roe1981},
HLLC solver \citep{Toro2009}, and
HLLE solver \citep{Einfeldt1991},
and three Riemann solvers for MHD, namely,
Roe's solver, HLLD solver \citep{Miyoshi2005}, and HLLE solver.
For the data reconstruction schemes,
we have implemented the
piecewise linear method \citep[PLM;][]{vanLeer1979} and
piecewise parabolic method \citep[PPM;][]{Woodward1984}, both of which
can be applied to either primitive or characteristic variables.
We have also implemented a variety of slope limiters, including the
generalized minmod limiter, van Leer-type limiter, van Albada-type
limiter, and a hybrid limiter combining the generalized minmod and
van Leer-type limiters \citep[for an introduction, see][]{Toro2009}.

All Riemann-solver-based schemes support
multispecies hydrodynamics by additionally solving the continuity
equations of an arbitrary number of chemical species,
which can be renormalized after every update to ensure that the sum
of their mass fractions equals unity.
The non-equilibrium chemical reaction network is updated by the library
\grackle\ \citep[][see \sref{subsec:grackle}]{Grackle}.

There are four standard fluid boundary conditions available in
\gamer, namely, periodic, outflow, inflow (i.e.\ user-defined), and
reflecting boundaries. See, for example, \citet{Enzo}, for
an introduction. The user-defined boundaries can be time-dependent.
The boundary conditions on different faces of the computational
domain can be set independently.

It is a well-known problem that the pressure of flows with very high
Mach number suffers from large truncation errors, which can be a serious
issue when the gas temperature is required (e.g.\ for calculating
chemistry and radiative cooling). The common remedy is to adopt the
so called `dual energy formalism', for which one evolves an
additional auxiliary variable, either thermal energy \citep{Bryan1995,Enzo}
or entropy \citep{Ryu1993,AREPO}, and uses that to calculate pressure or
temperature when required. \gamer\ adopts the gas entropy per unit
volume, $s \equiv P/\rho^{\gamma-1}$, where $P$ is gas pressure,
$\rho$ is gas mass density, and $\gamma$ is adiabatic index,
as the auxiliary variable since it is a conserved quantity outside
the shocks and therefore can be easily incorporated by regarding it as
one of the passively advected scalars. In the end of the fluid solver,
we check the ratio between the gas thermal energy $E_{\rm the}$ and
kinetic energy $E_{\rm kin}$ on each cell, where $E_{\rm the}$ is
estimated from the original total energy formulation
(i.e.\ $E_{\rm the} = E_{\rm tot} - E_{\rm kin}$ where $E_{\rm tot}$
is the gas total energy density). If this ratio
is below a given threshold, $E_{\rm the}/E_{\rm kin} < \xi$, where
$\xi \sim 10^{-2}$ typically, we correct $E_{\rm tot}$ using the entropy
calculated from the dual energy formalism. Otherwise, the auxiliary
entropy information on this cell is simply disregarded.

Unphysical results such as negative density and pressure may still
arise even with the help of the dual energy formalism, especially
when adopting a less diffusive hydrodynamic scheme (e.g.\ CTU scheme
and PPM reconstruction) in poorly resolved cold flows (e.g.\ when the
radiative cooling is very effective). If an unphysical result is
detected in a cell, \gamer\ adopts the following procedure to
try to remedy the problem.

\begin{enumerate}
\item If unphysical results occur in the intermediate region of the
      Roe's solver, we follow the approach adopted in \athena\ and
      switch to either HLLE, HLLC, or exact solver to recalculate the
      fluxes on the failed cell interfaces.
\item If unphysical results occur in the final update of the fluid
      solver, we recalculate the solutions with only first-order
      accuracy in space and time and in a dimensionally unsplit
      fashion.
\item If step (ii) still fails, we repeat step (ii) but with a
      dimensionally split update.
\item If step (iii) still fails, we reduce the time-step by a fixed
      ratio ($0.8$ by default) and recalculate all patches on the
      currently targeted level with this reduced time-step.
\item Repeat step (iv) until no unphysical results are detected or
      reaching a given minimum time-step threshold, for which the
      program is aborted.
\end{enumerate}

Note that steps (i) -- (iii) are applied only to the cells with
unphysical results, and steps (iv) and (v) are applied to all
patches on the currently targeted level. The latter is feasible
thanks to the adaptive time-step integration described in
\sref{subsec:AMR}. Moreover, the time-step in the next update is
automatically restored to the original value, and thus a maximum
time-step can be applied whenever possible to reach optimal
performance.

When a coarse level is synchronized with its child level, there are
two `fix-up' operations required to correct the coarse-grid data
in order to ensure the consistency between different levels.
First, for non-leaf coarse patches (i.e.\ patches with child patches),
their conserved quantities are replaced by the volume-weighted
average of the child patch data. This operation is sometimes referred
to as `restriction'. Second, for leaf coarse patches
(i.e.\ patches without child patches) near coarse-fine interfaces,
the conserved quantities on the coarse cells adjacent to the
interfaces are corrected by the differences between the coarse-grid
and fine-grid fluxes across the interfaces. These corrections ensure
conservation of conserved quantities to machine precision in pure
hydrodynamic simulations (see next section for the case with
self-gravity).

\subsection{Gravity}
\label{subsec:gravity}

Gravitational potential is evaluated by solving the discretized
Poisson equation subject to given boundary conditions. \gamer\
supports both periodic and isolated boundary conditions for gravity.
On the root level, we use the standard fast Fourier transform (FFT)
method with the FFTW package \citep{FFTW} to convolve mass density
with a proper Green's function in k-space and then transform back to
get real-space potential defined at the cell centers. For periodic
boundary conditions, the Green's function corresponds to a second-order
finite difference representation of the Laplacian operator
\citep{Hockney1988}. For isolated boundary conditions, the Green's
function takes into account proper zero-padding in the real space
\citep{Eastwood1979}, which is equivalent to regarding all cells as
point masses and solving their pairwise potential.

The gravitational potential in refined regions is calculated by solving the
discretized Poisson equation with the Laplacian operator replaced by
its second-order finite difference equivalent. We set the potential
boundary conditions by interpolating from the coarse patches
using quadratic interpolation in space (and linear interpolation in
time when adopting the adaptive time-step integration),
and then use relaxation methods to solve the discretized
Poisson equation to machine precision. \gamer\ by default uses the
successive over-relaxation \citep[SOR;][]{Press2007} method.
See \sref{subsec:gpu_implementation} for more discussions.

\begin{figure}
\centering
\includegraphics[width=\columnwidth]{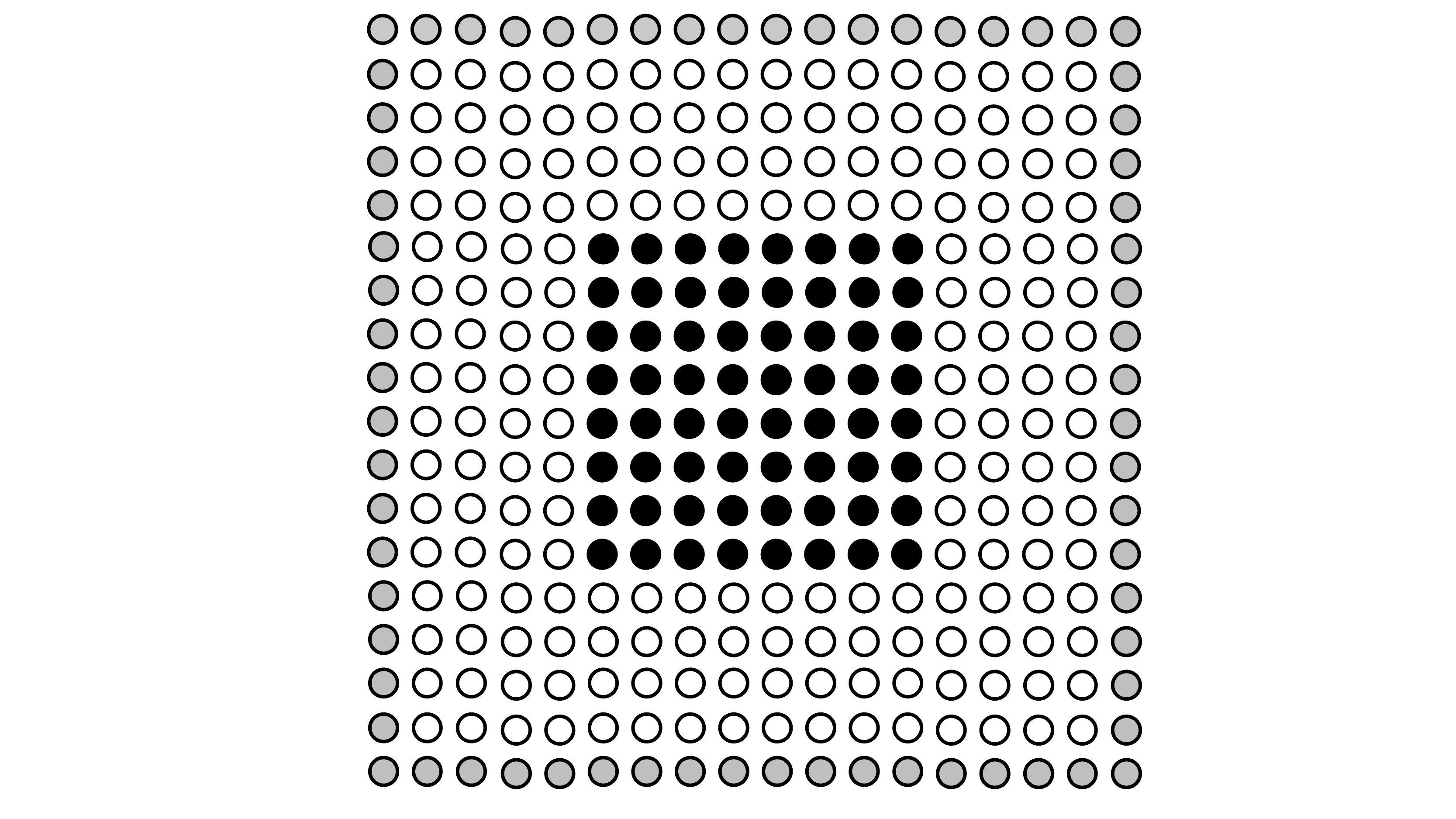}
\caption{
Two-dimensional example of the active and buffer zones adopted in
the Poisson solver used for refined regions. For each patch with $8^2$
active zones (central black-filled circles), we add 5 additional
buffer zones (open and grey-filled circles) around it to make the
potential smoother across the patch boundaries. The outermost cells
(grey-filled circles) provide the boundary conditions which are fixed
during the SOR iterations. Buffer zones are only
temporarily allocated for the patches being updated and are
discarded after being used for evaluating both the potential and
acceleration in the active zones.
}
\label{fig:poisson_grid}
\end{figure}

We calculate the gravitational potential of different patches on the
same refinement level independently in the sense that we do not
exchange solutions between nearby patches during the relaxation.
This approach leads to a significantly more efficient parallelization
compared to the multi-level Poisson solver adopted in \gamerone,
mainly because all patches on
the same level can be updated simultaneously.
Moreover, it can be applied straightforwardly to the adaptive
time-step integration since no correction from fine to coarse grids
is required.
However, it also results
in numerical errors accumulated on patch boundaries, including the
interfaces between patches both on the same level and on different
levels \citep[e.g.][]{Huang1999,Martin2000,Ricker2008}. To partially
alleviate this problem, we add 5 additional buffer zones around each
patch (see \fref{fig:poisson_grid}) and perform relaxation for the
entire $(8+2*5)^3=18^3$ grid, with the outermost cells being
fixed during the iterations. These buffer zones are only temporarily
allocated for the patches being updated and are discarded after
used for evaluating both potential and acceleration in the
active zones. This approach is found to
substantially reduce errors on patch boundaries, though not
completely eliminate them.
See Appendix \ref{sec:poisson} for details.
The comparisons with other codes based
on realistic astrophysical applications show satisfactory results
(see \sref{sec:test}). More complicated schemes dedicated to AMR
simulations \citep[e.g.][]{Ricker2008} will be investigated in the
future. Also note that although solving a large $18^3$ grid for each
$8^3$ patch might, at first glance, seem to have a great impact on the
performance, experiments show that our GPU Poisson and hydrodynamic
solvers achieve similar performance (see \sref{subsec:gpu_performance}).

Gravity is coupled to hydrodynamics in an operator-unsplit
predictor-corrector approach
identical to that implemented in \arepo\ \citep{AREPO} and
described by \citet{Muller1995},
which is second-order accurate.
First, we invoke the original hydrodynamic solver with the half-step
prediction of velocity incorporating gravity defined at the beginning
of the time-step. This gives us the proper mass density at the end
of the time-step, from which we can update gravity.
With the gravity both at the beginning and end of the time-step in hand,
we can correct momentum and then use the corrected momentum to
further correct energy, which completes the second-order accurate
updates of all fluid variables. Note that this procedure requires
only one invocation of Poisson solver per time-step.

For a given potential field, we evaluate the gravitational acceleration
at the cell centers via either two-point or four-point central
differencing along each direction. It can incorporate a user-defined,
time-dependent external acceleration. Note that the conservation of total
momentum and energy are prone to truncation errors when self-gravity is
included since it is not implemented in a conservative form
\citep[see e.g.][]{Jiang2013}.

\subsection{Particles}
\label{subsec:particles}

\gamer\ supports evolving active particles. We adopt the standard
`kick-drift-kick' (KDK) instead of `drift-kick-drift' (DKD) integrator,
since the former requires the gravitational acceleration at the
beginning and end of each time-step that is more consistent with the way
we incorporate gravity into hydrodynamic solvers
(see \sref{subsec:gravity}). Particles are associated with leaf
patches and updated together with these patches. They can be
dynamically created (e.g.\ for star formation) or removed (e.g.\ for
those leaving the computational domain) at runtime. The built-in
particle attributes include mass, positions, velocities,
accelerations, and time. We store the accelerations of each particle
to avoid redundant computations associated with the two kick
operators in the KDK scheme. We also record the physical time of each
particle since, for the adaptive time-step integration, particles are
not guaranteed to be synchronized with the associated leaf patches,
which will be discussed shortly. Particles can also carry an arbitrary
number of additional attributes (e.g.\ metallicity fraction).

When adopting the adaptive time-step integration, particles moving
across coarse-fine interfaces (after the `drift' operator) need to
be treated with special care since the coarse level is updated
before the fine level and with a larger time-step.
First, for particles moving from coarse to fine patches, we always
use the \emph{coarse}-grid gravity for the next kick operator to
avoid temporal extrapolation on the potential. Second, particles may
not be synchronized with their new associated leaf patches after
moving across coarse-fine interfaces. For particles moving from
coarse to fine patches, their physical time is always ahead of the
time associated with the new host patches. On the other hand, for
particles moving from fine to coarse patches, their physical time may
be behind the time of the new host patches. To solve this issue, we
record the physical time of all particles so that we can
synchronize each particle and its host patch in time by taking into
account their small time interval differences during the next particle
update.
It also allows us to `predict` particle positions at any
given time when required (e.g.\ for assigning particle mass onto
grids as discussed below).

Particles and fluid share the same gravitational force. Since \gamer\
adopts a grid-based gravity solver, one needs to not only deposit
particles onto grids to get the total mass density for the Poisson
solver, but also interpolate the cell-centered accelerations to particle
positions. We have implemented the standard nearest-grid-point (NGP),
cloud-in-cell (CIC), and triangular-shape cloud (TSC) schemes
\citep{Hockney1988} for this purpose. Special care needs to be taken
when assigning the buffer-zone densities (i.e.\ the open circles in
\fref{fig:poisson_grid}) outside the coarse-fine interfaces of fine
patches. Instead of depositing particle mass onto coarse grids and
then performing spatial interpolation, we deposit particles onto
the buffer zones of fine patches directly. This approach avoids the
possibility of double counting particles adjacent to the coarse-fine
interfaces. In addition, for the adaptive time-step scheme, particles
collected from both higher and lower levels need to be synchronized
with the targeted level before mass deposition.

There are several limitations in the current implementation of
particles. (i) Tracer particles following the motion of fluid are
not supported. (ii) Comoving coordinates for cosmological simulations
are not supported. (iii) There can only be a single particle type in
the sense that the number of particle attributes must be the same for
all particles. These restrictions will be removed in the near future.

\subsection{Chemistry and radiative processes}
\label{subsec:grackle}

\gamer\ supports multispecies hydrodynamics and
has incorporated the publicly available
\grackle\footnote{https://grackle.readthedocs.io.} chemistry and
radiative cooling/heating library \citep{Grackle}. \grackle\ supports
both a non-equilibrium solver with $6-12$ species and a tabulated
cooling function assuming ionization equilibrium for the primordial
chemistry and cooling. It also includes other features such as
tabulated metal cooling, photoelectric heating from dust, effective
cosmic microwave background (CMB) temperature floor, and various
ultraviolet (UV) background models. See \citet{Grackle} for details.

When updating a given AMR level, we transfer the species density and
internal energy of a fixed number of patches to \grackle\ at a single
time. It then evolves the chemical network with a semi-implicit method
(for the non-equilibrium solver), computes the radiative
cooling and heating rates, and updates the gas internal energy on a
cell-by-cell basis. The integrations are subcycled since the
characteristic time scales of chemistry and radiative processes
could be much smaller than those of hydrodynamics and particles.
We adopt an operator-split approach to couple the radiative cooling
and heating as energy source terms to a hydrodynamic scheme.

\subsection{Time-step}
\label{subsec:dt}

\gamer\ adopts various time-step criteria listed below:
\be
\Delta t_{\rm CFL1} <= C_{\rm CFL}\frac{\Delta h}{{\rm max}(|v_x|,|v_y|,|v_z|)+c_s},
\label{eq:dt_CFL1}
\ee
\be
\Delta t_{\rm CFL2} <= C_{\rm CFL}\frac{\Delta h}{|v_x|+|v_y|+|v_z|+3c_s},
\label{eq:dt_CFL2}
\ee
\be
\Delta t_{\rm acc} <= C_{\rm acc}\left(\frac{\Delta h}{{\rm max}(|a_x|,|a_y|,|a_z|)} \right)^{1/2},
\label{eq:dt_acc}
\ee
\be
\Delta t_{\rm par} <= C_{\rm par}\frac{\Delta h}{{\rm max}(|v_{{\rm par},x}|,|v_{{\rm par},y}|,|v_{{\rm par},z}|)},
\label{eq:dt_par}
\ee
where $\Delta h$ is cell spacing. For simplicity, here we assume that
all cells are cubic. $C_{\rm CFL}$, $C_{\rm acc}$, and $C_{\rm par}$
are the safety factors with typical values of $\about 0.5$.

Equations~(\ref{eq:dt_CFL1}) and~(\ref{eq:dt_CFL2}) specify the
Courant-Friedrichs-Lewy (CFL) condition of hydrodynamic schemes,
where $v_x$, $v_y$, $v_z$ are the fluid velocities and $c_s$ is the
sound speed. \eref{eq:dt_CFL1} applies to the RTVD and CTU schemes and
\eref{eq:dt_CFL2} applies to the MHM and VL schemes. Note that the CTU
scheme in \gamer, which invokes 6 Riemann solvers per cell per
time-step, requires $C_{\rm CFL} \le 0.5$, while the other three
schemes support $C_{\rm CFL} \le 1.0$.
\eref{eq:dt_acc} takes into account the accelerations resulting from
both self-gravity and external forces, where $a_x$, $a_y$, $a_z$ are
the accelerations of both fluid and particles. \eref{eq:dt_par}
prevents particles from traveling more than one cell width in a single
time-step, which is important for both improving accuracy and
simplifying particle manipulation.

For the shared time-step integration, we evaluate the time-step
constraints
of Equations~(\ref{eq:dt_CFL1})~--~(\ref{eq:dt_par})
on all levels and take the minimum value. For the
adaptive time-step integration, we only need to
evaluate these constraints for
the patches and particles on a given AMR level, with an additional constraint
that the physical time on a child level cannot be ahead of that
on its parent level. Moreover, as mentioned in \sref{subsec:AMR},
we allow the minimum time-step determined from
Equations~(\ref{eq:dt_CFL1})~--~(\ref{eq:dt_par}) to vary by a small fraction
(typically $\about 10\%$) to help synchronize adjacent levels.
Finally, as mentioned in \sref{subsec:hydro}, we further reduce the
time-step by a fixed ratio ($0.8$ by default) if the previously
adopted time-step led to unphysical results in the hydrodynamic
solver.

\subsection{Miscellaneous features}
\label{subsec:miscellaneous}

\subsubsection{Bitwise reproducibility}
\label{subsec:bitwise}

\gamer\ supports bitwise reproducibility, in the sense that the
round-off errors can be guaranteed to be the same when (i) running
simulations with different numbers of message passing interface (MPI) processes and OpenMP threads
(see \sref{subsec:hybrid}), and (ii) restarting simulations from checkpoint
files. This is a non-trivial task for any parallel AMR code,
especially with particles, since the order of all floating-point
operations needs to be carefully designed to be deterministic.
Specifically, \gamer\ performs the following additional calculations
when bitwise reproducibility is demanded:

\begin{itemize}
\item{
Recalculate gravitational potential before writing snapshots if
the potential data are not stored on disk. This ensures that the
potential and total mass density are fully consistent with the
adopted Poisson solver, which otherwise is not strictly guaranteed,
for example in non-leaf patches after applying the fine-to-coarse
data averaging (i.e.\ the `restriction' operation;
see \sref{subsec:hydro}) and in newly allocated patches whose potential
data are initialized by interpolation instead of Poisson solver.
}
\item{
Ensure that the differences between the coarse- and fine-grid
fluxes across the coarse-fine interfaces (which are used in the
flux correction; see \sref{subsec:hydro}) are computed in
a deterministic order even when the coarse- and fine-grid fluxes
are calculated separately by different MPI processes.
}
\item{
Ensure that the average total mass density in the entire
computational domain, which is required when solving the Poisson
equation in the comoving coordinates, is calculated in a
deterministic order.
}
\item{
Sort particles in the same patches by their spatial coordinates
before mass deposition.
}
\item{
Use the \texttt{static} OpenMP scheduling when necessary.
}
\end{itemize}

Note that, in general, \gamer\ does not support bitwise
reproducibility for simulations requiring random numbers. However,
it is achievable, for example, by having random number seeds be
a function of time and patch coordinates. We have implemented
it in the stochastic star formation model adopted in the isolated disk
galaxy simulations described in \sref{subsec:agora}.

Bitwise reproducibility is important for scientific reproducibility
and very helpful for debugging. However, it can also deteriorate
performance, especially for simulations requiring a large number of
particles due to the extra particle sorting. Therefore, we implement
this feature as a compile-time option, and disable it throughout
\sref{sec:test}. Also note that the bitwise reproducibility addressed
here does not apply to the results using different processors (e.g.\
different CPUs or GPUs), different compilers, or even different
compilation flags.

\subsubsection{Data analysis}
\label{subsec:analysis}

\gamer\ supports two file formats for writing simulation snapshots:
a simple binary format and HDF5\footnote{https://support.hdfgroup.org/HDF5.}.
HDF5 is our preferred format as it is more extensible,
portable, and easier for post-processing. We currently store each
snapshot into a single file and have different
MPI processes dump data serially. Writing data in parallel
and into multiple files will be investigated in the future.
Snapshots in both formats can be used for restarting simulations and
can be loaded in parallel.

The HDF5 snapshots of \gamer\ can be loaded by
\yt\footnote{http://yt-project.org.} \citep{yt},
a powerful, publicly available, python-based package for
analyzing and visualizing volumetric data. \yt\ supports multiple
simulation codes (e.g.\ \flash\ and \enzo), which is indispensable
for this work as it provides direct and fair comparisons between the
simulation results of different codes. It also allows one to share
the data analysis scripts to the community straightforwardly, which
greatly improves the scientific reproducibility. In addition, for
conducting inline analysis, we are experimenting with
\libyt\footnote{https://bitbucket.org/data-exp-lab/libyt.}, a runtime
interface for passing in-memory \gamer\ data structures to yt.

To help monitor the simulation status, \gamer\ supports recording a
rich set of simulation information after each root-level update
(i.e.\ after all levels are synchronized),
which can be used as diagnostics during as well as after
the simulations:

\begin{itemize}
\item{
Detailed timing analysis of all major routines
(such as hydrodynamics, gravity, particles, MPI, etc.).
}
\item{
Maximum and average memory consumption per MPI process.
}
\item{
Overall performance in terms of total cell updates and particle
updates per second.
}
\item{
Number of patches and particles on each level.
}
\item{
Estimation of load imbalance.
}
\item{
Errors in the conserved quantities (e.g.\ gas mass and momentum).
}
\item{
Message size and achieved bandwidth in various MPI calls.
}
\item{
Evolution time-steps on each level estimated from various
constraints (see \sref{subsec:dt}).
}
\end{itemize}

\subsubsection{Test problem infrastructure}
\label{subsec:test_problem}

The test problem infrastructure in \gamer\ is designed in a way that
each problem is self-contained and largely decoupled from the
complicated AMR structure and parallelization. To add a new simulation,
in most cases, one does not have to touch the existing source code
(except for defining a new problem identification). Instead, one
creates a new problem directory and specifies the gas and particle
initial conditions therein. The gas quantities can either be set by
an analytical function of space and time or loaded from a
binary file with a uniform-resolution array. The particle attributes
can also either be set dynamically or loaded from a binary file.
In addition, one can specify various problem-specific runtime
parameters and functionalities, for instance, refinement criteria,
time-step constraints, boundary conditions, and external accelerations.

\subsubsection{AMR+GPUs framework}
\label{subsec:framework}

Although \gamer\ is mainly designed for astrophysical hydrodynamic
simulations, it can also be adopted as a high-performance,
multi-science AMR framework powered by multi-GPU acceleration
\citep{Shukla2011}. It is because the AMR structure, parallelization,
and performance optimizations (see \sref{sec:optimization}) are
implemented carefully to be largely independent of the partial differential
equations (PDE) being solved. As a successful example, the code has been
extensively used for the wave dark matter ($\psi$DM) simulations
\citep{Schive2014a,Schive2014b,DeMartino2017}, where we solve the
Schr\"{o}dinger-Poisson equation by replacing the hydrodynamic solver
with a quantum-mechanical kinematic energy solver and by reusing the
same Poisson solver. The detailed implementation of this work will be
described elsewhere (Schive et al., in preparation).

\section{Performance optimizations}
\label{sec:optimization}

In this section, we focus on various performance optimization
strategies in \gamer, including the GPU implementation, hybrid
MPI/OpenMP/GPU parallelization, load balancing, and memory
management.

\subsection{GPU implementation}
\label{subsec:gpu_implementation}

We use CPUs to manipulate the AMR data structure and only port the
time-consuming routines to GPUs, currently including the hydrodynamic
solvers, Poisson solvers, and time-step calculations. This approach
takes advantage of both CPUs and GPUs, allowing a substantial
performance improvement compared to a calculation using only CPUs
without sacrificing the flexibility and extensibility of the code. In
addition, it allows us to store all the data in the CPU memory and
only temporarily transfer a small portion of data to the relatively small
GPU memory. We use CUDA \citep{CUDA} as the GPU programming interface.

We have implemented all the hydrodynamic solvers mentioned in
\sref{subsec:hydro} on GPUs, namely, the RTVD, MHM, VL, and CTU
schemes. See \citet{Schive2010,Schive2012} for the detailed implementation.
Note that the RTVD scheme takes advantage of the fast GPU shared
memory while the others only use the GPU global memory, since
the latter schemes are dimensionally unsplit which makes using
the small shared memory less straightforward. However, we still find
that the MHM, VL, and CTU schemes achieve significantly larger
performance speedups compared to the RTVD scheme, conceivably because
the Riemann-solver-based schemes have much higher arithmetic intensity
and thus are more GPU-friendly. See \sref{subsec:gpu_performance}
for the performance benchmarks of various CPU and GPU solvers.

The SOR solver mentioned in \sref{subsec:gravity} has been ported to GPUs.
We have abandoned the complicated scheme implemented in \gamerone\ that
utilizes \emph{both} the fast GPU shared memory and the per-thread
registers to reduce the shared memory usage. Modern GPUs have a
significantly larger shared memory (at least 48 KB per multiprocessor),
and thus we can simply store the gravitational potential of the entire
$18^3$ grid into the shared memory to boost the performance.
To reduce the communication between CPUs and
GPUs, we transfer the coarse-grid potential to GPUs, and then perform
spatial interpolation on the GPU to set both the boundary conditions and
initial guess of the potential solution for subsequent iterations. Obtaining an
initial guess of the potential on the entire $18^3$ grid, although
unnecessary, accelerates convergence.
Also note that we calculate the cell-centered gravitational
acceleration and use that to update fluid variables on GPUs
immediately after solving the potential, which helps reduce the amount
of data transferred between CPUs and GPUs.

We also use GPUs to compute the time-step constraints on grids,
i.e.\ Eqs. (\ref{eq:dt_CFL1}) -- (\ref{eq:dt_acc}), which otherwise
would take a surprisingly large fraction of simulation time (e.g.\
see the timing results of various operations shown in the end of
Sections~\ref{subsec:merger_performance} and~\ref{subsec:agora_performance}).
However, currently we have not ported any particle routines to GPUs,
which will be investigated in the future.

\gamer\ stores all the
data on the CPU's main memory, including both the AMR structure and physical data,
and only sends the data of patches being updated to GPUs.
Moreover, when updating a given level, it is usually unnecessary to
transfer all patches on this level to GPUs as long as the performance
has saturated. Therefore, the GPU memory consumption
can be greatly reduced, and, more importantly, is largely
independent of the simulation scale. However, one drawback of this
approach is that we need to transfer data between CPUs and GPUs
frequently, which can be expensive. To mitigate this issue, we
utilize CUDA streams \citep{CUDA} to overlap CPU-GPU communication
by both CPU and GPU computations
(see Fig. 5 in \citet{Schive2010} for an illustration), which can lead
to a factor of 2 speedup in the GPU solvers (see \sref{subsec:gpu_performance}).
In addition, as a result of the octree data structure, we can always group
eight sibling patches that share the same parent into a larger grid
(referred to as a `patch group'). Manipulating on patch groups instead
of individual patches reduces both the computation and communication
overhead associated with the ghost zones of each patch.

All GPU solvers in \gamer\ support single and double precision,
as a compile-time option. On high-end GPUs (e.g.\ P100 and V100),
single-precision performance is about 2 times faster than
double-precision performance, but this ratio can be noticeably
higher on older GPUs. We typically find that single precision provides
a satisfactory accuracy, as demonstrated in the comparison simulations
shown in Sections~\ref{subsec:merger_performance}
and~\ref{subsec:agora_performance}, except for applications requiring
either an extremely large dynamic range or resolving extremely small
perturbations. Therefore, we adopt single precision throughout this
paper unless otherwise specified.

\subsection{Hybrid MPI/OpenMP/GPU}
\label{subsec:hybrid}

In \gamer, only the most time-consuming routines are ported to GPUs,
and the code still uses CPUs extensively for various tasks, including,
for example, manipulating the AMR structure, checking the grid
refinement criteria, depositing particle mass onto grids, and updating
particle attributes. Moreover, even for those GPU-accelerated
routines, we still need to use CPUs to collect data from different
patches and fill the ghost zones of each patch (or patch group) by
either copying directly from sibling patches or interpolating from
coarse patches. Therefore, it is essential to efficiently exploit
both the CPU and GPU computing power in order to achieve optimal
overall performance.

\begin{figure*}
\centering
\includegraphics[width=\textwidth]{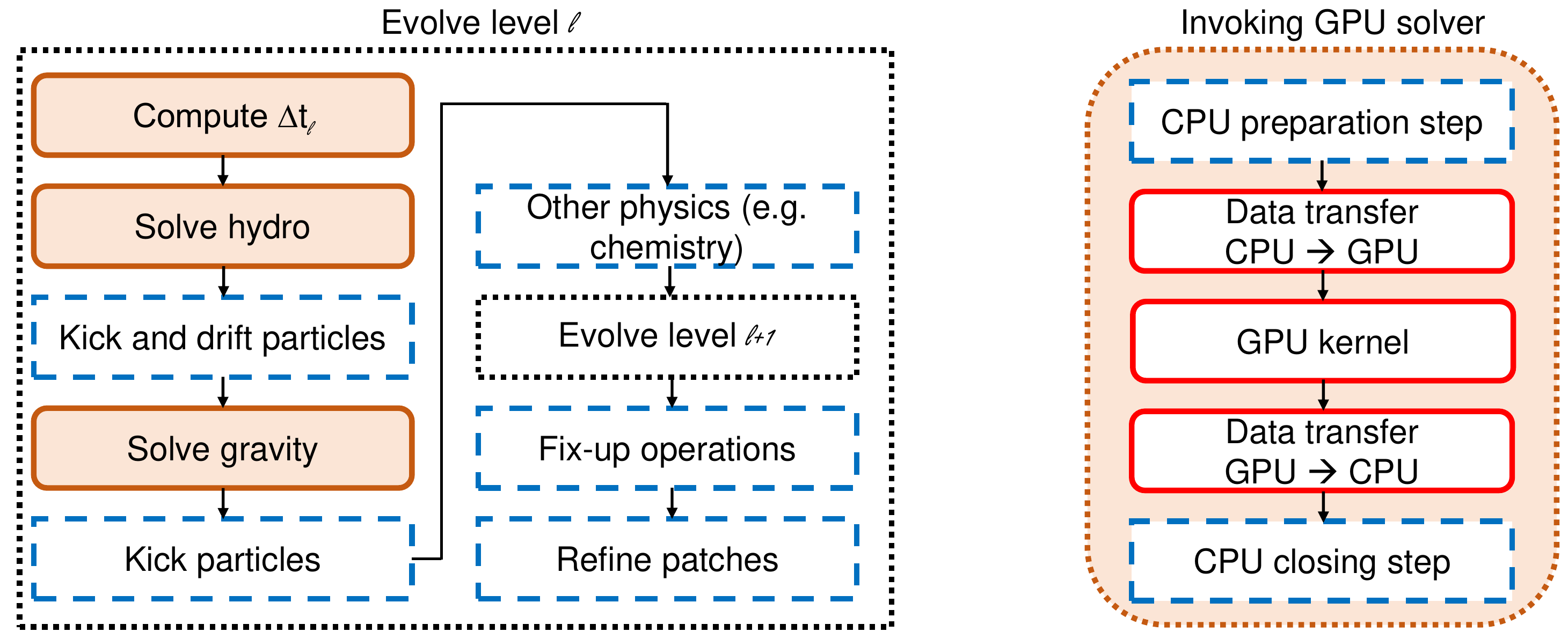}
\caption{
Illustration of the hybrid OpenMP/GPU parallelization in a single
MPI process.
Left panel: flowchart of the main loop for evolving a given AMR level
$l$, where dashed boxes are CPU-only operations parallelized with
OpenMP and solid boxes are GPU-accelerated solvers.
Right panel: procedure for invoking a GPU solver, such as computing
$\Delta t$ and solving hydrodynamics and gravity as indicated in the left
panel. Dashed boxes are CPU-only operations parallelized with OpenMP
(see Section 3.1 of \citet{Schive2010} for the definitions of CPU
preparation and closing steps) and solid boxes are GPU-related
operations. Note that both the GPU kernel execution and CPU-GPU
communication are asynchronous and can be overlapped with CPU
computation as long as they are targeting different patches.
See also Figs. 5 and 9 in \citet{Schive2010}.
}
\label{fig:flow_chart}
\end{figure*}

To this end, we have implemented a hybrid MPI/OpenMP/GPU parallelization
model. In addition to the GPU acceleration described in the previous
section, we further adopt OpenMP for intra-node parallelization
of all time-consuming CPU routines and MPI for inter-node
communication.
\fref{fig:flow_chart} shows an illustration.
This approach allows the code to fully exploit the
computing power in heterogeneous CPU/GPU supercomputers.
Moreover, the hybrid MPI/OpenMP implementation can significantly
improve the parallel scalability by reducing the amount of MPI
communication, especially when using a large number of nodes. It can
also reduce the CPU memory overhead associated with MPI buffers.

Note that the \grackle\ library used for solving chemistry and
radiative processes also supports OpenMP \citep{Grackle},
and therefore can be easily incorporated into the hybrid MPI/OpenMP
parallelization model adopted here. A GPU-accelerated version of
\grackle\ is under development.

To boost the performance further, we have utilized the asynchronous
feature of GPU kernels to let CPUs and GPUs work concurrently.
Specifically, we divide all patches on a given level to several
subsets, and use CPUs to prepare the input data of one subset
and GPUs to update a different subset simultaneously. See Fig. 9 in
\citet{Schive2010} for an illustration.
The number of patches in a single subset is chosen to saturate the
GPU performance (see \sref{subsec:gpu_performance}) and is set by
default to the product of the number of CUDA streams and the number
of multiprocessors in the adopted GPU.
We find that substantial
performance improvement up to a factor of 2 can be achieved by
this approach, especially for simulations where the CPU and GPU
computation times are comparable.

There is no restriction on the number of MPI processes per node and the
number of OpenMP threads associated with each MPI process. Typically,
we set the number of MPI processes equal to the total number of GPUs,
and then determine the number of OpenMP threads from the ratio
between CPU cores and GPUs. However,
empirically we have found that launching multiple MPI processes
to access the same GPU using the CUDA Multi-Process Service (MPS)
can improve the performance.
It is also important to take into account thread affinity
and non-uniform memory access (NUMA). Generally, it is recommended
to have OpenMP threads running in the same NUMA domain to improve
memory affinity. But one needs to experiment with different
configurations to fine-tune the overall performance.

Note that \gamer\ can also run in a `CPU-only' mode, since for all
GPU solvers we have implemented their CPU counterparts. These `CPU'
solvers are parallelized with OpenMP, with different threads
calculating different patches (or patch groups). This parallelization
method is found to be very efficient since the computational workload
associated with each patch is not only balanced
(when disregarding particles) but also generally much larger than the
OpenMP overhead. Moreover, the same MPI implementation can be applied
to both GPU-accelerated and CPU-only simulations. Therefore, \gamer\
is also suitable for CPU-only supercomputers, particularly for those
with a larger number of cores per node
(e.g.\ Intel Xeon Phi Knights Landing; KNL), for which hybrid MPI/OpenMP
is essential to get optimal performance. We will further investigate
and optimize this promising feature in the future.

\subsection{Load balancing}
\label{subsec:load_balancing}

Load balancing is crucial for parallel scalability. For a given level,
we use a Hilbert space-filling curve to map the three-dimensional (3D)
coordinates of all patches on this level onto a one-dimensional (1D)
curve. We assign a weight to each patch, which estimates its
computational workload, and then cut the curve into $N_{\rm process}$
segments with approximately equal weights, where $N_{\rm process}$ is
the total number of MPI processes. Load balancing can then be achieved
by having different MPI processes calculate patches on different curve
segments.

Since some PDE solvers use patch groups
(a patch group is defined as the set of eight sibling patches
that have the same parent in the octree data structure)
instead of patches as the
basic unit (e.g.\ hydrodynamic solvers), we require the eight patches
belonging to the same patch group to be located in the same MPI process.
However, we do not require parent and child patches to be located in
the same process,
which allows patches on different levels to be distributed on different
MPI processes in a completely independent fashion. Accordingly, we can
adopt different and independent Hilbert curves on different levels and
then use the Hilbert curve of a given AMR level to assign unique 1D
indices for all patches on that level. By doing so, the load balancing
can be achieved on a level-by-level basis. This feature is particularly
important for the adaptive time-step integration, since in that case
patches on different levels in general cannot be evolved simultaneously.
In addition, the Hilbert curve mapping between 3D and 1D space
preserves locality, meaning that, in general, patches close along
a 1D curve are also close in 3D space.
This property reduces MPI communication and thus is very important
for achieving higher parallel scalability.

We do not duplicate the entire AMR hierarchy on each MPI process.
Instead, for each process, we only allocate its `real' patches that
store the physical data (i.e.\ fluid variables and particles)
to be updated by this process, and the `buffer' patches necessary for
filling the ghost zones of these real patches by either direct copy
or interpolation. Specifically, for each real patch, we examine its
sibling, parent, and parent-sibling patches
(i.e.\ sibling patches of a parent patch), and allocate the
corresponding buffer patches locally if any of these patches exist in
the complete AMR hierarchy but do not live on the same MPI process as
the targeted real patch. These buffer patches allow each process to
correctly identify the nearby AMR structure (e.g.\ finding the
coarse-fine interfaces) and to retain a copy of physical data
received from other MPI processes for filling the ghost zones of real
patches. By avoiding duplication of the entire AMR hierarchy, we can
reduce CPU memory consumption and improve parallel scalability,
particularly when running extremely large parallel simulations.

\begin{figure}
\centering
\includegraphics[width=\columnwidth]{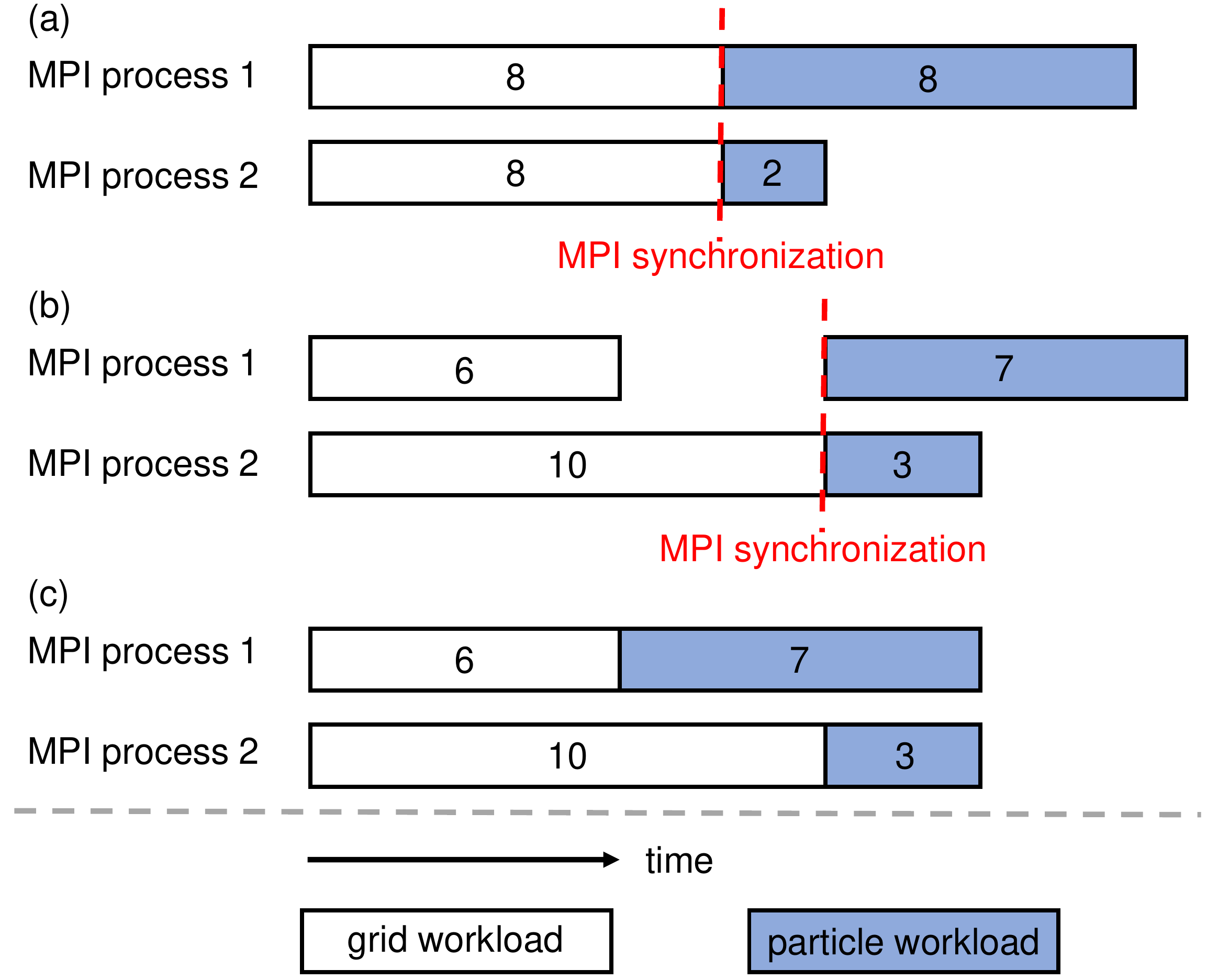}
\caption{
Illustration of load balancing in the simulations with particles.
Open and filled rectangles represent the computational workloads
associated with grids and particles, respectively. Case (a) does not
take into account particle weights, and thus the total workloads
of two MPI processes are imbalanced. Case (b) considers both
grid and particle weights. However, there is an undesirable MPI
synchronization between grid and particle routines, leading to
even worse performance compared to case (a). Case (c) further
eliminates this MPI synchronization and gives the optimal
performance.
}
\label{fig:particle_load_balancing}
\end{figure}

For the simulations without particles, it is relatively
straightforward to achieve good load balancing since the
computational workloads of different patches are roughly the same.
However, it becomes much more challenging when particles are involved,
not only because different patches now may have different workloads
due to the different numbers of particles associated with them, but
because there may be unavoidable MPI synchronization between grid
and particle routines. See \fref{fig:particle_load_balancing} for an
illustration. To partially alleviate this problem, we assign a
relative weight to each particle, $W_{\rm par}$, which is normalized
to the weight per cell, when estimating the total weight of each patch
for load balancing. Moreover, we arrange the order of various grid
routines (e.g.\ hydrodynamic and gravity solvers), particle routines
(e.g.\ mass deposition and particle update), and MPI communication
(e.g.\ filling the grid data of buffer patches and transferring
particles between neighboring patches) carefully so as to minimize
the MPI synchronization between grid and particle routines.
We typically adopt $W_{\rm par} = 1.0-2.0$. The optimal value depends
on the adopted physics, for example, whether or not the radiative
library \grackle\ is included. See
Sections~\ref{subsec:merger_performance} and~\ref{subsec:agora_performance}
for some comparisons between the simulation performance with and
without applying these optimizations.
Also note that chemistry and radiative cooling/heating (e.g.\ when
using \grackle) may also lead to widely different costs from cell to
cell and deteriorate load balancing.

Note that, when estimating the load-balancing weight of a non-leaf
patch, it is necessary to take into account all particles occupying
the same space as this targeted patch, even though these particles
live on leaf patches.
It is because we still need to temporarily transfer the masses and
positions of particles from leaf to non-leaf patches when calculating
the total mass density on a lower level. Ignoring the weights of
higher-level particles may result in an undesirable situation where
a large fraction of particles are temporarily transferred to a small
fraction of
MPI processes when depositing particles onto lower-level grids,
which could lead to severe load imbalance and memory exhaustion.
See also \sref{sec:summary} for a possible alternative solution.

Also note that it may be impossible to completely eliminate the MPI
communication between grid and particle routines. For example, after
updating particle positions, one must transfer the particles moving
across patch boundaries to neighboring patches before invoking the
Poisson solver. We will investigate other more advanced approaches,
for example, using MPI non-blocking communication
\citep[e.g.\ Enzo;][]{Enzo} or task-based parallelism model
(e.g.\ \textsc{athena++}\footnote{http://princetonuniversity.github.io/athena.})
in the future.

As a side note, we reuse large MPI send and receive buffers when
applicable. In other words, we do not deallocate these buffers after
performing a MPI communication. Instead, we reuse them for the next
communication provided that their sizes are large enough. We find that
this approach improves the achieved bandwidth in some circumstances.

\subsection{Memory management}
\label{subsec:memory}

AMR simulations require allocating and deallocating grids and
rebuilding the data structure frequently, which can easily lead to
memory fragmentation that deteriorates performance and exhausts
memory. \gamer\ supports using a local memory pool for each MPI process to solve this problem.
However, unlike conventional methods, it does not require
preallocating this memory pool. Instead, it relies on the \emph{reuse} of
allocated memory. Specifically, when removing patches from the AMR
hierarchy after the derefinement operation, we simply mark these
patches as `inactive' but do not actually free the memory associated
with them. These inactive patches can then be `reactivated' later
to serve as new patches, which can be either real or buffer patches,
after the refinement operation.
In comparison to the method of preallocating a large memory pool
(which is also supported in \gamer), this approach is more flexible
since it eliminates the need for users to guess in advance the maximum
number of patches required.
On the contrary, the size of the memory pool will be adjusted
automatically to fit the requirement.

We also allocate a separate memory pool for each MPI process to store all particle
attributes, and have leaf patches only record the particle
indices associated with them. By doing so, when particles moving
between patches living on the same MPI process, we only have to
update the particle index list of relevant leaf patches and do
not have to touch the particle memory pool. In addition, for particles
traveling to different host processes, we simply mark these particles
as `inactive' in their original host processes but do not actually free the memory associated
with them, which is similar to the method adopted in the grid memory
pool. These inactive particles can then be reactivated later to
represent, for example, particles migrating from other processes or
new particles triggered by star formation. Moreover, to further
minimize memory management due to particle movement, we preallocate
small memory buffers ($10\%$ of the total number of particles tracked
by a given MPI process by default) for both the particle memory
pool and the particle index list of each leaf patch. For example,
for a process owning $1000$ particles, we can preallocate a particle
memory pool with the size of $1100$ particles to reduce frequent memory
reallocation caused by the migration of a small number of particles.
Also note that the size of the particle memory pool will be adjusted
automatically based on the number of particles currently hosted by
each process, and thus users do not have to guess in advance
the maximum number of particles per process during the entire
simulation.

The ghost zones associated with each patch can lead to severe memory
overhead. For example, the CTU scheme with PPM reconstruction needs
3 ghost zones. Accordingly, for a patch consisting of $8^3$ cells,
the total memory consumption including the ghost zones is
$\about 5.4$ times larger than that without the ghost zones.
To solve this problem, \gamer\ does not permanently allocate the
ghost zones of all patches. Instead, it only temporarily allocates the
ghost zones for the patches being transferred to GPUs, the number of
which can be fixed (typically a few thousand) and is independent of
the simulation scale. This approach also solves the issue of the
relatively small GPU memory since the code does not need to send
all patches to GPUs at a single time.

\section{Code Tests}
\label{sec:test}

In this section, we conduct various tests to demonstrate the performance and
accuracy of \gamer. Since the numerical algorithms of \gamer\
(see \sref{sec:algorithm}) have been tested extensively by many astrophysical codes
\citep[e.g.][]{FLASH,Athena,Enzo}, we do not repeat the analysis of these
standard numerical tests here (see, however, \citet{Schive2010, Schive2012}
for the standard tests conducted previously). Instead, we directly
compare \gamer\ with two widely adopted codes, namely, \flash\ \citep{FLASH}
and \enzo\ \citep{Enzo}, based on more complicated and realistic astrophysical
applications, which arguably provides much more direct and convincing
results.

This section is organized as follows.
We first measure the performance of individual GPU solvers
(\sref{subsec:gpu_performance}) and the weak scaling with and without AMR
in a three-dimensional Kelvin-Helmholtz instability test
(\sref{subsec:weak_scaling}). We then compare the accuracy and
strong scaling performance of \gamer\ with \flash\ in binary cluster merger
simulations (\sref{subsec:merger}), and with \enzo\ in isolated disk
galaxy simulations (\sref{subsec:agora}).

\subsection{Performance of GPU solvers}
\label{subsec:gpu_performance}

The key feature of \gamer\ is GPU acceleration. So we first measure the
performance of individual GPU solvers,
which is largely independent of the adopted test problems.
This performance includes the time of
transferring data between CPU and GPU but excludes the time of all
other CPU operations related to AMR and MPI
communication. It thus represents the optimal performance of \gamer,
which can only be approached in certain particular cases
(e.g.\ large uniform-grid simulations) according to Amdahl's law \citep{Amdahl1967}. This information can be useful for
assessing the performance deterioration in more complicated simulations.

\begin{figure*}
\centering
\includegraphics[width=12cm]{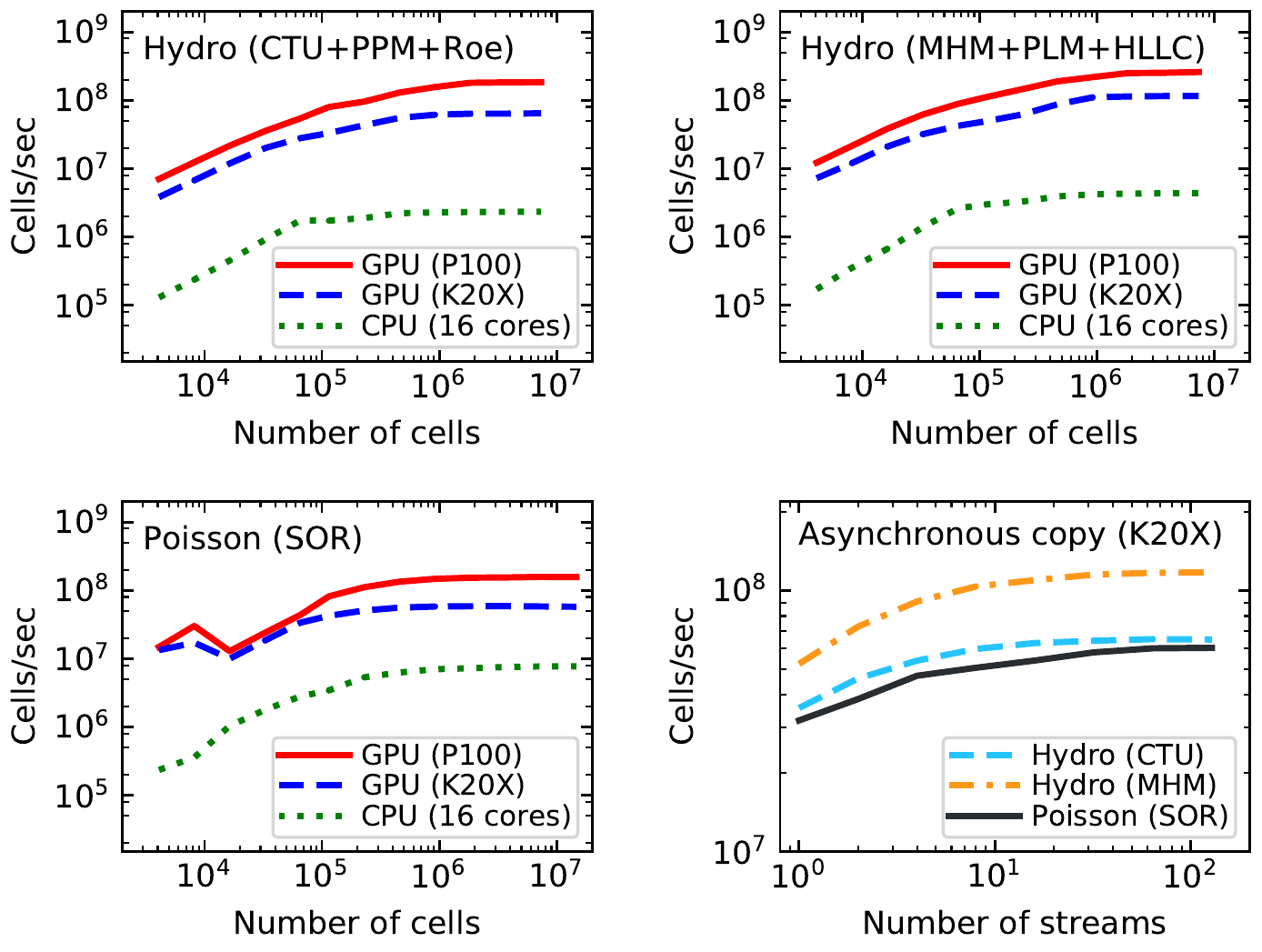}
\caption{
Performance in cell updates per second for individual GPU solvers and their
CPU counterparts. The first three panels show the performance of the
hydrodynamic solver with the CTU scheme, PPM reconstruction, and Roe's
solver (upper left), the hydrodynamic solver with the MHM
scheme, PLM reconstruction, and HLLC solver (upper right), and
the Poisson solver with the SOR scheme (lower left) as a function of the
number of cells updated at a time. The GPU performance is measured on both
NVIDIA Tesla K20X and P100-PCIe GPUs, and the CPU performance is measured
on a 16-core AMD Opteron 6276 processor (using all 16 cores). The P100 GPU achieves a performance
of $\about 2\times10^8 \CellsPerSec$ in all three solvers. We also measure the
K20X GPU performance as a function of the number of CUDA streams (lower right),
demonstrating a factor of 2 speedup when utilizing the asynchronous data
transfer between CPU and GPU.
}
\label{fig:gpu_performance}
\end{figure*}

\fref{fig:gpu_performance} shows the performance in cell updates per
second for individual GPU solvers and their CPU counterparts using exactly
the same numerical schemes. The performance is measured on a Blue Waters
XK node with an NVIDIA Tesla K20X GPU and a 16-core AMD Opteron 6276 CPU.
We also measure the performance on an NVIDIA Tesla P100-PCIe GPU.
The key findings can be summarized as follows.

\begin{itemize}
\item{
The K20X GPU is measured to be $27$ and $8$ times faster than
the 16-core CPU for the hydrodynamic and Poisson solvers, respectively.
Furthermore, the P100 GPU is measured to be $2.3-3.0$ times faster
than the K20X GPU, achieving $\about 2\times10^8 \CellsPerSec$ in all three
solvers.
}
\item{
The GPU performance already begins to saturate when updating only $\about 10^6$
cells at a time. It is typically much smaller than the total number of
cells computed by each MPI process in a real astrophysical application,
suggesting that we only need to transfer a small fraction of cells to GPU
at a time to fully exploit the GPU acceleration. This important property
allows for (i) efficient overlapping between CPU and GPU computations
(see \sref{subsec:hybrid}) and (ii) efficient overlapping between CPU-GPU
communication and CPU/GPU computations.
}
\item{
The GPU performance increases by a factor of 2 by taking advantage of
the asynchronous data transfer between CPU and GPU with CUDA streams.
The performance saturates when using more than $\about 10-20$ streams.
}
\item{
The GPU hydrodynamic and Poisson solvers exhibit comparable
performance ($\about 5 \times 10^7$ -- $2 \times 10^8$ cell updates per second).
}
\item{
The performance of CPU hydrodynamic solvers seems to be relatively low
compared to other codes, for example, \athena\ and
\textsc{ramses}, partially because of the
different CPU and the smaller patch size ($8^3$ cells) adopted in
this work. More quantitative comparisons and further optimizations will be investigated.
}
\end{itemize}

The small patch size ($8^3$ cells) adopted throughout this paper
results in a considerable computational overhead associated with the
ghost zones of each patch (although it allows for more flexible grid
refinement and efficient load balancing).
Increasing the patch size to $16^3$ cells is found to
improve the performance of the hydrodynamic GPU solvers by $\about 20\%$.
In addition, compared to the single-precision performance, adopting double
precision on the P100 GPU is measured to be $2.1$ and $2.6$ times slower for
the hydrodynamic and Poisson solvers, respectively. The Poisson solver shows
greater performance degradation due to the larger number of iterations
required to converge to the machine precision and the limited amount of GPU
shared memory.

\subsection{Weak scaling}
\label{subsec:weak_scaling}

We now measure the overall performance of \gamer, starting by showing the
weak scaling from a three-dimensional Kelvin-Helmholtz (KH) instability test.
The weak scaling is useful for demonstrating algorithmic scalability and
is particularly important for uniform-grid
simulations, for example, in the study of non-gravitating turbulence.

\begin{figure}
\centering
\includegraphics[width=\columnwidth]{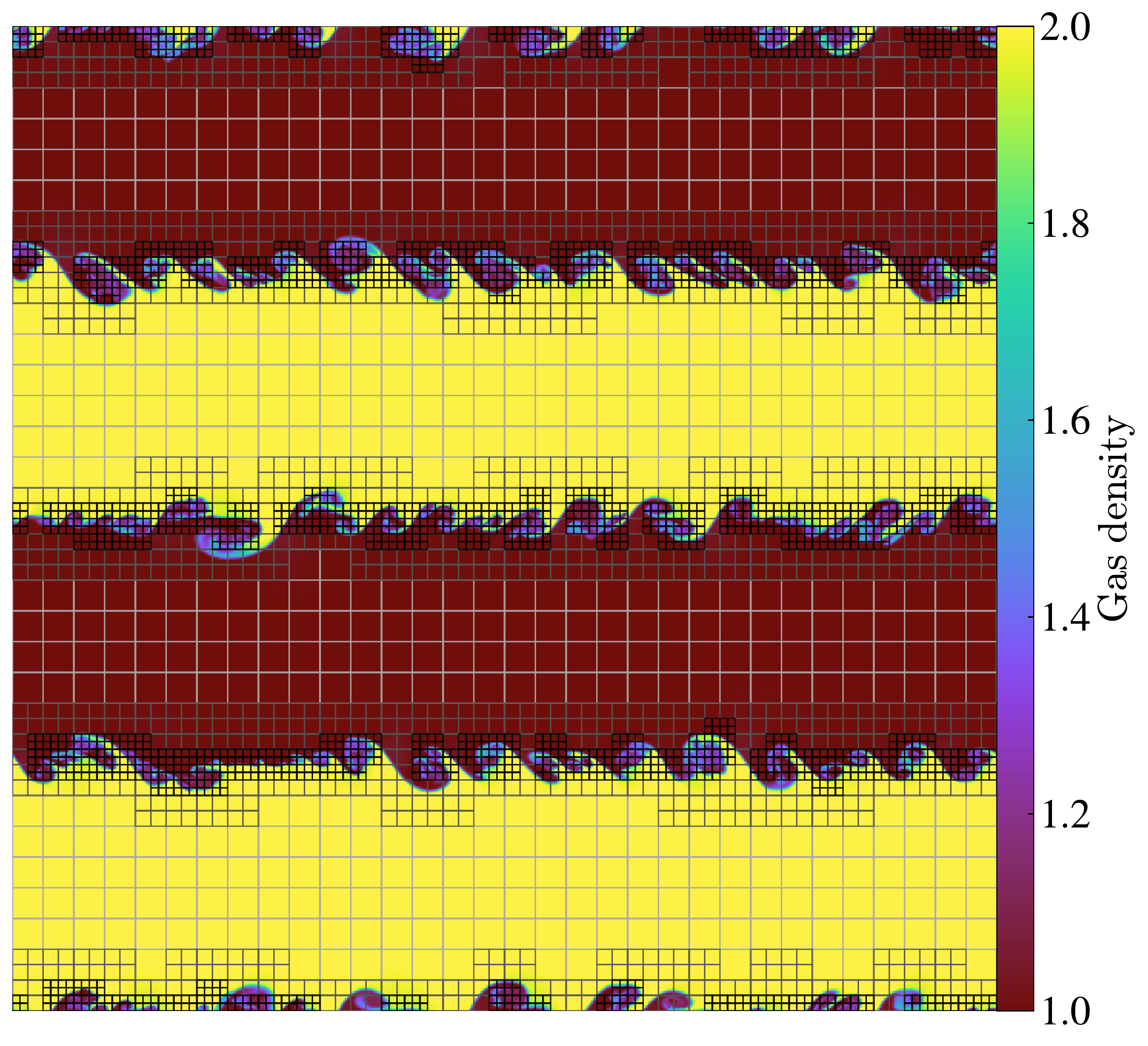}
\caption{
Density slice and grid patches in a single node in the KH instability
weak scaling test at $t=0.5$.
}
\label{fig:KH_grid}
\end{figure}

The simulation setup is as follows. Each node computes a periodic domain
of unit length on a side. We set the gas density $\rho = 2$ and velocity
$v_x = 0.5$ in the regions $z<0.25$ and $0.5<z<0.75$ and have $\rho = 1$
and $v_x = -0.5$ otherwise, leading to four surfaces of contact
discontinuities per node. Velocity perturbations with an amplitude of
$10^{-2}$ and a white noise spectrum are added along all three directions to trigger the
instabilities and make it a three-dimensional test. The gas has a uniform
pressure $P = 2.5$ and an adiabatic index $\gamma = 1.4$.
All simulations are conducted from $t=0$ to $0.5$. We adopt the
CTU scheme with PPM reconstruction and Roe's solver.

We measure the performance of both uniform-grid and AMR simulations
using $1-4096$ nodes on Blue Waters, where we use one MPI process
and 16 OpenMP threads per node. For the uniform-grid test,
each node computes a $640^3$ grid, resulting in an overall resolution
as high as $10{,}240^3$ with 4096 nodes. For the AMR test, each node
computes a $128^3$ root grid with three refinement levels, where we adopt
flow vorticity as the refinement criterion
and enable the adaptive time-stepping.
\fref{fig:KH_grid} shows a density slice perpendicular to the shear-flow
plane at $t=0.5$ in a single node, with the grid patches overlaid.

\begin{figure}
\centering
\includegraphics[width=\columnwidth]{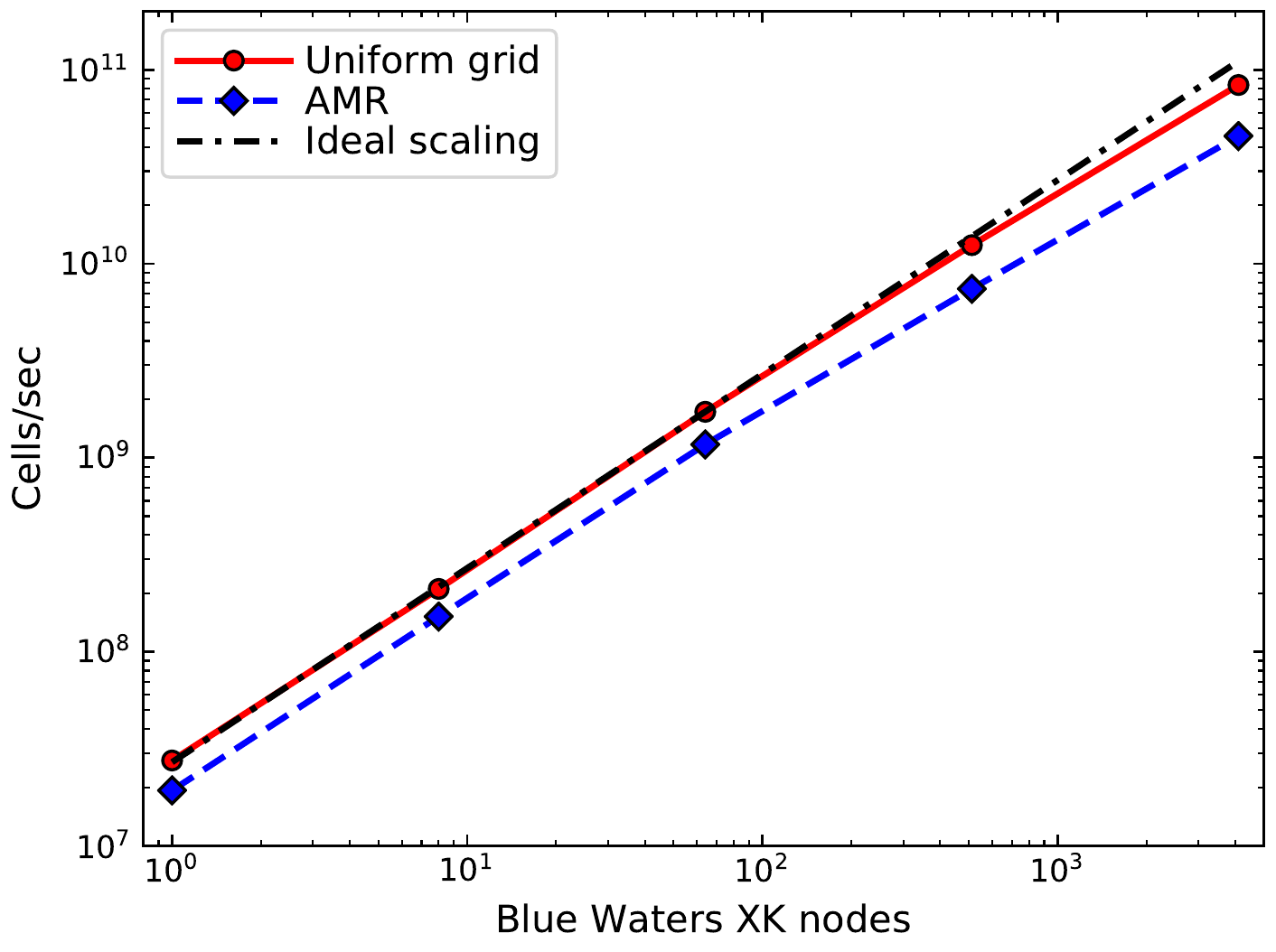}
\caption{
Weak scaling for the KH instability test, showing the total number
of cell updates per second as a function of the number of XK nodes on Blue Waters.
Each XK node is composed of one NVIDIA Tesla K20X GPU and one 16-core AMD
Opteron 6276 CPU. We measure the performance of both uniform-grid
($640^3$ cells per node; solid line) and AMR ($128^3$ root grid per node with three
refinement levels; dashed line) simulations.
Note that for the uniform-grid test with 4096 nodes, we achieve a
resolution of $10{,}240^3$ cells and an overall performance of
$8.3\times10^{10} \CellsPerSec$.
The dashed-dotted line represents the ideal scaling.
See \fref{fig:weak_scaling_metrics} for the detailed performance metrics
of this test.
}
\label{fig:weak_scaling}
\end{figure}

\begin{figure*}
\centering
\includegraphics[width=12cm]{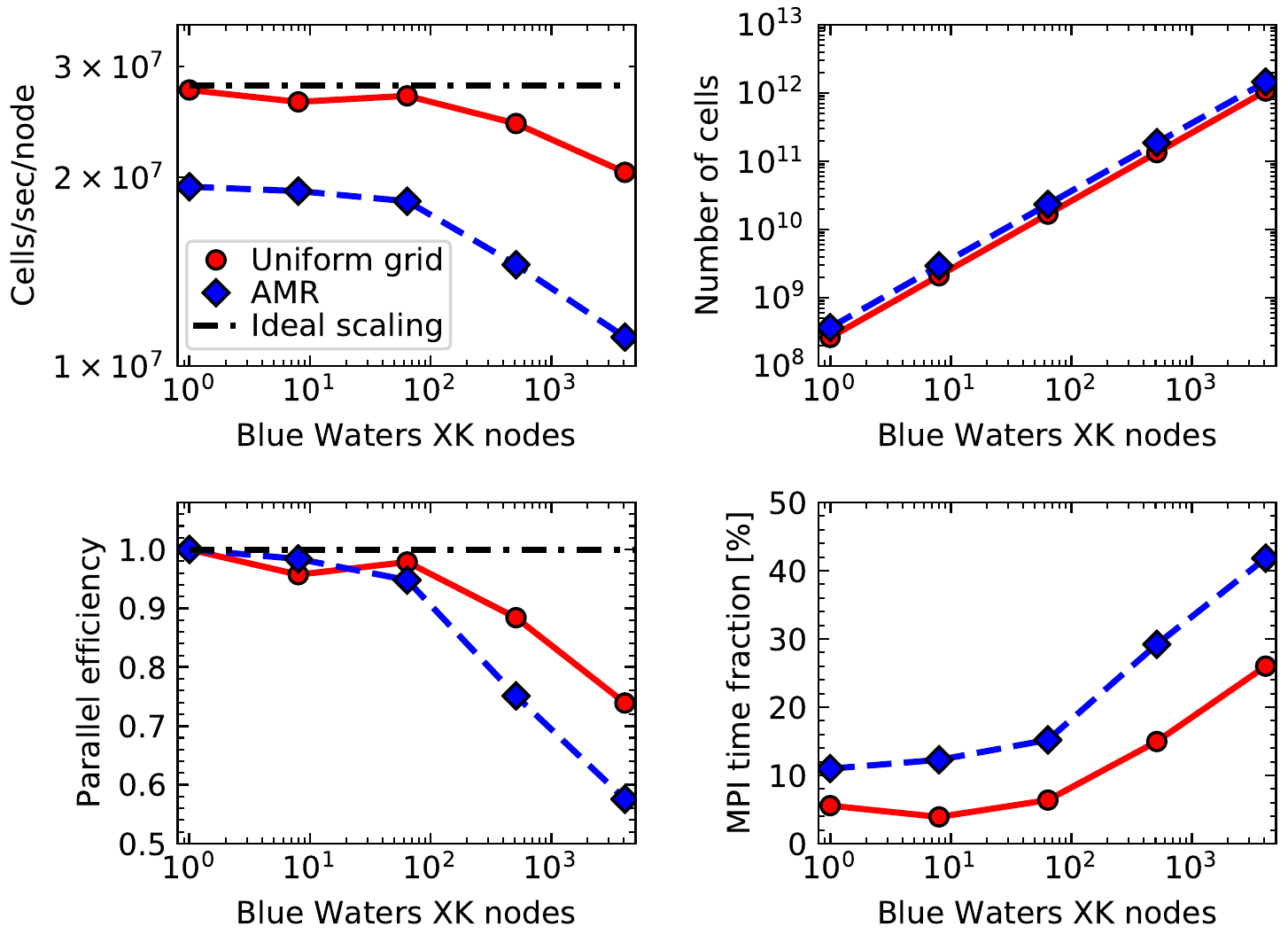}
\caption{
Performance metrics of the KH instability weak scaling test.
This is complementary to and uses the same symbols as
\fref{fig:weak_scaling}. Different panels show the
number of cell updates per second per node (upper left),
total number of cells (upper right),
parallel efficiency (lower left), and
fraction of time spent on MPI communication (lower right).
}
\label{fig:weak_scaling_metrics}
\end{figure*}

\fref{fig:weak_scaling} shows the weak scaling of the KH instability test,
and \fref{fig:weak_scaling_metrics} records the corresponding performance
metrics, including the number of cell updates per second per node,
total number of cells, parallel efficiency, and the fraction of
time spent on MPI communication as a function of the number of nodes.
The parallel efficiency of weak scaling is defined as
\be
P_{\rm{weak}}(\Nnode) = \frac{T(1)}{T(\Nnode)},
\label{eq:pe_weak}
\ee
where $T(\Nnode)$ is the simulation wall time using $\Nnode$ nodes.
Both the uniform-grid and AMR tests exhibit reasonably good scalability for
$\Nnode=1-4096$.
Note that, thanks to the hybrid MPI/OpenMP/GPU
parallelization, we are able to fully exploit both 4096 GPUs and
65,536 CPU cores simultaneously, and achieve a peak performance of
$8.3\times10^{10} \CellsPerSec$ and $P_{\rm{weak}}(4096) = 74\%$
in the uniform-grid test and
$4.6\times10^{10} \CellsPerSec$ and $P_{\rm{weak}}(4096) = 58\%$
in the AMR test.
A noticeably higher fraction of time in MPI is found in the case with AMR
(see the lower-right panel of \fref{fig:weak_scaling_metrics}), thus
partially explaining the relatively lower parallel efficiency achieved.
The total CPU memory consumption for $\Nnode=4096$ is
$\about 53{\,\rm TB}$ and $\about 74{\,\rm TB}$ for the uniform-grid and
AMR tests, respectively.

\subsection{Galaxy cluster merger: GAMER-2 vs. FLASH}
\label{subsec:merger}

Mergers play an important role in the formation of galaxy clusters,
driving shocks and turbulence that heat up the intracluster medium
\citep[e.g.][]{Gaspari2013,Banerjee2014,Lau2017},
provide additional support against gravity
\citep[e.g.][]{Nagai2007,Khatri2016}, and accelerate
relativistic particles emitting radio waves
\citep[e.g.][]{Brunetti2007,Eckert2017}. Numerical simulations of
the galaxy cluster merger are challenging partially due to the large
dynamic range required to both capture the large-scale effects of the
cluster merger and to resolve the properties of turbulence down to at least
the kpc scale. The lack of sufficient resolution will produce non-negligible
numerical viscosity that alters the turbulence cascade and biases the mass
estimates of clusters.

In this section, we conduct cluster merger simulations with both \gamer\
and \flash, and demonstrate that the physical results produced by the
two codes agree very well with each other, and \gamer\ outperforms
\flash\ by almost two orders of magnitude. We first describe the
simulation setup, with particular emphasis on the similarities and
differences of the two codes. We then check the consistency of the
physical results, and finally compare the strong scaling performance.

\subsubsection{Simulation setup}
\label{subsec:merger_setup}

We simulate a head-on merger of two equal-mass clusters for $10\Gyr$.
The simulation setup is identical to that of \citet{ZuHone2011}.
The two clusters have initial separation of $3.1 \Mpc$ and initial relative velocity
of $1352 \kms$. The simulation domain is cubic with a length
$L=14.26 \Mpc$. Each cluster has a virial mass
$M_{\rm 200}=6\times10^{14} \Msun$ and a gas mass fraction $f_g=0.1056$.
We adopt an Navarro-Frenk-White \citep[NFW;][]{NFW} profile with a
concentration parameter $c=4.5$
as the total density profile, and calculate the gas density profile under
the assumptions of spherical symmetry, hydrostatic equilibrium, and a
power-law entropy profile. The dark matter particle velocities are realized
by sampling the velocity distribution function directly via solving the
Eddington formula \citep{Eddington1916} instead of assuming
a Maxwellian distribution. See \citet{ZuHone2011} for the detailed
implementation.

\renewcommand{\arraystretch}{1.5}
\setlength{\tabcolsep}{0.3cm}
\begin{table*}
\caption{
Comparison of the numerical setup between \gamer\ and \flash\
in the galaxy cluster merger simulations.
}
\centering
\begin{tabular}{m{3.0cm}m{6.5cm}m{6.5cm}}
\hline\hline

& \gamer\ & \flash \\
\hline

AMR implementation
& fixed patch size of $8^3$ cells,
no permanent allocation for patch ghost zones
& fixed patch size of $8^3$ cells,
patch ghost zones are allocated permanently$^a$ \\
\hline

Grid resolution
& root grid $128^3$, 4 refinement levels, maximum resolution $\Delta h = 7.0 \kpc$
& same as \gamer \\
\hline

Particle resolution
& number of particles $N_{\rm p}=5\times10^6$ (for each cluster),
mass resolution $m_{\rm p} = 1.4\times10^8\Msun$
& same as \gamer \\
\hline

Fluid solver
& CTU scheme with 6 Riemann solvers per cell, PPM reconstruction,
Roe's solver, van Leer slope limiter
& same as \gamer\ except for a revised CTU scheme
requiring only 3 Riemann solvers per cell$^b$ \\
\hline

Poisson solver
& successive over-relaxation (SOR)
& multigrid$^c$ \\
\hline

Particle solver
& CIC interpolation, kick-drift-kick particle update
& CIC interpolation, variable time-step leapfrog particle update \\
\hline

Boundary condition
& fluid solver: outflow \newline
Poisson solver: isolated
& same as \gamer \\
\hline

Refinement
& (1) L\"{o}hner's error estimator on gas density, pressure, and
      temperature with a refinement threshold of 0.8 and a minimum
      gas density threshold of $10^{-28}\,{\rm g/cm^3}$ \newline
  (2) maximum number of particles in a patch: 100
& same as \gamer \\
\hline

Derefinement
& no explicit derefinement criteria
& (1) L\"{o}hner's error estimator with a derefinement threshold of 0.2 \newline
  (2) minimum number of particles in a patch: 12 \\
\hline

Time-step
& $C_{\rm par} = 0.8$, $C_{\rm CFL} = 0.5$
& $C_{\rm par} = 0.8$, $C_{\rm CFL} = 0.6$ \\
\hline

Parallelization
& hybrid MPI/OpenMP/GPU
& MPI and CPU-only \\
\hline

Load balancing
& Hilbert space-filling curve
& Morton space-filling curve \\
\hline

Time integration
& adaptive time-step
& shared time-step \\
\hline

Floating-point format
& single precision
& double precision$^d$ \\
\hline

\end{tabular}
\\[0.2cm]
\raggedright \footnotesize
$^a$ \flash\ supports the `NO-PERMANENT-GUARDCELLS' (npg) mode,
which however does not work well with active particles. We have
therefore disabled this functionality. \\
$^b$ \citet{Lee2013}. \\
$^c$ \citet{Ricker2008}. \\
$^d$ Single precision is not officially supported in \flash. But we
have demonstrated that this discrepancy does not affect the physical
results here. See discussions in Sections~\ref{subsec:merger_accuracy}
and~\ref{subsec:merger_performance}. \\
[0.2cm]
\label{table:merger_setup}
\end{table*}

\tref{table:merger_setup} summarizes the similarities and differences
of the numerical setup adopted by \gamer\ and \flash\ in these
comparison simulations. Here we elaborate on the major differences.

\begin{itemize}
\item{
\gamer\ adopts an adaptive time-step integration where higher levels can
have smaller time-steps. Moreover, the time-step ratio between two
adjacent levels does not need to be a constant (see \sref{subsec:AMR}).
By contrast, \flash\ adopts a shared time-step integration where all
levels share the same time-step.
}
\item{
\gamer\ adopts the original CTU scheme \citep{Colella1990, Athena}
requiring 6 Riemann solvers per cell per time-step and
$C_{\rm CFL}\le0.5$. In comparison, \flash\ adopts a revised CTU scheme
\citep{Lee2013} requiring only 3 Riemann solvers per cell per
time-step and allowing for a larger CFL number $C_{\rm CFL}\le1.0$,
which in principle should be significantly faster than \gamer.
We, however, adopt $C_{\rm CFL}=0.6$ for \flash\ since it crashed
with $C_{\rm CFL}>0.6$.
}
\item{
For the Poisson solver, \gamer\ uses the SOR scheme and adds 5
additional buffer zones around each patch to make the potential smoother
across the patch boundaries (see \fref{fig:poisson_grid}). In comparison, \flash\
adopts a finite-volume multigrid scheme that aims to minimize the
global residual \citep{Ricker2008}, which, in general, should be more
accurate but also more computationally expensive.
}
\item{
Unlike \flash, \gamer\ does not implement explicit grid derefinement
criteria (see \sref{subsec:AMR}). However, we have verified that the grid
distribution of the two codes are very similar in this test
(the difference in the numbers of maximum level cells is less than
$\about 20\%$, e.g.\ see \tref{table:merger_grids}).
}
\item{
For the floating-point accuracy, \gamer\ uses single precision while
\flash\ uses double precision, because single precision is not officially
supported in \flash.
This discrepancy makes our performance comparison in favor of \gamer,
which could be unfair in this sense. However, we also demonstrate in the
next section that the physical results obtained by the two codes are very
consistent, suggesting that double precision may not be necessary for
this test.
}
\end{itemize}

\subsubsection{Accuracy comparison}
\label{subsec:merger_accuracy}

\begin{figure*}
\centering
\includegraphics[width=\textwidth]{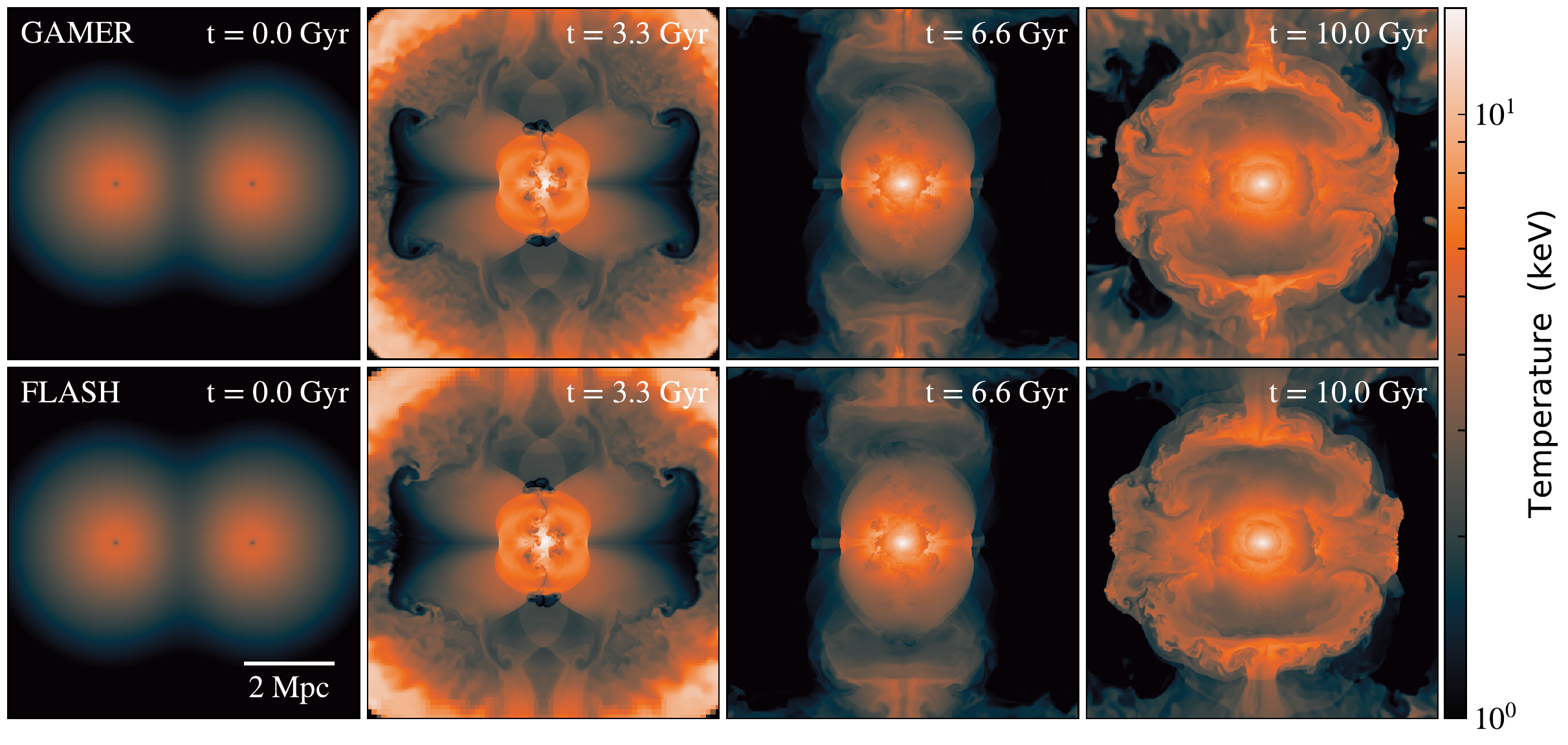}
\caption{
Slices of gas temperature through the cluster center at four different
epochs in the galaxy cluster merger simulations. Each panel is $8 \Mpc$
on a side. The results obtained by \gamer\ (upper panels) and \flash\
(lower panels) are found to be in very good agreement with each other.
See \fref{fig:merger_profile} for more quantitative comparisons
of the radial profiles.
}
\label{fig:merger_temperature_slice}
\end{figure*}

\fref{fig:merger_temperature_slice} shows the slices of gas temperature
through the cluster center at $t=0.0$, $3.3$, $6.6$, and $10.0 \Gyr$
obtained by \gamer\ and \flash. The first core passage occurs at
$t \sim 1.5 \Gyr$, after which the oscillating dark matter cores
continue driving shocks and turbulence that heat up the intracluster
medium, eventually forming a high temperature and constant entropy core.
The results of \gamer\ (upper panels) and \flash\
(lower panels) are verified to be very consistent with each other.

\begin{figure}
\centering
\includegraphics[width=\columnwidth]{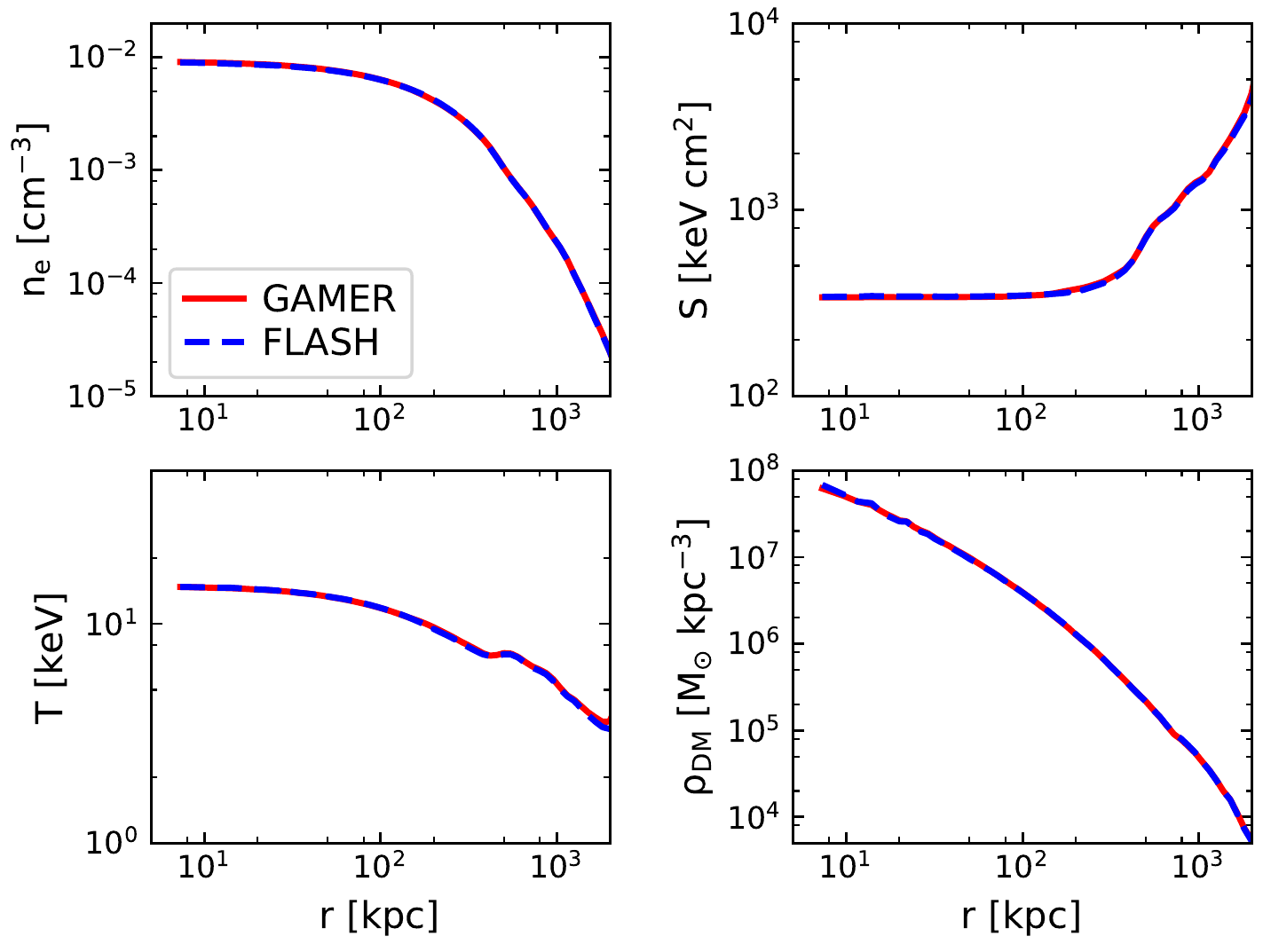}
\caption{
Radial profiles at $t=10\Gyr$ in the galaxy cluster merger simulations.
Different panels show the electron number density (upper left),
gas entropy (upper right), gas temperature (lower left), and dark matter
mass density (lower right). A remarkable agreement is observed between
the results of \gamer\ (solid lines) and \flash\ (dashed lines).
}
\label{fig:merger_profile}
\end{figure}

\fref{fig:merger_profile} shows the radial profiles of the electron number
density $n_e$, gas entropy $S$, gas temperature $T$, and dark matter
mass density $\rho_{\rm DM}$ at $t=10 \Gyr$, where gas is assumed to
be fully ionized and the gas specific entropy is defined as
$S \equiv k_B T n_e^{-2/3}$ with $k_B$ the Boltzmann constant.
A constant entropy core can be clearly identified within $\about 300 \kpc$.
Most strikingly, the results obtained by \gamer\ and \flash\ are
found to be literally indistinguishable. It demonstrates the consistent
numerical setup we adopt for this code comparison experiment, including,
for example, the initial condition, boundary conditions, spatial and
temporal resolution, AMR implementation, and grid refinement criteria.
It also indicates that the different numerical
schemes between the two codes described in \sref{subsec:merger_setup},
for example, the time integration, fluid and Poisson solvers, and
floating-point accuracy, do not have a significant impact here.

\subsubsection{Performance comparison}
\label{subsec:merger_performance}

Based on the very consistent physical results between \gamer\
and \flash, as shown in Figs.~\ref{fig:merger_temperature_slice}
and~\ref{fig:merger_profile}, here we compare their strong scaling performance
on Blue Waters. In order to have a fair
comparison between the codes with and without GPU-acceleration, we run
\gamer\ on the XK nodes while \flash\ on the XE nodes: each XK node
is composed of one GPU (NVIDIA Tesla K20X) and one 16-core CPU
(AMD Opteron 6276), and each XE node is composed of \emph{two}
16-core CPUs. In addition, since there are 2 NUMA domains per
XK node, each of which shares
an 8 MB L3 cache, for \gamer\ we launch 2 MPI processes per node and
7\footnote{We use 7 instead of 8 OpenMP threads per MPI process
since using 8 threads somehow binds two threads to the same CPU core.}
OpenMP threads per MPI process in order to improve
memory affinity by having all threads running in the same NUMA domain.
The two MPI processes running on the same node shares the same GPU by taking
advantage of the CUDA MPS. For \flash, we
launch 32 MPI processes per XE node.

\begin{figure}
\centering
\includegraphics[width=\columnwidth]{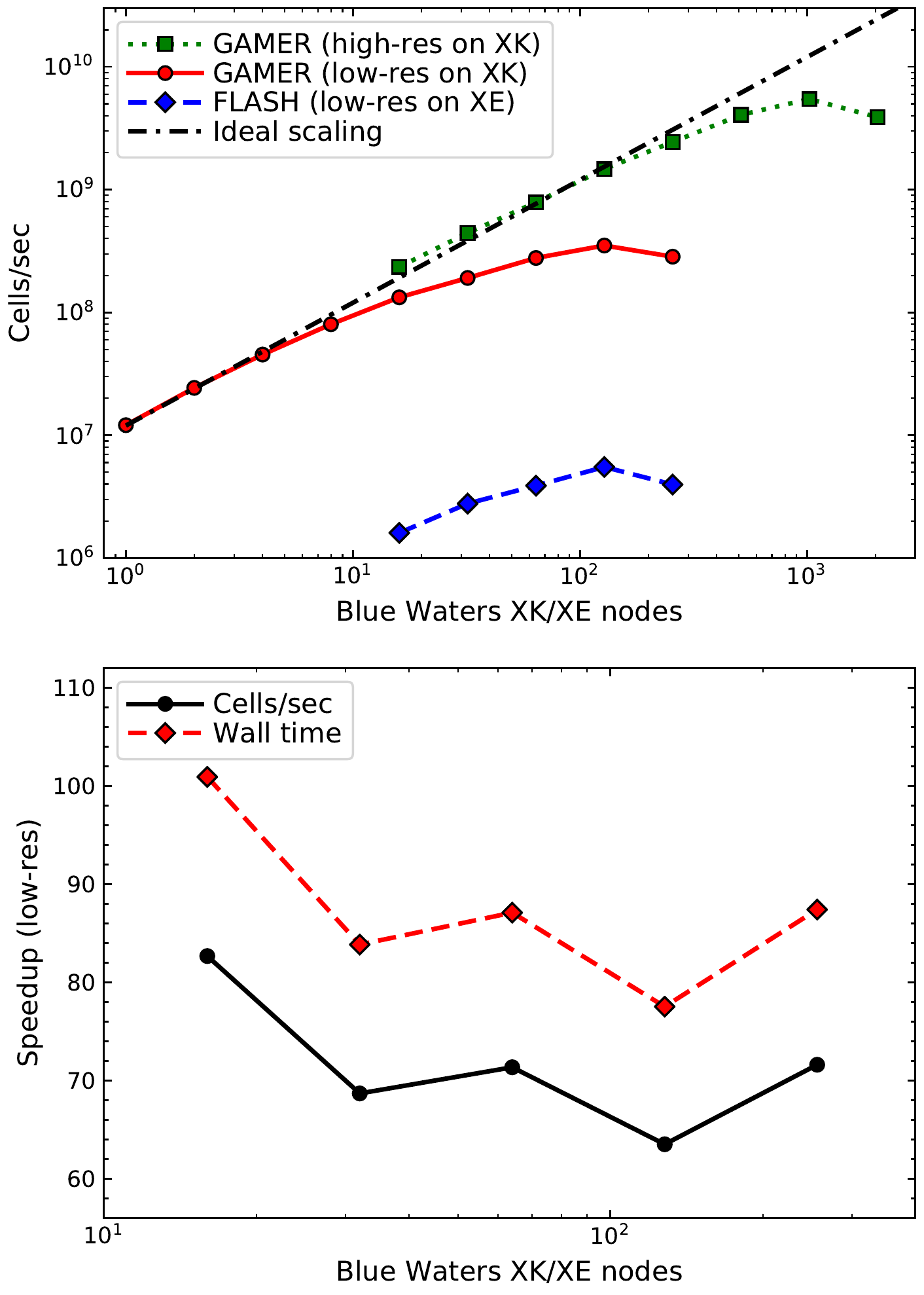}
\caption{
Strong scaling of the galaxy cluster merger simulations run with \gamer\
and \flash\ on Blue Waters. \gamer\ runs on the XK nodes
composed of one GPU (NVIDIA Tesla K20X) and one 16-core CPU (AMD Opteron
6276) per node, while \flash\ runs on the XE nodes composed of two 16-core
CPUs per node. The upper panel shows the total number of cell updates
per second for (i) the lower-resolution tests ($\Delta h \sim 7.0 \kpc$)
of \gamer\ (solid line) and \flash\ (dashed line) using up to 256 nodes,
and (ii) the higher-resolution test of \gamer\
($\Delta h \sim 0.87 \kpc$, dotted line) using
$16-2048$ nodes. Note that the minimum number of nodes adopted in
\flash\ is 16 instead of 1 due to its much larger memory consumption
(see text for details).
The dashed-dotted line represents the ideal scaling.
The lower panel shows the speedup of \gamer\ over \flash\ in the
lower-resolution tests in terms of either cell updates per second
(solid line) or total wall time (dashed line). Both cases reveal
nearly two orders of magnitude speedup.
See \fref{fig:merger_strong_scaling_metrics} for the detailed
performance metrics of this test.
Note that \gamer\ uses single-precision but \flash\ uses
double-precision arithmetic.
}
\label{fig:merger_strong_scaling}
\end{figure}

\fref{fig:merger_strong_scaling} shows the strong scaling of the cluster
merger simulations. We first conduct the lower-resolution tests
($\Delta h \sim 7.0 \kpc$, the same as that adopted in
\sref{subsec:merger_accuracy}) for both \gamer\ and \flash\ using up to
256 nodes. Note that the minimum number of nodes adopted in \flash\ is
16 instead of 1 due to its much larger memory consumption, which will be
discussed at the end of this section.
We measure the performance in $t=4-6 \Gyr$, during
which there are $\about 2.1\times10^8$ cells in total and
$\about 7.1\times10^7$ cells on the maximum refinement level.
When using the same number of nodes, the speedup of \gamer\ over \flash\
is measured to be $78 - 101$ and $64 - 83$ in terms of total wall time
and cell updates per second, respectively. For example, for $\Nnode=16$,
\gamer\ achieves $8.3\times10^6$ cell updates per second per XK node, and
\flash\ achieves $1.0\times10^5$ cell updates per second per XE node
(corresponding to $3.1\times10^3$ cell updates per second per CPU core).
Moreover, this speedup ratio only drops by $\about 23\%$ when increasing
$\Nnode$ from 16 to 128, and the performance of both codes starts to
decline when $\Nnode>128$. It suggests that \gamer\ and \flash\ exhibit
similar parallel scalability, despite the fact that the
computational time of \gamer\ has been greatly reduced with GPUs.

The different speedups measured from the total wall time and cell updates
per second are due to several factors. The cell updates per second depends
mostly on the performance of individual PDE solvers, which itself is
related to the CPU/GPU performance, the adopted floating-point accuracy,
and the numerical schemes adopted in the PDE solvers.
In comparison, the speedup in term of total wall time is arguably more
comprehensive since it takes into account not only
the performance of PDE solvers but also many other factors, such as
the time integration scheme, the evolution time-step, and the number of
cells on each level. See \sref{subsec:merger_setup}, especially
\tref{table:merger_setup}, for the summary of different numerical setup
between \gamer\ and \flash.

\begin{table}
\setlength{\tabcolsep}{6pt}
\caption{
Comparison of the volume-filling fractions on the refinement
levels between \gamer\ and \flash\ in the lower-resolution galaxy
cluster merger simulations at $t=5\Gyr$.
}
\centering
\begin{tabular}{crrrrr}
\hline\hline

Level & $\Delta h/\kpc$ & \multicolumn{2}{c}{Filling fraction} & \multicolumn{2}{c}{Number of cells} \\
      &                 & \gamer & \flash                      & \gamer & \flash \\
\hline

1  &  55.70  &  56.45\%  &  38.65\%  &  $9.5\times10^6$  &  $6.5\times10^6$ \\
2  &  27.85  &  31.62\%  &  23.08\%  &  $4.2\times10^7$  &  $3.1\times10^7$ \\
3  &  13.93  &   7.13\%  &   6.48\%  &  $7.7\times10^7$  &  $7.0\times10^7$ \\
4  &   6.96  &   0.86\%  &   0.75\%  &  $7.4\times10^7$  &  $6.4\times10^7$ \\
\hline

\end{tabular}
\\[0.2cm]
\label{table:merger_grids}
\end{table}

In short, \flash\ uses a more efficient
fluid solver and a more accurate but also more computationally expensive
Poisson solver. The CFL safety factor adopted for \flash\
($C_{\rm CFL}=0.6$) is larger than \gamer\ ($C_{\rm CFL}=0.5$), but
the average time-steps on the maximum level are found to be very similar
due to the same time-step criterion for updating particles.
More precisely, the minimum time-steps in both codes are set by the
fastest-moving particles on the maximum refinement level, which always
move faster than the characteristic hydrodynamic speed which sets the
hydrodynamic time-step. \gamer\
is found to allocate
$\about 30\% - 50\%$ more cells on lower levels and
$\about 10\% - 20\%$ more cells on higher levels
than \flash\ (e.g.\  see \tref{table:merger_grids}), mainly because
\gamer\ tends to pre-allocate patches earlier than \flash\ due to the
implementation of large flag buffers (see \sref{subsec:AMR}).
This issue of over-allocation in \gamer\ is relatively more serious
on lower levels since there are fewer patches on these levels.
However, it is not a serious problem since \gamer\ adopts an adaptive
time-step integration that allows patches on lower levels to have
larger time-steps.
It is also the main reason why the speedup in terms of total wall time
is $\about 20\%$ higher than that in terms of cell updates per second.
Last but not least, we remind the reader that we adopt single
precision for \gamer\ and double precision for \flash\ in this test.
In principle, using single precision for \flash\ could improve
performance, however single-precision calculations are not currently
supported by \flash.

To test the scalability of \gamer\ further, we also conduct
higher-resolution runs ($\Delta h \sim 0.87 \kpc$, 8 times higher than
the lower-resolution counterpart) using $16 - 2048$ XK nodes. We measure
the performance from $t=4.0$ to $4.5 \Gyr$, during which there are
$\about 8.8\times10^9$ cells in total and $\about 7.1\times10^9$ cells
on the maximum refinement level. The total number of particles is fixed
to $10^7$, the same as the lower-resolution test. The total memory
consumption for $\Nnode=2048$ at $t=4.5 \Gyr$ is $\about 2.2 {\,\rm TB}$.
\fref{fig:merger_strong_scaling} (upper panel, dotted line) shows
the strong scaling of this higher-resolution test, exhibiting a much
better scaling than its lower-resolution counterpart. It demonstrates
that \gamer\ can scale well to thousands of GPUs in a realistic
astrophysical application. It also indicates that the parallel scalability
may be sensitive to the load imbalance resulting from particles,
which will be investigated shortly in this section.

Before giving a more detailed analysis of strong scaling, we first
introduce two quantities useful for quantifying the parallel scalability.
The `parallel efficiency' of strong scaling is defined as
\be
P_{\rm{strong}}(\Nnode) = \frac{T(\Nnoderef) \Nnoderef}{T(\Nnode) \Nnode},
\label{eq:pe_strong}
\ee
where $T(\Nnode)$ is the simulation wall time using $\Nnode$ nodes.
Generally speaking, the parallel efficiency also depends on the reference
performance $T(\Nnoderef)$. For a proper comparison, we adopt
$\Nnoderef=16$ for the lower-resolution tests of both codes, even though
the minimum $\Nnode$ in the \gamer\ runs is 1 instead of 16.
For the higher-resolution test we adopt $\Nnode=16$.
We also introduce another quantity to quantify the scalability, namely,
the `doubling efficiency', which is defined as
\be
D_{\rm{strong}}(\Nnode) = \frac{T(\Nnode/2)}{T(\Nnode)}-1,
\label{eq:de_strong}
\ee
This quantity corresponds to the performance gain when doubling the
computational resource from $\Nnode/2$ to $\Nnode$, which is arguably
more intuitive than the conventional parallel efficiency and has the
advantage of being independent of the minimum $\Nnode$ adopted.
For example, one may have $D_{\rm{strong}}(\Nnode) = 0.6$ for
$\Nnode=2-2048$, which suggests reasonable scalability since a
performance speedup of 1.6 is always obtained when doubling the number
of nodes. However, the corresponding parallel efficiency is as low as
$P_{\rm{strong}}(2048) = (1.6/2)^{11} \sim 9\%$, which could be misleading.

\begin{figure*}
\centering
\includegraphics[width=12cm]{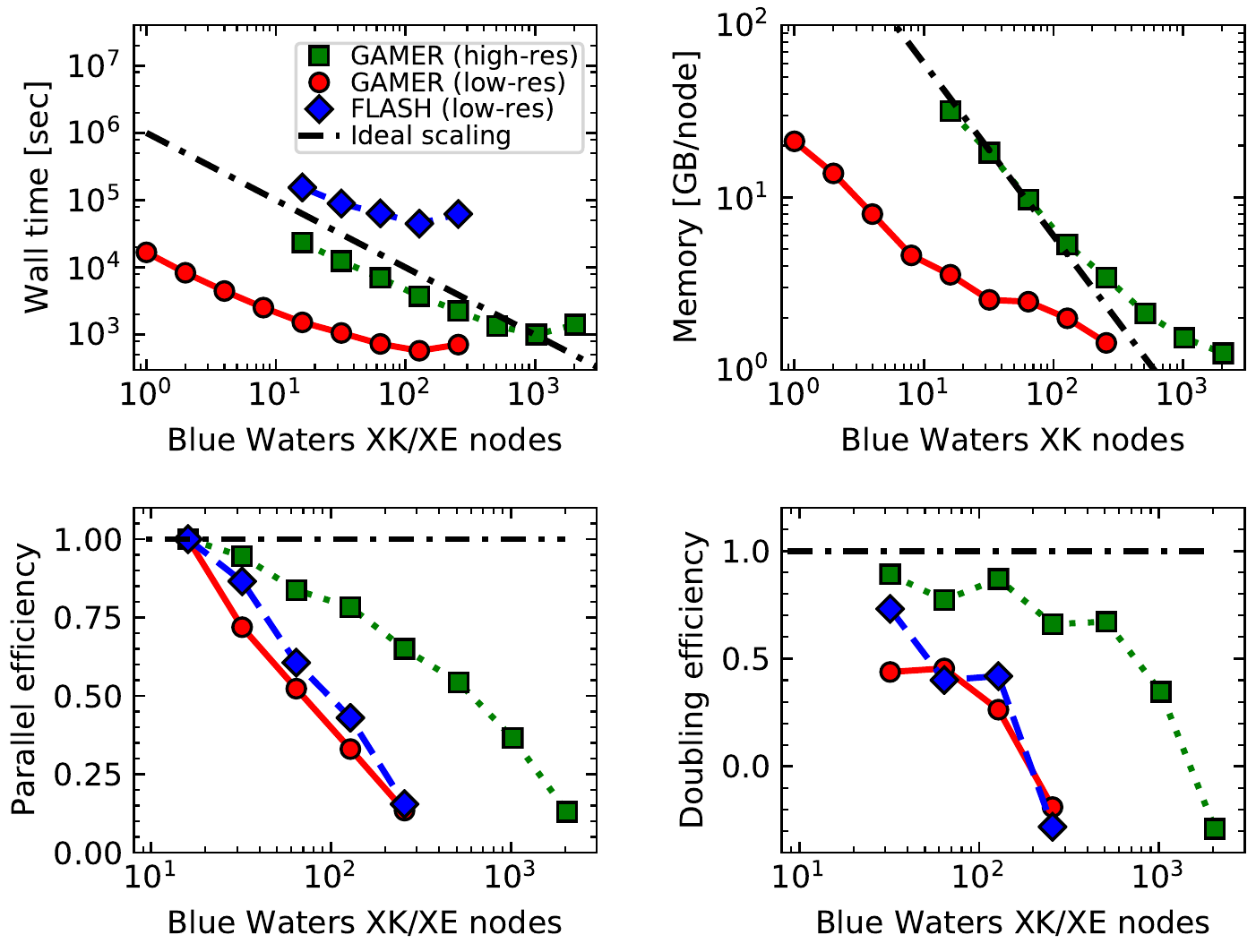}
\caption{
Performance metrics of the strong scaling of the cluster merger simulations.
This is complementary to and uses the same symbols as
\fref{fig:merger_strong_scaling}. Different panels show the
total wall time (upper left),
maximum CPU memory consumption per MPI process (upper right),
parallel efficiency (lower left), and
doubling efficiency (lower right).
See Equations~(\ref{eq:pe_strong}) and~(\ref{eq:de_strong}) for the
definitions of parallel and doubling efficiencies in strong scaling.
Note that the minimum number of nodes adopted in \flash\ is 16 instead
of 1 due to its much larger memory consumption. Therefore, for a proper
comparison, we adopt $\Nnoderef=16$ in \eref{eq:pe_strong}
for both codes when calculating the parallel efficiency.
\gamer\ and \flash\ exhibit similar parallel scalability,
even though \gamer\ is about two orders of magnitude faster.
We do not show the maximum per-process memory consumption of \flash\
because it is mainly determined by the size of the preallocated
memory pool set manually.
}
\label{fig:merger_strong_scaling_metrics}
\end{figure*}

\fref{fig:merger_strong_scaling_metrics} shows the performance metrics
of the cluster merger simulations, including the total wall time,
maximum CPU memory consumption per MPI process, parallel efficiency,
and doubling efficiency as a function of the number of XK and XE
nodes for \gamer\ and \flash, respectively. Most importantly, as
demonstrated by both the parallel and doubling efficiencies, the two
codes exhibit similar parallel scalability, especially for $\Nnode>32$.
It is consistent with the finding of an almost constant speedup
of \gamer\ over \flash\ for $\Nnode>32$, as shown in the lower panel
of \fref{fig:merger_strong_scaling}.

In addition, the higher-resolution test of \gamer\ shows a better
scaling than its lower-resolution counterpart. The scalability is
reasonably good in a large range of $\Nnode$ from 16 to 1024.
The parallel efficiency is measured to be
$78\%$ for $\Nnode=128$ and
$37\%$ for $\Nnode=1024$, and the doubling efficiency is measured to be
$87\%$ for $\Nnode=128$ and $35\%$ for $\Nnode=1024$.

Also, note that the CPU memory consumption per MPI process in \gamer\
deviates from the ideal scaling, especially when increasing the
number of nodes. It is most likely due to the allocation of buffer
patches and MPI buffers.

\begin{figure}
\centering
\includegraphics[width=\columnwidth]{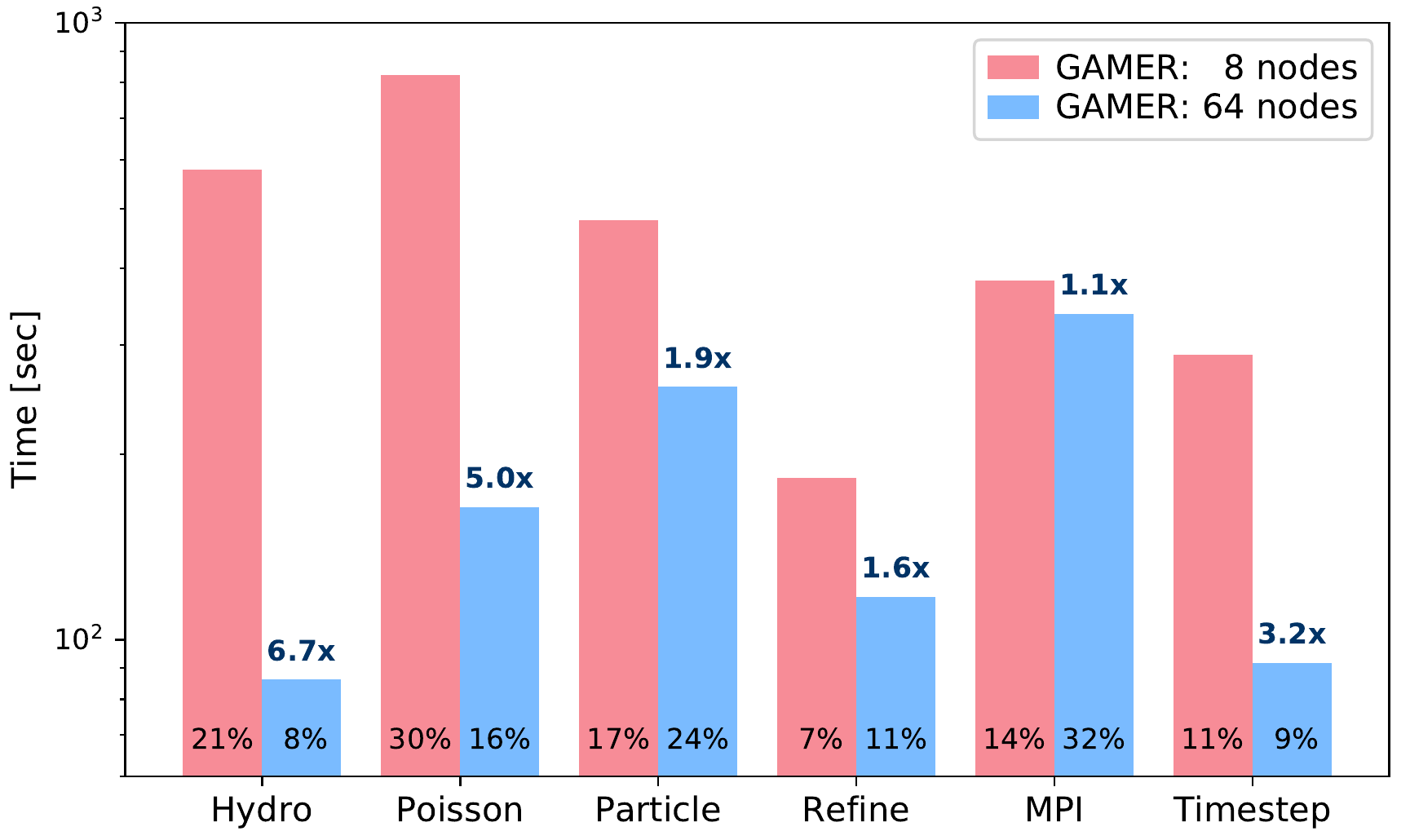}
\caption{
Wall time of various operations measured from the lower-resolution
cluster merger simulations with \gamer\ using 8 and 64 nodes. For each
operation, we also show the fraction of time in the two runs and the
speedup gained from increasing $\Nnode=8$ to $64$. The `MPI' time
includes transferring both grids and particles for the `Hydro',
`Poisson', and `Particle' routines but excludes the MPI
communication during the grid refinement, which is included in the
`Refine' time. The `Poisson' time includes depositing particle
mass onto grids with the CIC interpolation. It shows that the
grid PDE solvers (i.e.\ the hydrodynamic and Poisson solvers) are
the major performance bottlenecks for $\Nnode=8$, while the particle
routines and MPI communication become the bottlenecks for $\Nnode=64$
as these operations exhibit poor scalability.
Note that, for better clarification, the performance shown here does
not consider particle weights for load balancing, which is therefore
different from the optimized performance shown in
Figs.~\ref{fig:merger_strong_scaling} and~\ref{fig:merger_strong_scaling_metrics}.
See text for details.
}
\label{fig:merger_time_fraction}
\end{figure}

To determine the performance bottleneck in \gamer, especially for
large $\Nnode$, we show in \fref{fig:merger_time_fraction} the wall time
of various operations in the lower-resolution runs using 8 and 64 nodes.
It is found that the grid PDE solvers, namely, the hydrodynamic and
Poisson solvers, are the major performance bottlenecks for $\Nnode=8$,
while for $\Nnode=64$ the bottlenecks shift to the particle routines
and MPI communication. From $\Nnode=8$ to $64$, the fraction of time spent
on the hydrodynamic and Poisson solvers decrease from $21\%$ and $30\%$
to $8\%$ and $16\%$, respectively; in comparison, the fraction of time
spent on the particle routines and MPI communication increase from
$17\%$ and $14\%$ to $24\%$ and $32\%$, respectively.

There are several things to note about these results. First of all,
the average MPI message size per process for $\Nnode=64$ is found to be
as small as $\about 1$ MB, which thus suffers from a relatively larger
latency. Second, for better clarification, the performance shown
in \fref{fig:merger_time_fraction} is measured from a separate set of
simulations that (i) adds an explicit MPI synchronization between grid
and particle routines in order for a proper timing analysis,
and (ii) does not consider the particle load-balancing
weights when estimating the workload of each patch
(i.e.\ $W_{\rm par}=0.0$). It partially explains the poor
scalability observed in the particle routines and MPI communication,
the latter of which also includes transferring particles. For $\Nnode=64$, it is
found that by removing that extra MPI synchronization between grid and
particle routines, the overall performance is improved by $\about 37\%$.
Having $W_{\rm par}=2.0$ further improves the performance by $\about 10\%$.
The performance shown in Figs.~\ref{fig:merger_strong_scaling}
and~\ref{fig:merger_strong_scaling_metrics} has incorporated these
optimizations. These findings reveal the importance of balancing the
workload of both grids and particles simultaneously, as discussed in
\sref{subsec:load_balancing}.

Finally we compare the CPU memory consumption between \gamer\ and \flash.
We find that \flash\ consumes about an order of magnitude more
memory than \gamer\ when using the same number of nodes, which is why
the minimum $\Nnode$ adopted for \flash\ is 16 instead of 1 as for \gamer.
This is mainly because \flash\ allocates the ghost zones of all patches
permanently (see footnote \textit{a} in \tref{table:merger_setup})
but \gamer\ does not (see \sref{subsec:memory}). For a patch with $8^3$
interior cells and 4 ghost zones on each side (which is the number of
ghost zones adopted in \flash), the total memory consumption including
the ghost zones is 8 times larger than that without the ghost zones.
In addition, \flash\ uses double precision arithmetic which doubles
the memory consumption. On the other hand, \gamer\ adopts the adaptive
time-step integration requiring storing all the grid data at two
different physical times for the temporal interpolation, which
also roughly doubles the memory consumption. Last but not least,
unlike \flash, \gamer\ does not pre-allocate a memory pool for all
blocks that will be used during the simulation
(see \sref{subsec:memory}).

\subsection{AGORA isolated disk galaxy: GAMER-2 vs. Enzo}
\label{subsec:agora}

Simulations of the gas, stars, and dark matter in an idealized isolated disk
galaxy present a unique numerical challenge for astrophysical simulation
codes. These simulations combine self-gravity, gas dynamics, particle dynamics
with particles existing at a range of masses, radiative cooling, and star
formation. In addition, gas temperatures may reach as low as $10\,{\rm K}$, but
have a velocity relative to the simulation box of hundreds of kilometers per
second, requiring the use of a dual energy formalism to avoid spurious
temperature fluctuations and negative temperatures due to truncation errors. On
top of the bulk circular velocity, the gas also exhibits supersonic turbulence
driven by gravitational instability \citep{Goldbaum2015}. Despite these
challenges, isolated disk galaxy simulations are commonly used to understand
more complicated zoom-in simulations \citep{AGORA}, galaxy merger simulations
\citep{Robertson2006} and as a testbed for physics modules that will be used in
more realistic simulations. These simulations also allow {\it ab initio\/}
exploration of the dynamics of the interstellar medium (ISM) of a galaxy much
like the Milky Way \citep{Goldbaum2016}, enabling direct comparison with
observations of the ISM of our own Galaxy.

In this section, we simulate a Milky Way-sized isolated disk galaxy
with star formation using both \gamer\ and \enzo, from which we
demonstrate that, not only the physical results obtained by the two codes
are in good agreement, but \gamer\ outperforms \enzo\ by almost one order
of magnitude. Similar to the previous section, we first describe the
simulation setup, with particular emphasis on the similarities and
differences of the two codes. We then check the consistency of the
physical results, and finally compare the strong scaling performance.

\subsubsection{Simulation setup}
\label{subsec:agora_setup}

The simulation setup of our isolated disk galaxy simulations closely
follow \citet{Goldbaum2015} and the AGORA High-resolution Galaxy
Simulations Comparison Project \citep{AGORA}. So we only provide a
short summary here.

The initial condition is composed of three components: a dark matter halo,
a galactic disk consists of both gas and stars, and a stellar bulge.
(i) The dark matter halo follows an NFW profile with a virial mass
$M_{\rm 200}=1.1\times10^{12} \Msun$, a concentration parameter $c=10$,
and a spin parameter $\lambda=0.04$. (ii) The disk follows an exponential
profile with a total disk mass $M_{\rm d}=4.3\times10^{10} \Msun$, a scale
length $r_{\rm d}=3.4 \kpc$, a scale height $z_{\rm d}=0.34 \kpc$, and
a gas mass fraction $f_{\rm d}=M_{\rm d,gas}/M_{\rm d}=0.2$. The gas disk
has an initial metal mass fraction
$Z_{\rm d}=M_{\rm d,metal}/M_{\rm d,gas}=1.3\times10^{-2}$ and an
initial temperature $T_{\rm d}=10^4 \kelvin$. The circular velocity is
set such that the disk is in centrifugal equilibrium. (iii) The stellar
bulge is modeled as a Hernquist profile \citep{Hernquist1990}
with a mass $M_{\rm b}=4.3\times10^9 \Msun$. The initial conditions of both
dark matter and stellar particles as well as the gas rotation curve
can be downloaded from the AGORA Project\footnote{http://goo.gl/8JzbIJ.
Note that we use the high-resolution files while \citet{AGORA} use the low-resolution files.}
The simulation has a cubic domain with a length $L=1.31 \Mpc$ and is
evolved for $500 \Myr$.

The \grackle\ library is used
to solve the chemistry and radiative processes in this work. We adopt
the equilibrium solver in \grackle\ using tabulated cooling and heating
rates, which are functions of gas density and temperature. These tables
incorporate both primordial and metal cooling as well as a UV background.
We also include the photoelectric heating from dust and an effective
CMB temperature floor. The metal field is
treated as a passive scalar advected along with the gas, and thus the
metal fraction can vary in space and time. To avoid artificial
fragmentation, we follow \citet{Goldbaum2015} and \citet{AGORA} and employ a
pressure floor in the hydrodynamic solver to ensure that the local Jeans
length is resolved by at least 4 cells on the maximum refinement level.

The full details of the subgrid model for stochastic star formation can
be found in Section 2.4 of \citet{Goldbaum2015}. Specifically, we adopt
a gas density threshold $n_{\rm H,thres}=20\,{\rm cm}^{-3}$ and a star
formation efficiency $f_\star=1\%$. We also impose a minimum star
particle mass $m_\star=2\times10^3 \Msun$. No star formation feedback
is included in this work.

\renewcommand{\arraystretch}{1.5}
\setlength{\tabcolsep}{0.3cm}
\begin{table*}
\caption{
Comparison of the numerical setup between \gamer\ and \enzo\
in the isolated disk galaxy simulations.
}
\centering
\begin{tabular}{m{3.0cm}m{6.5cm}m{6.5cm}}
\hline\hline

& \gamer\ & \enzo \\
\hline

AMR implementation
& fixed patch size of $8^3$ cells,
no permanent allocation of patch ghost zones
& patch size is not fixed,
patch ghost zones are allocated permanently \\
\hline

Grid resolution
& root grid $64^3$, 10 refinement levels, maximum resolution $\Delta h = 20 \pc$
& same as \gamer \\
\hline

Particle resolution
& halo:  $N_{\rm p}=1   \times10^7$, $m_{\rm p} = 1.3\times10^5\Msun$ \newline
stellar disk:  $N_{\rm p}=1   \times10^7$, $m_{\rm p} = 3.4\times10^3\Msun$ \newline
  bulge: $N_{\rm p}=1.25\times10^6$, $m_{\rm p} = 3.4\times10^3\Msun$
& same as \gamer \\
\hline

Fluid solver
& dimensionally unsplit MUSCL-Hancock scheme with PPM reconstruction
(requiring 3 Riemann solvers per cell), HLLC solver, hybrid van Leer
and generalized minmod slope limiter, dual energy formalism solving
the entropy equation
& dimensionally split direct Eulerian approach with PPM reconstruction
(requiring 3 Riemann solvers per cell), HLLC solver, hybrid van Leer
and generalized minmod slope limiter, dual energy formalism solving
the internal energy equation \\
\hline

Poisson solver
& successive over-relaxation (SOR)
& multigrid \\
\hline

Particle solver
& CIC interpolation, kick-drift-kick particle update
& CIC interpolation, drift-kick-drift particle update \\
\hline

Boundary condition
& fluid solver: outflow \newline
Poisson solver: isolated
& fluid solver: periodic$^a$ \newline
Poisson solver: isolated \\
\hline

Refinement
& (1) maximum particle mass in a cell: $1.0\times10^6 \Msun$ \newline
  (2) maximum gas      mass in a cell: $3.4\times10^2 \Msun$ \newline
  (3) resolving Jeans length by at least 64 cells \newline
  (4) 5 additional levels of statically refined regions above the
      root grid, enclosing volumes that are successively smaller by
      a factor of 8
& same as \gamer \\
\hline
Derefinement
& no explicit derefinement criteria
& same as \gamer \\
\hline

Time-step
& $C_{\rm par} = 0.5$, $C_{\rm CFL} = 0.5$
& same as \gamer \\
\hline

Parallelization
& hybrid MPI/OpenMP/GPU
& MPI and GPU acceleration$^b$ \\
\hline

Load balancing
& Hilbert space-filling curve
& same as \gamer \\
\hline

Time integration
& adaptive time-step$^c$
& same as \gamer \\
\hline

Floating-point format
& single precision
& same as \gamer \\
\hline

\end{tabular}
\\[0.2cm]
\raggedright \footnotesize
$^a$ The difference in the boundary conditions of the fluid solver is
     found to have negligible effect in this work.\\
$^b$ In \enzo, currently only the fluid and MHD solvers have been
     ported to GPUs.\\
$^c$ Both \gamer\ and \enzo\ do not restrict the time-step ratio
     between two adjacent levels to be a constant. \\
[0.2cm]
\label{table:agora_setup}
\end{table*}

\tref{table:agora_setup} summarizes the similarities and differences
of the numerical setup adopted by \gamer\ and \enzo\ in these
comparison simulations. Here we elaborate on the major differences.

\begin{itemize}
\item{
\gamer\ restricts all patches to have exactly the same size ($8^3$ cells
in this work), which has several advantages. For example, the AMR hierarchy
can be manipulated efficiently with an octree data structure.
The memory allocation is more predictable, which eases the issue of
memory fragmentation and maximizes memory reuse (see \sref{subsec:memory}).
The smaller patch size conceivably leads to a more efficient use
of both CPU and GPU caches. In addition, it is more straightforward
to optimize load balancing.
In comparison, \enzo\ allows all patches to have different
sizes. It reduces the number of cells that are unnecessarily refined
compared to \gamer. The relatively larger patches also reduce both the
computation and communication overhead associated with the ghost zones
of each patch. However, arguably, reconstructing the AMR hierarchy becomes
more expensive in this approach due to the more complicated grid
structure, which might deteriorate the parallel scalability, especially
in massively parallel simulations.
}
\item{
For the Poisson solver, \gamer\ uses the SOR scheme suitable for smaller
grids (see \sref{subsec:gpu_implementation}), while \enzo\ adopts the
multigrid scheme suitable for larger grids due to its higher convergence
rate. In addition, \gamer\ has a fixed patch size of $8^3$ cells and
adds 5 additional buffer zones around each patch to make the potential
smoother across the patch boundaries (see \fref{fig:poisson_grid}).
In comparison, \enzo\, has grid patches generally
larger than $8^3$ and also allocates a slightly larger buffer zones of 6
around each patch. Besides, it applies an iterative procedure to exchange
potential between sibling grids to improve the accuracy further,
which has not been implemented into \gamer. In this work, we adopt
10 such iterations in the \enzo\ simulations, which is measured to
increase the simulation time by $\about 10\%$.
}
\end{itemize}

\subsubsection{Accuracy comparison}
\label{subsec:agora_accuracy}

\begin{figure*}
\centering
\includegraphics[width=\textwidth]{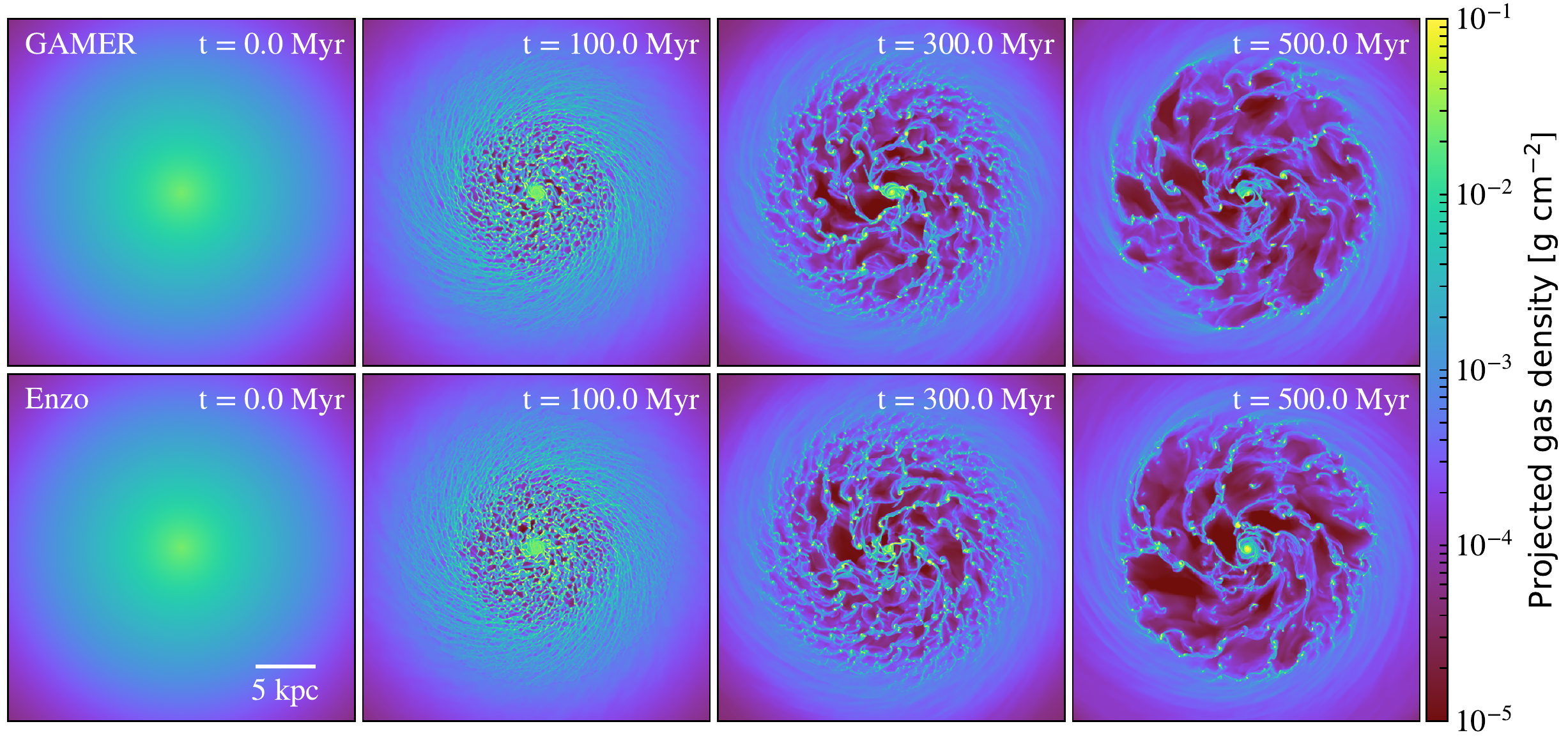}
\caption{
Face-on projection of gas density at four different epochs in the
isolated disk galaxy simulations. Each panel is $30 \kpc$ on a side.
The simulations with \gamer\ (upper panels) and \enzo\ (lower panels)
show very similar filamentary structures. Subtle differences are
expected to some extent because of the stochastic star formation
and the different numerical implementations
(see \tref{table:agora_setup}). See Figs.~\ref{fig:agora_profile}~--~\ref{fig:agora_sfr}
for more quantitative comparisons between the
two codes. At late times, a significant fraction of gas
has collapsed and merged into large gravitationally bound clouds and
there are no prominent spiral arms, mainly because we do not
include star formation feedback in this work
\citep[see Fig. 2 in][]{Goldbaum2016}.
}
\label{fig:agora_density}
\end{figure*}

\fref{fig:agora_density} shows the face-on projection of gas density
in the central $30 \kpc$ region at $t=0$, $100$, $300$, and $500 \Myr$
obtained by \gamer\ and \enzo. The filamentary structures form quickly
due to self-gravity, radiative cooling, and shear flow. These filaments
then continuously collapse into gravitationally bound clouds and trigger
star formation. We notice that, at later epochs of the simulations, a
significant fraction of gas has collapsed and merged into large clouds
and formed unrealistically massive star clusters, and there is no
prominent spiral structure. These results are inconsistent with the
smooth disk and prominent spiral arms observed in disk galaxies,
and are mainly due to the lack of star formation feedback in this work
that leads to over-cooling of gas
\citep[see Fig. 2 in][]{Goldbaum2016}.
Active galactic nucleus feedback is also expected to strongly change
the thermodynamics and kinematics of the multiphase gas
\citep[e.g.][]{Gaspari2018}.
It is, however, not a concern
here since we focus on the comparison between different codes.
\fref{fig:agora_density} shows that the gross morphological features
obtained by \gamer\ and \enzo\ agree well with each other.
Subtle differences are expected to some extent because of the stochastic
star formation and the different numerical implementations
(see \tref{table:agora_setup}). More quantitative comparisons are
provided below.

\begin{figure}
\centering
\includegraphics[width=\columnwidth]{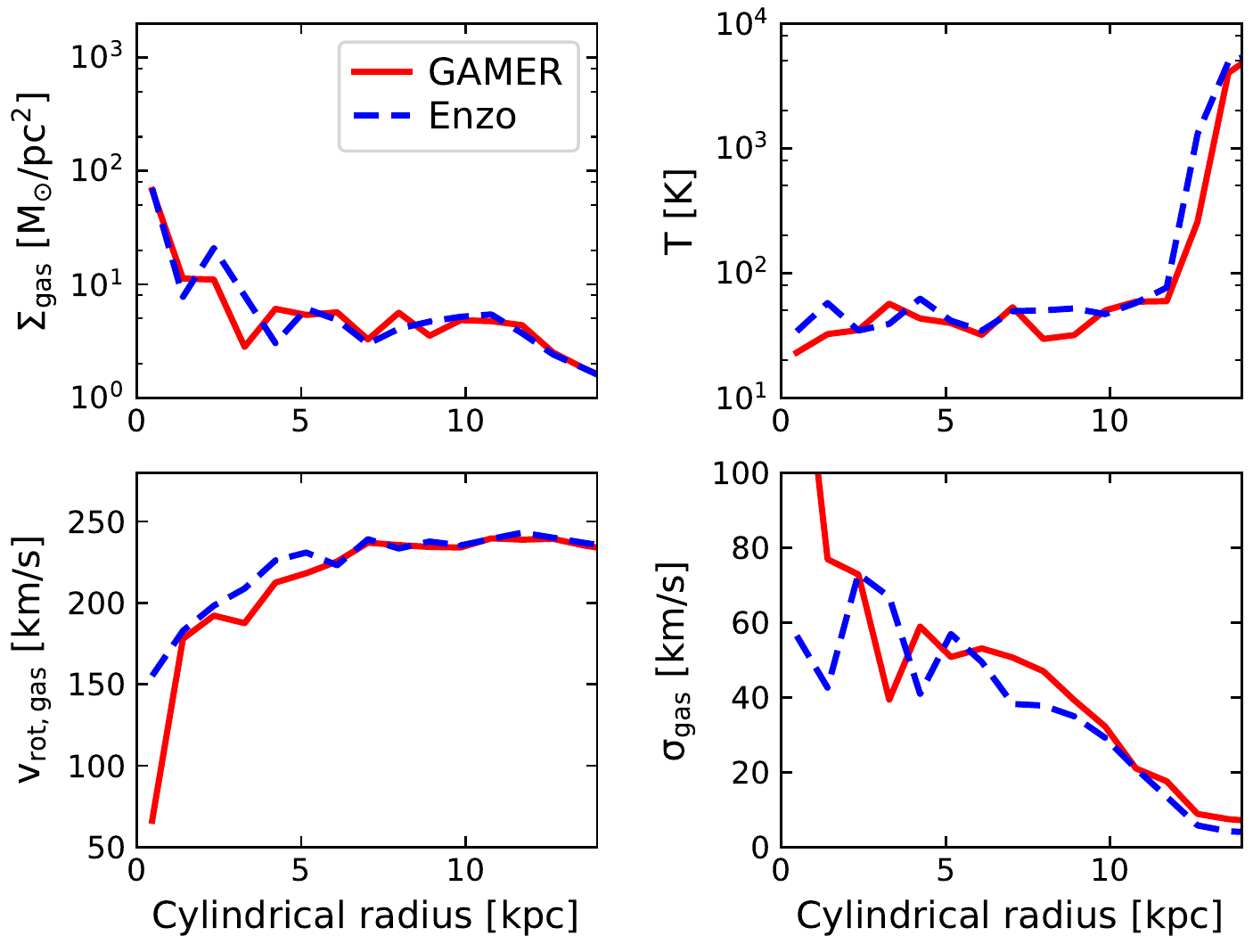}
\caption{
Azimuthally averaged profiles at $t=500 \Myr$ in the isolated disk
galaxy simulations. Different panels show the gas surface density
(upper left), gas temperature (upper right), gas rotation velocity
(lower left), and gas velocity dispersion (lower right).
The results of \gamer\ (red lines) and \enzo\ (blue lines) are
in good agreement with each other.
A relatively large discrepancy appears in the innermost region,
which is somewhat expected since the central region is sensitive to
both the determination of the galactic center and local mergers.
}
\label{fig:agora_profile}
\end{figure}

\fref{fig:agora_profile} shows the azimuthally averaged profiles
of various gas properties at $t=500 \Myr$, including the surface density,
temperature, rotation velocity, and velocity dispersion.
Following \citet{AGORA}, we set the galactic center to the location of
peak gas density within $1 \kpc$ from the center of gas mass.
All profiles exhibit clear oscillation, which become more prominent in
time as an increasing fraction of gas collapses into massive clouds.
The temperature within $\about 12 \kpc$ drops significantly from the
initial temperature of $10^4 \kelvin$ to below $\about 100 \kelvin$ due
to the balance between efficient metal line cooling and UV heating.
Substantial turbulent velocity dispersion, primarily driven by
gravitational instability, develops quickly and increases toward the
galactic center.
Importantly, we find very good agreement between \gamer\ and \enzo\
in all four profiles. A relatively large discrepancy appears in
the innermost region, which is somewhat expected since the central
region is sensitive to both the determination of the galactic center
and local mergers.

\begin{figure}
\centering
\includegraphics[width=\columnwidth]{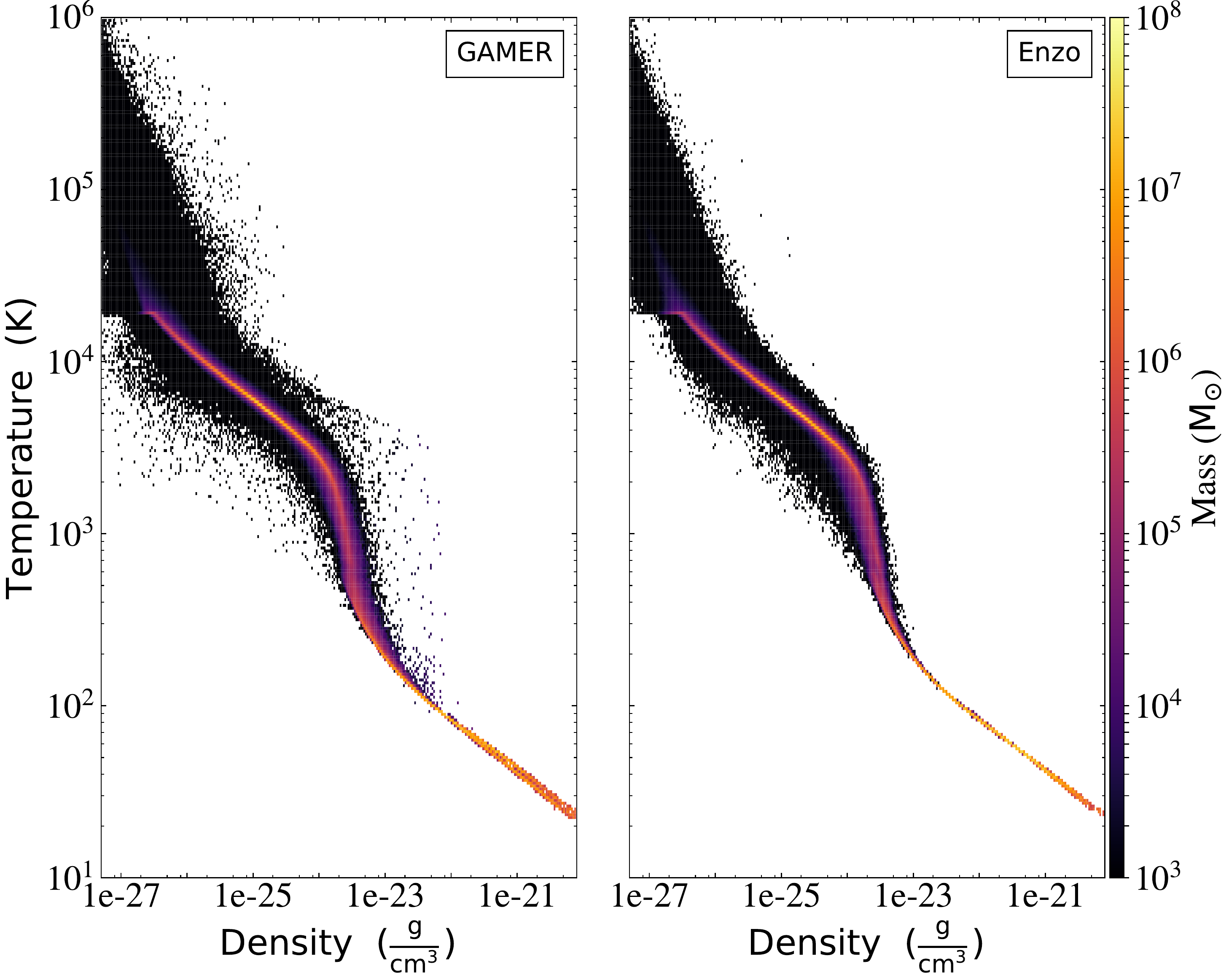}
\caption{
Probability distribution function of gas in the density-temperature
plane of the isolated disk galaxy simulations at $t=500 \Myr$.
Color bar represents the gas mass in each bin. The results obtained
by \gamer\ (left panel) and \enzo\ (right panel) are in very good
agreement thanks to the common library \grackle\ adopted for
calculating the chemistry and radiative processes.
}
\label{fig:agora_temp_dens}
\end{figure}

\fref{fig:agora_temp_dens} shows the probability distribution function
of gas in the density-temperature plane at $t=500 \Myr$. A clear branch
toward the high-density, low-temperature corner can be easily
identified, resulting from the balance between the various cooling and
heating mechanisms adopted in this work (see \sref{subsec:agora_setup}).
The low-density, high-temperature component in the upper left corner
corresponds to the gaseous halo included in the consideration of
numerical stability \citep{AGORA}. This figure further validates
the consistency of the gaseous thermodynamic properties between our
\gamer\ and \enzo\ simulations, thanks to the common chemistry and
radiative library \grackle. A relatively large scatter is found in the
\gamer\ simulation, which however constitutes a negligible fraction of
the total gas mass.

\begin{figure}
\centering
\includegraphics[width=\columnwidth]{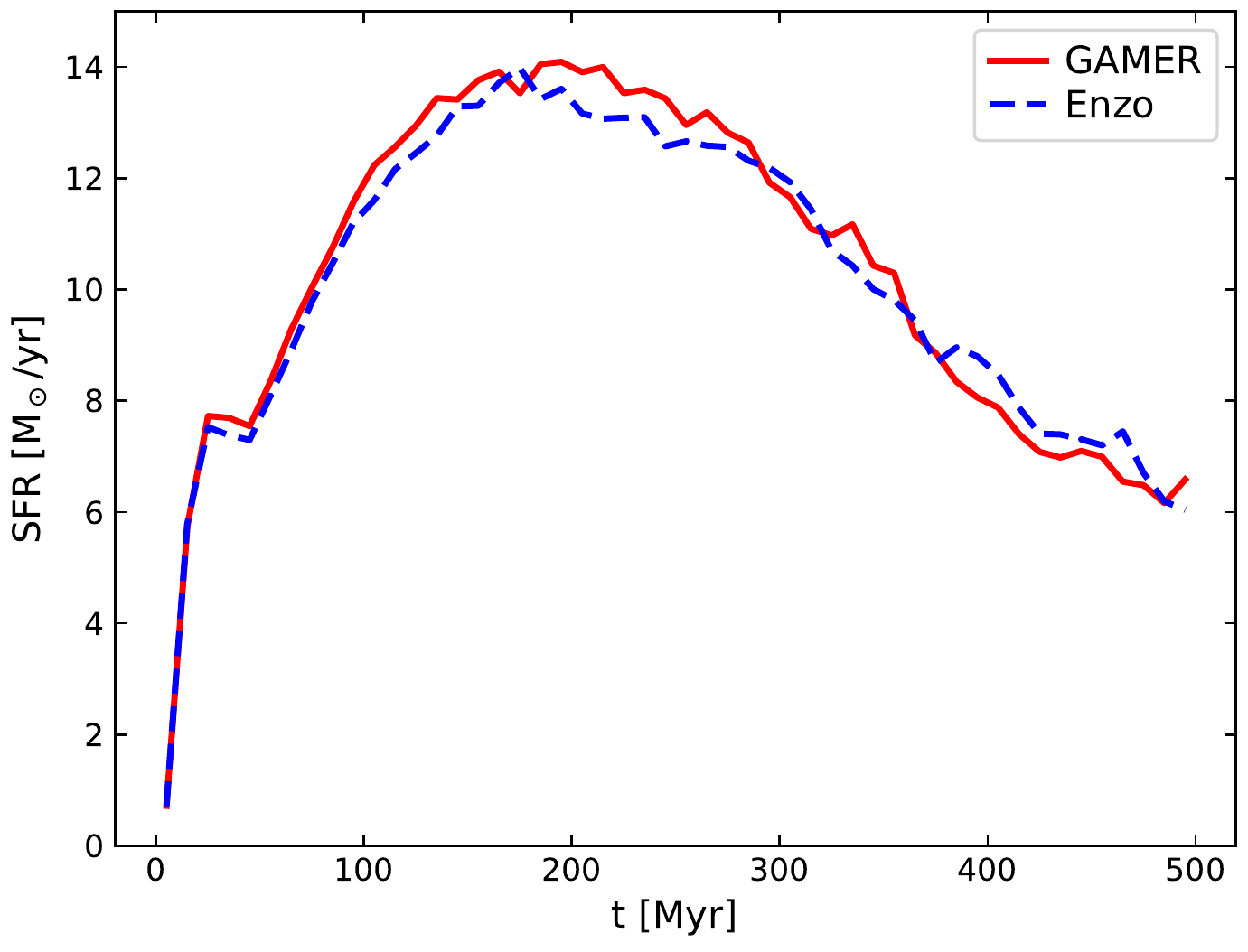}
\caption{
Star formation rate as a function of time in the isolated disk galaxy
simulations. We find very good agreement between \gamer\ (red line)
and \enzo\ (blue line).
}
\label{fig:agora_sfr}
\end{figure}

\fref{fig:agora_sfr} shows the star formation rate (SFR) as a function
of time, for which again \gamer\ and \enzo\ agree very well with
each other. The SFR reaches $\about 10 \Msun/{\rm yr}$ after
$t \gtrsim 100 \Myr$, consistent with the result of \citet{Goldbaum2015}
which also does not include star formation feedback, and is about a factor
of 5 higher than the SFR obtained in the simulations with feedback
\citep{AGORA,Goldbaum2016}. Interestingly, the consistency between
\gamer\ and \enzo\ shown in this figure seems to be better than
that found in the comparison of grid codes in the AGORA comparison project
\citep[see Fig. 26 in][]{AGORA}. It remains to be investigated whether
this level of consistency could be achieved after including feedback.

The agreement between the simulations results of \gamer\ and \enzo,
as verified in Figs.~\ref{fig:agora_density}~--~\ref{fig:agora_sfr},
demonstrates the consistent numerical setup adopted for this code
comparison experiment, including, for example, the initial condition,
spatial and temporal resolution, and grid refinement criteria. It also
suggests that the differences between the two codes described in
\sref{subsec:agora_setup}, for example, the AMR implementation, fluid
and Poisson solvers, and particle integration, do not have a serious
impact here. These facts strengthen the results of performance comparison
shown in the next section.

\subsubsection{Performance comparison}
\label{subsec:agora_performance}

Here we compare strong scaling performance between \gamer\ and \enzo\
measured on Blue Waters. Since \enzo\ also supports GPU acceleration,
although only for the hydrodynamic and MHD solvers, we run both codes
on the XK nodes, each of which is composed of one GPU (NVIDIA Tesla K20X)
and one 16-core CPU (AMD Opteron 6276). Similar to the cluster merger
simulations described in \sref{subsec:merger_performance}, for \gamer\
we launch 2 MPI processes per node and 7 OpenMP threads per MPI process
to improve memory affinity, and use CUDA MPS to allow these two processes
to share the same GPU. For \enzo, we launch 16 MPI processes per node
since it does not support hybrid MPI/OpenMP parallelization.
In addition, for $\Nnode=1-4$, we disable
\texttt{SubgridSizeAutoAdjust}\footnote{See https://enzo.readthedocs.io/en/latest/index.html.}
and set \texttt{MaximumSubgridSize=8192} to avoid exhausting the
GPU memory due to the too large grid size. Changing
\texttt{MaximumSubgridSize} by a factor of 2 reduces the
performance of $\Nnode=4$ by $\about 10-20\%$.
For $\Nnode=8-128$, we enable \texttt{SubgridSizeAutoAdjust} and have
\texttt{OptimalSubgridsPerProcessor=16}, which are found to generally
achieve the best performance in our tests. The Hilbert space-filling
curve is adopted for load balancing in both \gamer\ and
\enzo\footnote{It corresponds to \texttt{LoadBalancing=4}, which is not
officially supported but works well in our tests.}.
We measure the performance in a relatively short period of
$t=300 - 305 \Myr$, which is representative enough since both the
total number of cells and evolution time-step are found to be quite
stable after $t \gtrsim 100 \Myr$. At $t=300 \Myr$, there are
$\about 2.2\times10^8$ cells in total, $\about 20\%$ of which are on
the maximum refinement level.

\begin{figure}
\centering
\includegraphics[width=\columnwidth]{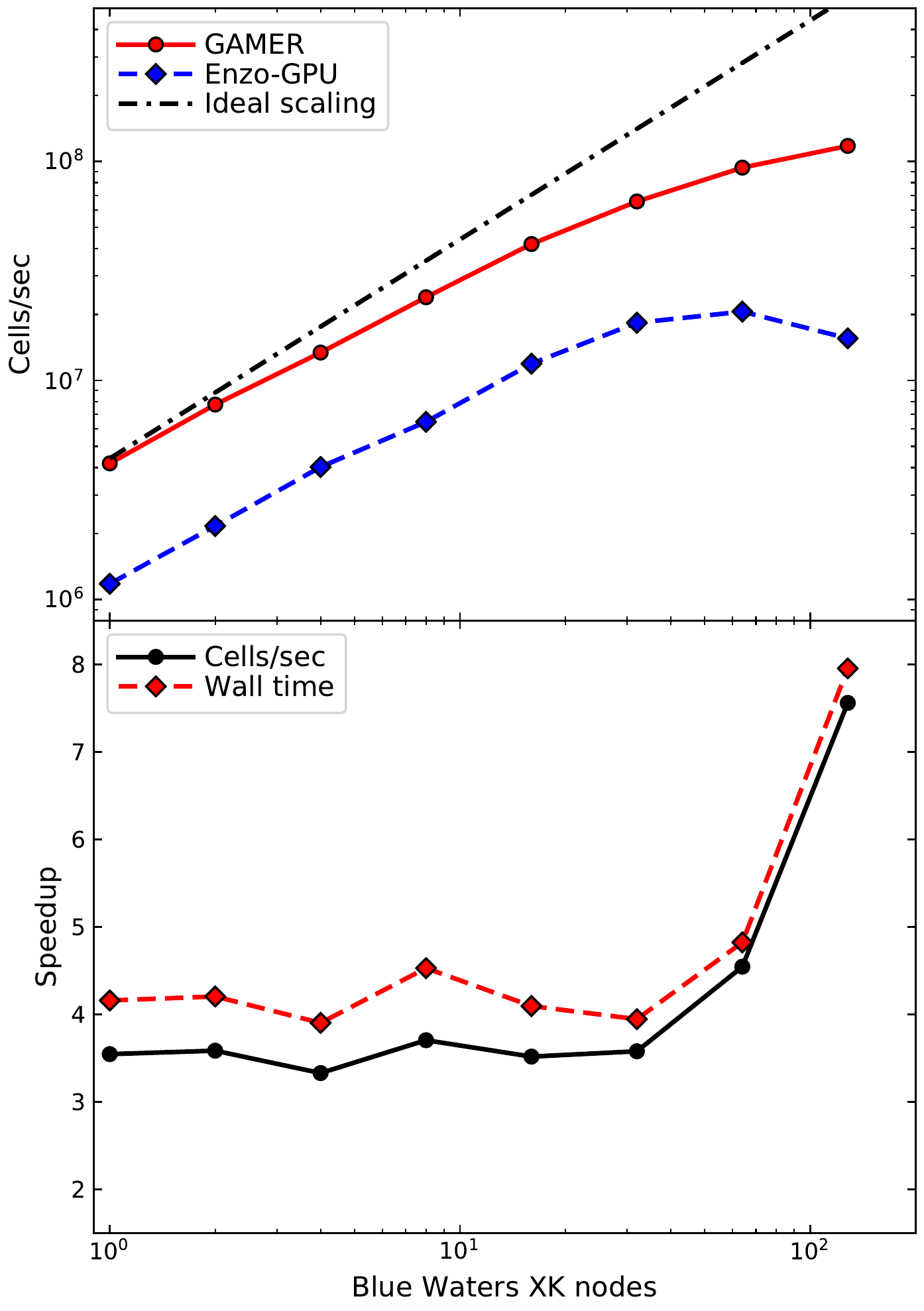}
\caption{
Strong scaling of the isolated disk galaxy simulations run with \gamer\
(solid line) and \enzo\ (dashed line) using $1-128$ nodes.
Both codes support GPU acceleration and run on the Blue Waters XK nodes,
each of which is composed of one GPU (NVIDIA Tesla K20X) and one
16-core CPU (AMD Opteron 6276). The upper panel shows the total number
of cell updates per second, where the dashed-dotted line represents the
ideal scaling. The lower panel shows the speedup of \gamer\ over \enzo\
in terms of both cell updates per second (solid line) and total
wall time (dashed line). \gamer\ is measured to be $\about 4-8$ times
faster than \enzo. More importantly, this speedup ratio is approximately
a constant for $\Nnode \le 32$ and increases for $\Nnode > 32$, suggesting
that \gamer\ exhibits better parallel scalability than \enzo.
See \fref{fig:agora_strong_scaling_metrics} for the detailed
performance metrics of this test.
}
\label{fig:agora_strong_scaling}
\end{figure}

\fref{fig:agora_strong_scaling} shows the strong scaling of the
isolated disk galaxy simulations. The speedup of \gamer\ over \enzo\
is measured to be about $4-8$ in terms of both total wall time
and cell updates per second. For example, for $\Nnode=64$,
\gamer\ and \enzo\ achieve $1.5\times10^6$ and $3.2\times10^5$
cell updates per second per node, respectively.
This result is encouraging, especially because both codes take
advantage of GPU acceleration.
More importantly,
this speedup ratio is approximately a constant for $\Nnode \le 32$
and \emph{increases} for $\Nnode > 32$. The overall performance
of \enzo\ starts to drop for $\Nnode > 64$, while that of \gamer\
starts to drops for $\Nnode > 128$. These results suggest that \gamer\
not only runs faster but also scales better than \enzo.

\begin{table}
\setlength{\tabcolsep}{4pt}
\caption{
Comparison of the volume-filling fractions on higher refinement
levels between \gamer\ and \enzo\ in the isolated disk galaxy
simulations at $t=300\Myr$.
}
\centering
\begin{tabular}{crccrr}
\hline\hline

Level & $\Delta h/\pc$ & \multicolumn{2}{c}{Filling fraction} & \multicolumn{2}{c}{Number of cells} \\
      &                 & \gamer & \enzo                       & \gamer & \enzo \\
\hline

 7  &  160.0  &  $2.4\times10^{-3}$\%  &  $1.0\times10^{-3}$\%  &  $1.3\times10^7$  &  $5.7\times10^6$ \\
 8  &   80.0  &  $4.3\times10^{-4}$\%  &  $2.0\times10^{-4}$\%  &  $1.9\times10^7$  &  $8.8\times10^6$ \\
 9  &   40.0  &  $1.2\times10^{-4}$\%  &  $6.3\times10^{-5}$\%  &  $4.2\times10^7$  &  $2.2\times10^7$ \\
10  &   20.0  &  $1.5\times10^{-5}$\%  &  $1.5\times10^{-5}$\%  &  $4.2\times10^7$  &  $4.1\times10^7$ \\
\hline

\end{tabular}
\\[0.2cm]
\label{table:agora_grids}
\end{table}

In \fref{fig:agora_strong_scaling}, we show the performance speedups
in terms of both total wall time and cell updates per second.
The former arguably provides a more comprehensive comparison because
it considers not only the performance of all PDE solvers but also many
other factors such as the AMR implementation and evolution time-step.
In this test, the speedup in terms of total wall time is measured
to be $\about 5-20\%$ higher than that in terms of cell updates
per second, partially because \gamer\ generally requires less extra
updates for synchronizing nearby AMR levels
(see \sref{subsec:AMR}, especially \fref{fig:adaptive_dt}).
On the other hand, although \gamer\ and \enzo\ allocate roughly the same
number of cells on the maximum refinement level $l=10$
(the difference is less than $3\%$), we find that \gamer\ typically allocates
$\about 50\% - 150\%$ more cells than \enzo\ on other high levels
(e.g.\  $l=7-9$, see \tref{table:agora_grids}).
It is mainly because \gamer\ restricts all
patches to have the same size which results in over-refinement,
especially along the direction perpendicular to the galactic disk
and on the levels with cell sizes much larger
than the disk scale height. This issue, however, does not pose a
serious problem here since lower levels are updated much less
frequently thanks to the adaptive time-step integration.

\begin{figure*}
\centering
\includegraphics[width=12cm]{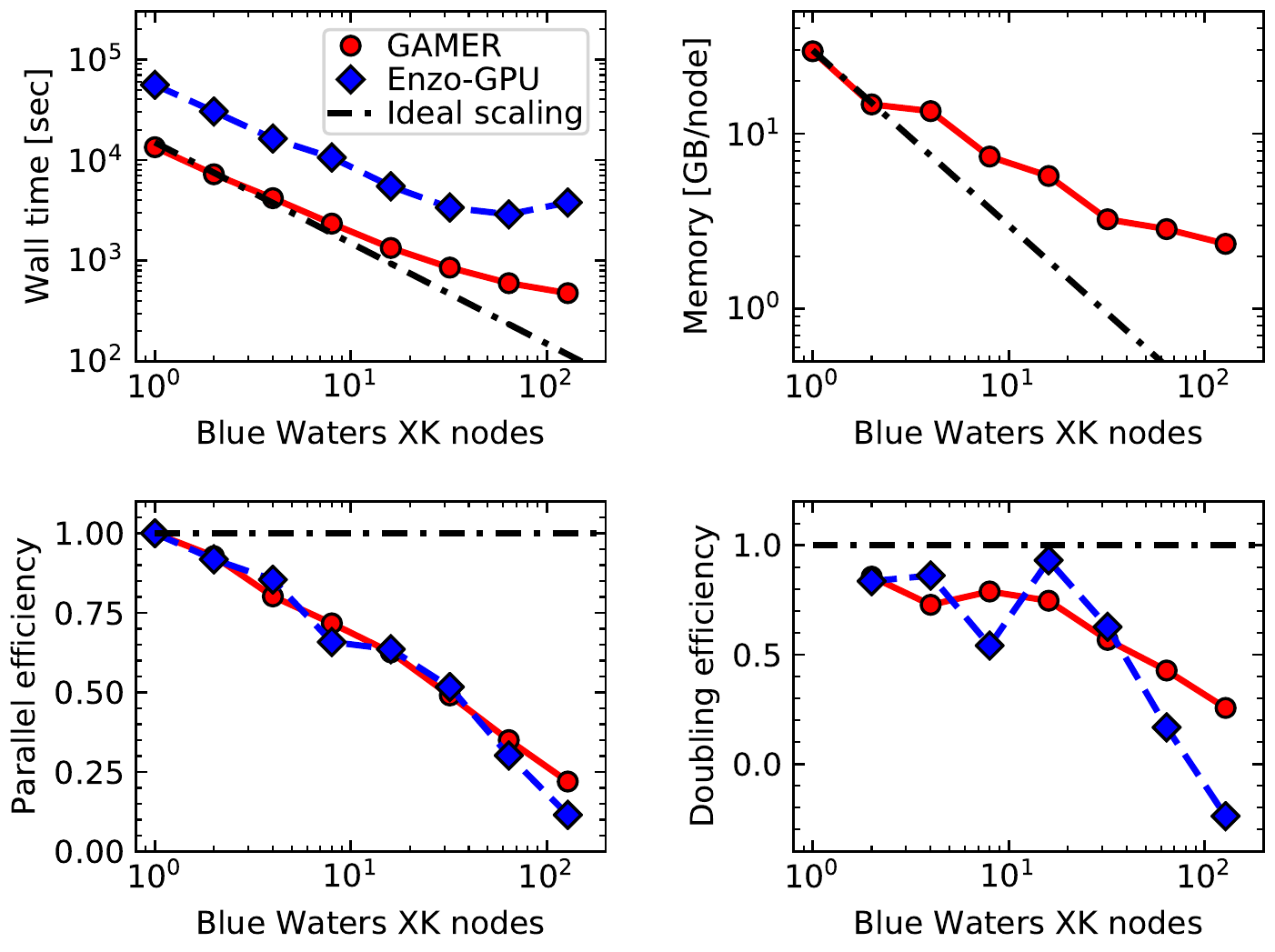}
\caption{
Performance metrics of the strong scaling of the isolated disk galaxy
simulations. This is complementary to and uses the same symbols as
\fref{fig:agora_strong_scaling}. Different panels show the
total wall time (upper left),
maximum CPU memory consumption per MPI process (upper right),
parallel efficiency (lower left), and
doubling efficiency (lower right).
See Equations~(\ref{eq:pe_strong}) and~(\ref{eq:de_strong}) for the
definitions of parallel and doubling efficiencies in strong scaling.
\gamer\ and \enzo\ exhibit very similar parallel
scalability for $\Nnode \le 32$, and \gamer\ scales noticeably better
than \enzo\ for $\Nnode > 32$.
The maximum per-process memory consumption of \enzo\ is not shown
since the data are not available.
}
\label{fig:agora_strong_scaling_metrics}
\end{figure*}

\fref{fig:agora_strong_scaling_metrics} shows the performance metrics
of the isolated disk galaxy simulations, including the total wall time,
maximum CPU memory consumption per MPI process, parallel efficiency,
and doubling efficiency as a function of the number of XK nodes.
Most importantly, both the parallel and doubling efficiencies
demonstrate that the two codes exhibit very similar parallel
scalability for $\Nnode \le 32$, and, furthermore, \gamer\ scales
noticeably better than \enzo\ for $\Nnode > 32$, consistent with
\fref{fig:agora_strong_scaling}.

We notice that the doubling efficiency of \enzo\ oscillates for
$\Nnode=4-16$, likely because, as described in the beginning of this
section, we enable \texttt{SubgridSizeAutoAdjust} for $\Nnode > 4$
to improve performance and scalability. Also note that the CPU memory
consumption per MPI process in \gamer\ deviates from the ideal scaling,
especially when increasing the number of nodes, most likely due to
the allocation of buffer patches and MPI buffers.

\begin{figure}
\centering
\includegraphics[width=\columnwidth]{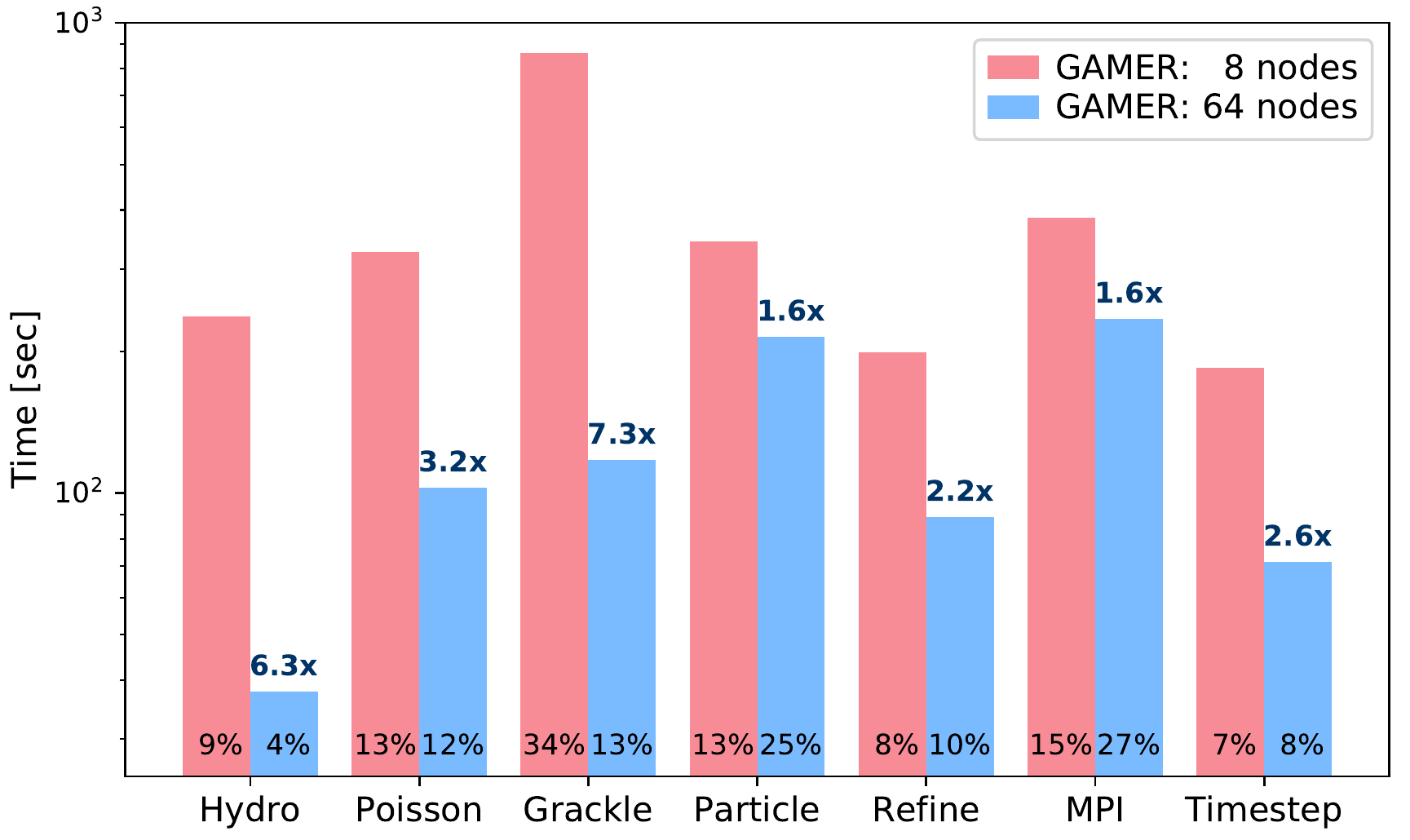}
\caption{
Wall time of various operations measured from the isolated disk
galaxy simulations with \gamer\ using 8 and 64 nodes. For each
operation, we also show the fraction of time in the two runs and the
speedup gained from increasing $\Nnode=8$ to $64$. The `MPI' time
includes transferring both grids and particles for the `Hydro',
`Poisson', and `Particle' routines but excludes the MPI
communication during the grid refinement, which is included in the
`Refine' time. The `Poisson' time includes depositing particle
mass onto grids with the CIC interpolation. It shows that the
chemistry and cooling library \grackle\ is the major performance
bottleneck for $\Nnode=8$, while the particle routines and MPI
communication become the bottlenecks for $\Nnode=64$ because these
operations exhibit poor scalability. Note that, for better
clarification, the performance shown here does not consider particle
weights for load balancing, which is therefore different from the
optimized performance shown in Figs.~\ref{fig:agora_strong_scaling}
and~\ref{fig:agora_strong_scaling_metrics}. See text for details.
Also note that a GPU-supported \grackle\ is currently under
development, which would likely lead to larger speedups.
}
\label{fig:agora_time_fraction}
\end{figure}

To determine the performance bottleneck of \gamer\ in the isolated
disk galaxy simulations, especially for large $\Nnode$, we compare in
\fref{fig:agora_time_fraction} the wall time of various operations
in the $\Nnode=8$ and $64$ runs. We find that, for $\Nnode=8$,
the chemistry and cooling library \grackle\ is the major performance
bottleneck. In contrast, for $\Nnode=64$, the bottlenecks move to
the particle routines and MPI communication, which is similar to what
we find in the cluster merger simulations
(see \fref{fig:merger_time_fraction}). It is mainly because, currently,
the load balancing in \gamer\ is better optimized for grid solvers like
\grackle\ than for particle related routines. From $\Nnode=8$ to $64$,
the fraction of time spent on \grackle\ decreases from $34\%$ to $13\%$,
while that on the particle routines and MPI communication increase from
$13\%$ and $15\%$ to $25\%$ and $27\%$, respectively.
Also note that a GPU-supported \grackle\ is currently under
development, which would likely lead to larger speedups.

There are several things to note about these results, which are similar
to the discussions given in the cluster merger simulations.
First of all, the average MPI message size per process for $\Nnode=64$
is found to be as small as $\about 1$ MB, which is smaller than that
for $\Nnode=8$ by a factor of few. The $\Nnode=64$ run thereby
suffers from a relatively larger communication latency. Second, for
better clarification, the performance shown in \fref{fig:agora_time_fraction}
is measured from a separate set of simulations that (i) adds an explicit
MPI synchronization between grid and particle routines in order for a proper
timing analysis, and (ii) disregards particle weights in load balancing
(i.e.\ $W_{\rm par}=0.0$), both of which deteriorate the
scalability. For $\Nnode=64$, we find that the overall performance
is improved by $\about 22\%$ after removing that extra MPI synchronization
between grid and particle routines, and by another $\about 23\%$ after
adopting $W_{\rm par}=1.0$. These optimizations have been incorporated into
the strong scaling shown in Figs.~\ref{fig:agora_strong_scaling} and
~\ref{fig:agora_strong_scaling_metrics}.
Third, note that the time fraction of the Poisson solver shown in
\fref{fig:agora_time_fraction} includes the time for depositing particle
mass onto grids, which is why it scales worse than the
other two grid solvers (i.e.\ hydrodynamic and \grackle\ solvers).

\section{Summary and Future Work}
\label{sec:summary}

\renewcommand{\arraystretch}{1.5}
\setlength{\tabcolsep}{0.3cm}
\begin{table*}
\caption{
Summary of the major differences between \gamer\ and \gamerone.
}
\centering
\begin{tabular}{m{2.8cm}m{5.5cm}m{5.5cm}m{1.6cm}}
\hline\hline

Features & \gamer\ & \gamerone & References \\
\hline

Adaptive time-step
& fully supported without requiring the time-step ratio between two
  adjacent levels to be a constant
& only supported in pure hydrodynamic simulations and the time-step ratio
   between two adjacent levels must be 2
& $\S$ \ref{subsec:AMR} \\
\hline

Fluid solvers
& both dimensional split, Riemann-solver-free scheme (RTVD) and
  dimensional unsplit, Riemann-solver-based schemes (MHM, VL, CTU)
  with PLM/PPM data reconstructions and various Riemann solvers
& RTVD only
& $\S$ \ref{subsec:hydro} \\
\hline

Dual energy formalism
& supported
& unsupported
& $\S$ \ref{subsec:hydro} \\
\hline

MHD
& supported using the CTU+CT scheme
& not supported
& Zhang et al. submitted \\
\hline

Poisson solver
& adding 5 buffer zones around each refinement-level patch
& multi-level solver that eliminates the pseudomass sheets on the
  patch interfaces
& $\S$ \ref{subsec:gravity} \\
\hline

Gravity in hydro
& operator-unsplit approach
& operator-split approach
& $\S$ \ref{subsec:gravity} \\
\hline

Boundary conditions
& fluid: periodic, outflow, reflecting, inflow (i.e.\ user-defined) \newline
  gravity: periodic, isolated
& fluid: periodic \newline
  gravity: periodic
& $\S$ \ref{subsec:hydro}, $\S$ \ref{subsec:gravity} \\
\hline

Particles
& supported
& unsupported
& $\S$ \ref{subsec:particles} \\
\hline

\grackle
& supported
& unsupported
& $\S$ \ref{subsec:grackle} \\
\hline

Bitwise reproducibility
& supported
& unsupported
& $\S$ \ref{subsec:bitwise} \\
\hline

HDF5 output
& supported
& unsupported
& $\S$ \ref{subsec:analysis} \\
\hline

\yt\ data analysis
& supported
& unsupported
& $\S$ \ref{subsec:analysis} \\
\hline

Test problem infrastructure
& supported
& unsupported
& $\S$ \ref{subsec:test_problem} \\
\hline

AMR+GPUs framework
& supported (e.g.\ $\psi$DM simulations)
& unsupported
& $\S$ \ref{subsec:framework} \\
\hline

Hybrid MPI/OpenMP
& supported
& unsupported
& $\S$ \ref{subsec:hybrid} \\
\hline

Parallelization
& Hilbert curve for load balancing
& rectangular domain decomposition
& $\S$ \ref{subsec:load_balancing} \\
\hline

Memory pool
& supported
& unsupported
& $\S$ \ref{subsec:memory} \\
\hline

\end{tabular}
\\[0.2cm]
\label{table:gamer1_vs_gamer2}
\end{table*}

In this paper, we have presented \gamer, a significant revision of
the GPU-accelerated adaptive mesh refinement code \gamerone\
\citep{Schive2010}. It includes much richer functionality and
incorporates significant improvements in accuracy, stability,
performance, and scalability. \tref{table:gamer1_vs_gamer2}
summarizes the major differences between \gamer\ and \gamerone.

To reveal the optimal performance of \gamer, we first measure the
performance of individual GPU solvers and show that both the
hydrodynamic and Poisson solvers achieve a single-precision performance
of $\about 2\times10^8 \CellsPerSec$ on an NVIDIA Tesla P100 GPU.
We also measure the weak scaling performance with and without AMR
in a three-dimensional Kelvin-Helmholtz instability test on the
Blue Waters supercomputer using $1 - 4096$ XK nodes, each of which
is composed of one NVIDIA Tesla K20X GPU and one 16-core AMD
Opteron 6276 CPU. By taking advantage of the hybrid MPI/OpenMP/GPU
parallelization, we are able to fully exploit both 4096 GPUs and
65,536 CPU cores simultaneously, and achieve a peak performance of
$8.3\times10^{10} \CellsPerSec$ and $4.6\times10^{10} \CellsPerSec$
and a parallel efficiency of $74\%$ and $58\%$ in the uniform-grid
and AMR tests, respectively. Note that the simulation reaches an
overall resolution as high as $10{,}240^3$ cells with 4096 nodes in
the uniform-grid test.

To further provide clear and convincing demonstrations of the accuracy
and performance of \gamer, we directly compare \gamer\ with two widely
adopted AMR codes, \flash\ and \enzo, based on realistic astrophysical
simulations running on Blue Waters. First, we compare \gamer\ with
\flash\ in binary cluster merger simulations, which closely follow the
numerical setup of \citet{ZuHone2011} and involve hydrodynamics,
self-gravity, and dark matter. We show that the physical results
obtained by the two codes are in excellent agreement, and \gamer\ is
measured to be $78 - 101$ times faster than \flash\ in strong scaling
tests using $1 - 256$ nodes. More importantly, both codes exhibit
similar parallel scalability, despite the fact that the computational
time of \gamer\ has been greatly reduced by exploiting GPUs. We also
measure the strong scaling of \gamer\ from 16 to 2048 nodes using a
set of higher-resolution simulations, and obtain a parallel efficiency
of $78\%$ with 128 nodes and $37\%$ with 1024 nodes.

Second, we compare \gamer\ with \enzo\ in isolated disk galaxy
simulations, which closely follow the numerical setup of
\citet{Goldbaum2015} and the AGORA High-resolution Galaxy
Simulations Comparison Project \citep{AGORA} but with a spatial
resolution of $20 \pc$. These simulations involve a richer set of
physical modules, including hydrodynamics, self-gravity, dark matter,
advection of metals, radiative cooling and heating, and stochastic
star formation. Again, we find very good agreement between the
physical results obtained by \gamer\ and \enzo. To compare the
performance, we also enable GPU acceleration in \enzo\ for the
hydrodynamic solver. Even so, \gamer\ is still measured to be
$\about 4-8$ times faster than \enzo\ in strong scaling tests using
$1 - 128$ nodes.
It may be partially due to the fact that Enzo currently does not
support asynchronous GPU kernel execution and CPU-GPU communication.
Further investigation will be conducted in the future.
More importantly, this speedup ratio is
approximately a constant of $4-5$ with $1 - 32$ nodes and increases
to $5-8$ when using more than 32 nodes, suggesting that \gamer\ not
only runs faster but also scales noticeably better than \enzo.

\gamer\ has supported the following features to improve the parallel
scalability:

\begin{itemize}
\item Hybrid OpenMP/MPI parallelization (see \sref{subsec:hybrid}).
      It reduces inter-node communication and therefore improves the
      parallel scalability, especially when using a large number of
      nodes.
\item Fixed patch size. It greatly simplifies
      the parallel manipulation of AMR hierarchy and load balancing,
      especially in massively parallel simulations.
      Moreover, we do not require duplicating the entire AMR
      hierarchy on each MPI process (see \sref{subsec:load_balancing}).
\item Level-by-level load balancing with Hilbert space-filling
      curves (see \sref{subsec:load_balancing}).
      Especially, we take into account the particle weights in
      load balancing (see \fref{fig:particle_load_balancing}), and
      further minimize the MPI synchronization between grid and
      particle routines.
\end{itemize}

\gamer\ allocates memory pools for both grid and particle data to
alleviate the issue of memory fragmentation and to maximize memory reuse
(see \sref{subsec:memory}). Moreover, the code minimizes the GPU memory
requirement by storing all the data on the CPU's main memory and
transferring only a small and fixed amount of patch data to GPUs
(typically several hundreds of MB to a few GB per GPU) at a time.

We have identified several performance bottlenecks from the detailed
timing analysis conducted in this work (e.g.\ see
Figs.~\ref{fig:merger_time_fraction} and~\ref{fig:agora_time_fraction}),
including load imbalance due to particles, \grackle\ library, MPI
communication, and CPU performance when preparing the input data
for GPU solvers. To improve performance further, we are currently
investigating the following optimizations:

\begin{itemize}
\item Transferring the deposited particle mass density on grids
      instead of individual particles when calculating the total mass
      density on levels other than the maximum level. This will
      greatly reduce the MPI communication for particles and also
      improve load balancing.
\item Porting some of the particle routines to GPUs.
\item MPI non-blocking communication. It will allow overlapping
      MPI communication by both CPU and GPU computations.
\item GPU-accelerated \grackle. \grackle\ computes the chemistry and
      radiative processes on a cell-by-cell basis, which should
      be very GPU-friendly because no synchronization and data exchange
      between different cells are required. We have obtained an
      order of magnitude speedup in preliminary tests.
\item Optimization of CPU routines. One important optimization in
      \gamer\ is to allow CPUs and GPUs to work concurrently
      (see \sref{subsec:hybrid}). Accordingly, depending on the CPU
      and GPU specifications, we find that the performance
      bottleneck may occur in CPUs when invoking a GPU kernel (since we
      still rely on CPUs to prepare the input data for GPUs). It is
      therefore essential to optimize the CPU performance further by,
      for example, improving the OpenMP parallel efficiency, porting
      more operations to GPUs, and optimizing memory access.
\end{itemize}

Given the excellent performance reported here, it is then
essential to extend the functionality of \gamer\ so that it can be
applied to a broader range of applications. The following new
features are being or will be investigated in the near future.
In addition, since \gamer\ is an open-source code, we are also
looking forward to contributions from the community.

\begin{itemize}
\item New particle features including tracer particles, comoving
      coordinates, and multiple species. We also plan to store
      `relative' instead of absolute particle positions, which can
      be very helpful for simulations demanding an extremely large
      dynamic range.
\item Non-Cartesian coordinates.
\item Non-ideal hydrodynamics and non-ideal MHD.
\item Radiative transfer.
\item Parallel I/O.
\item Testing framework for ensuring the correctness of the code.
\end{itemize}

Note that \gamer\ can also run in a `CPU-only' mode using a hybrid
MPI/OpenMP parallelization, for which we simply replace all GPU
solvers by their CPU counterparts parallelized with OpenMP and use
the same MPI implementation as in the GPU-accelerated code.
Therefore, \gamer\ is also suitable for CPU-only supercomputers,
especially for those with a larger number of cores per node like
Intel Xeon Phi. Hybrid MPI/OpenMP is essential to achieve optimal
performance in such systems, and it may require further optimization
about, for example, thread affinity, thread load balancing, and OpenMP
nested parallelism.

Finally, we emphasize that the great performance and scalability of
\gamer\ demonstrated here in both binary cluster merger and isolated
disk galaxy simulations allow one to study various astrophysical
phenomena requiring resolutions that are not realistically attainable
previously. For example, for the cluster merger simulations, we have
obtained preliminary results from simulations with sub-kpc resolution,
which will enable us to reduce the numerical viscosity significantly
and to investigate the properties of the turbulent
cascade down to a scale where the
effects of a physical viscosity are expected to become relevant.
It is also possible to increase the spatial resolution of isolated
disk galaxy simulations to $\about 5\pc$, which will produce a
dynamically evolving interstellar medium, undergoing repeated cycles
of collapse, star formation, feedback, rarefaction, and re-collapse
which have been extremely difficult to fully resolve in a
global galactic scale simulation over galactic dynamical times.

\section{Acknowledgement}
\label{sec:acknowledgement}

H.S. would like to thank Edward Seidel, Gabrielle Allen, and
Tzihong Chiueh for their great support on this project, and
Roland Haas and Brian O'Shea for stimulating discussions.
H.S. would also like to thank Sandor Molnar for helping implement
the galaxy cluster merger simulations
and Hsiang-Chih Hwang for helping implement particles into \gamer.
H.S. and M.G. are grateful to James Stone for insightful discussions.
The authors are also grateful to Britton Smith for helping incorporate
\grackle\ into \gamer.
Finally, we want to thank the referee, Michael Norman, for a
constructive report that helped to improve the paper.
This publication is supported in part by the Gordon and Betty Moore
Foundation's Data-Driven Discovery Initiative through Grant GBMF4561
to Matthew Turk, and is based upon work supported by the National
Science Foundation under Grant No. ACI-1535651.
This research is part of the Blue Waters sustained-petascale computing
project, which is supported by the National Science Foundation
(awards OCI-0725070 and ACI-1238993) and the state of Illinois.
Blue Waters is a joint effort of the University of Illinois at
Urbana-Champaign and its National Center for Supercomputing
Applications (NCSA). This work also used computational resources
provided by the Innovative Systems Lab (ISL) at NCSA.
M.G. is supported by NASA through Einstein Postdoctoral Fellowship
Award Number PF5-160137 issued by the Chandra X-ray Observatory
Center, which is operated by the SAO for and on behalf of NASA under
contract NAS8-03060. Support for this work was also provided by
Chandra grant GO7-18121X.

\bibliographystyle{mnras}
\bibliography{ref}

\appendix

\section{Poisson Solver}
\label{sec:poisson}

The Poisson solver of \gamer\ on the refined patches is
substantially different from that of \gamerone. To smooth out the
gravitational potential across patch boundaries, \gamer\ adds several
buffer zones around each patch (see \fref{fig:poisson_grid}) while
\gamerone\ adopts a multi-level relaxation scheme to reduce the
pseudomass sheets on the patch boundaries \citep{Schive2010}.
To compare their accuracy, we calculate the potential of a Hernquist
profile \citep{Hernquist1990}:
\be
\rho(r) = \frac{\rho_0}{r/r_0(1+r/r_0)^3},
\label{eq:hernquist_density}
\ee
where $r_0$ and $\rho_0$ are the characteristic radius and density,
respectively. This profile has a finite mass $M=2\pi r_0^3 \rho_0$
and an analytical form of potential:
\be
\phi_{\rm anal}(r) = -\frac{GM}{r + r_0},
\label{eq:hernquist_potential}
\ee
where $G$ is the gravitational constant. We adopt $G=r_0=\rho_0=1$.
The computational domain is cubic with a length $L=100$ and a $64^3$
root grid. A cell on level $l$ is flagged for refinement if its
density exceeds $10^{-2}\times4^l$, and we enable six refinement levels
to well resolve $r_0$ by a maximum resolution of
$\about 2.4\times10^{-2}$. Isolated boundary conditions for gravity
are adopted.

\begin{figure*}
\centering
\includegraphics[width=15cm]{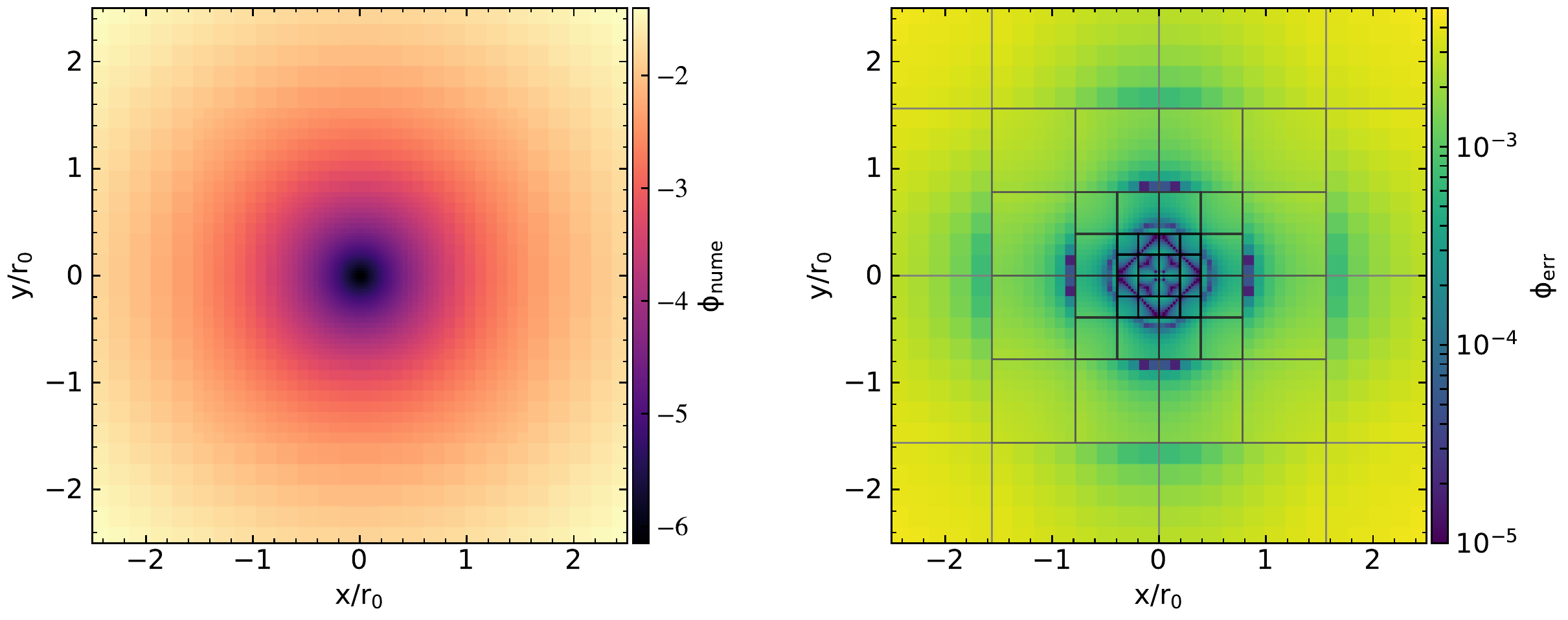}
\caption{
Central slice of the gravitational potential of a Hernquist profile.
The left panel shows the numerical results evaluated by \gamer\ using
5 buffer zones and 6 refinement levels. The right panel shows the
corresponding relative errors by comparing with the analytical
solution, overlaid with AMR patch outlines.
}
\label{fig:poisson_error_slice}
\end{figure*}

\fref{fig:poisson_error_slice} shows the gravitational potential
on a central $5\,r_0$ slice evaluated by \gamer\ using 5 buffer zones.
The left and right panels show the numerical results $\phi_{\rm nume}$
and the corresponding relative errors,
$\phi_{\rm err} \equiv |(\phi_{\rm nume}-\phi_{\rm anal})/\phi_{\rm anal})|$,
respectively. The relative errors within $r \lesssim r_0$ are found to
be as low as on the order of $10^{-3}-10^{-5}$, although numerical
artefacts introduced by the patch interfaces are still present.

\begin{figure}
\centering
\includegraphics[width=\columnwidth]{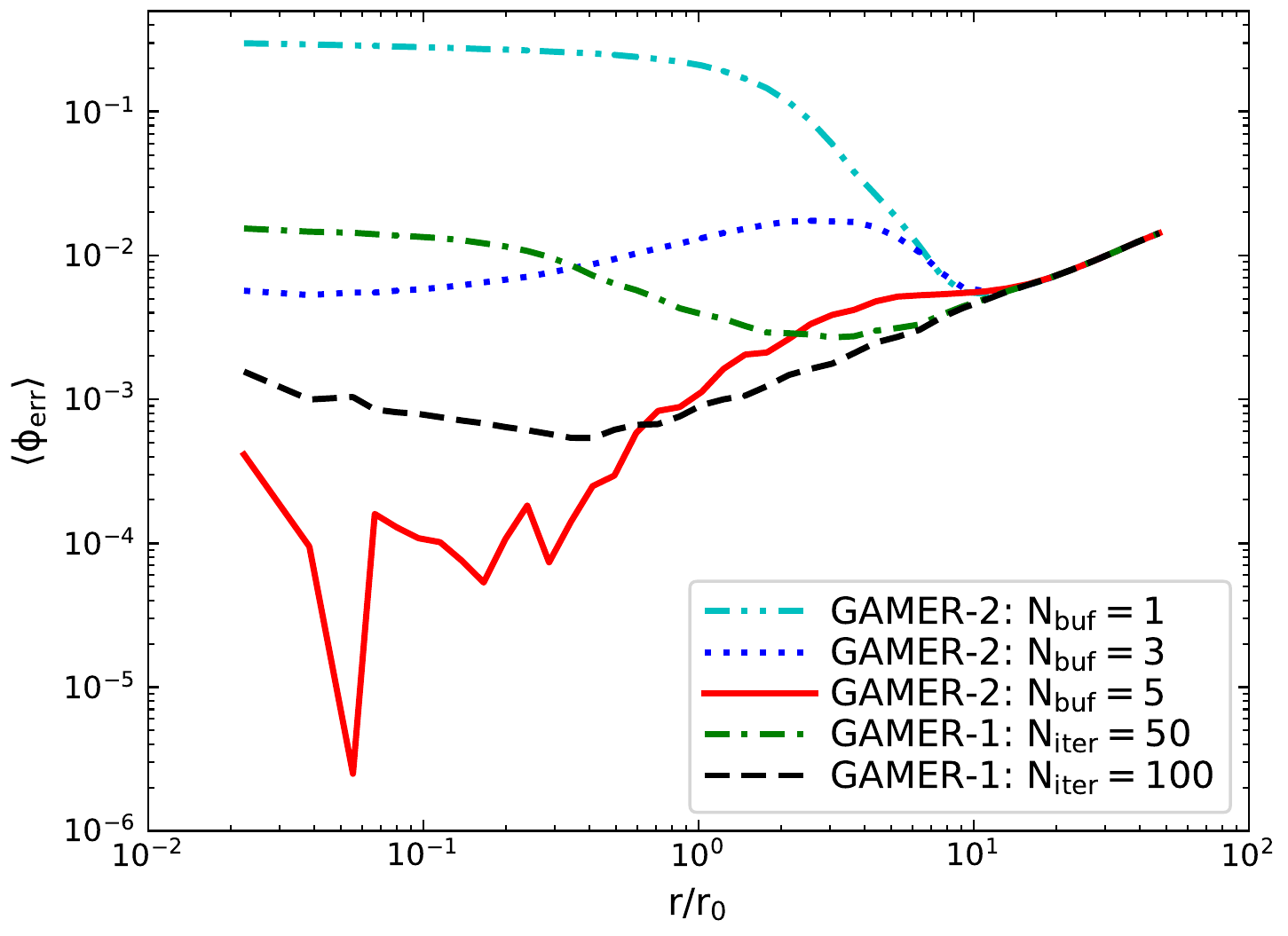}
\caption{
Volume-weighted numerical errors as a function of radius for computing the
potential of a Hernquist profile. We compare the errors of different
schemes: \gamer\ with 1 (dashed-double-dotted line), 3 (dotted line),
and 5 (solid line) buffer zones, and \gamerone\ with 50
(dashed-dotted line) and 100 (dashed line) V-cycle iterations and
sibling relaxation steps (see \citet{Schive2010} for details).
\gamer\ with 5 buffer zones is found to provide the most accurate
solution within $r \lesssim r_0$.
}
\label{fig:poisson_error_profile}
\end{figure}

\fref{fig:poisson_error_profile} shows the volume-weighted radial
profile of the numerical errors of different schemes. We compare
\gamer\ with $\Nbuf=1, 3,$ and $5$ buffer zones and \gamerone\
with $\Niter=50$ and $100$ V-cycle iterations and sibling
relaxation steps (see \citet{Schive2010} for details). It shows that
in these cases the numerical accuracy improves with a larger number
of $\Nbuf$ or $\Niter$, and \gamer\ with $\Nbuf=5$ provides the most
accurate solution within $r \lesssim r_0$. We also find
that $\Nbuf > 5$ does not improve accuracy further. In addition, the
gravitational potential of a patch of $8^3$ cells and 5 ghost zones
on each side consumes $\about 46$ KB memory per patch in double
precision, which can just fit into the small but fast shared memory
of modern GPUs. Therefore, we adopt $\Nbuf=5$ by default.

\bsp
\label{lastpage}
\end{document}